\documentclass[12pt]{article}
\usepackage{formatting}
\usepackage{shortcuts}

\makeatletter
\newcommand\semilarge{\@setfontsize\semilarge{22.72}{27.38}}
\makeatother

\newcommand{\iqft}[0]{$[\text{QFT}]$}
\newcommand{\ieftc}[0]{$[\text{EFT}]$}
\newcommand{\ieftp}[0]{$[\text{EFT}_{\text{proj}}]$}

\begin{document}

\begin{titlepage}
\begin{center}
\phantom{ }
\vspace{1.25cm}

 {\bf \Large{The Baby Universe is Fine and the CFT Knows It:}} \\[.2cm] {\bf \Large{On Holography for Closed Universes}}
\vskip 1.4cm
\large{Stefano Antonini$\,^{\vbarcircle}$, Pratik Rath$\,^{\vbarcircle}$, Martin Sasieta$\,^{\circleeye}
$,\\[.2cm] Brian Swingle$\,^{\circleeye}
$, Alejandro Vilar L\'{o}pez$\,^{\fisheye}$}
\vskip 1cm

$^{\vbarcircle}$\,{\small
{\textit{Leinweber Institute for Theoretical Physics and Department of Physics,}}
\vskip -.1cm
{\textit{ University of California, Berkeley, California 94720, USA}}}\\[.4cm]
$^{\circleeye}$\,{\small{\textit{Martin Fisher School of Physics, Brandeis University, }}
\vskip -.1cm
{\textit{Waltham, Massachusetts 02453, USA}}}\\[.4cm]
$^{\fisheye}$\,{\small{\textit{Department of Physics and Astronomy, University of British Columbia,}}
\vskip -.1cm
{\textit{6224 Agricultural Road, Vancouver, B.C. V6T 1Z1, Canada}}}\\[.4cm]

\vskip .5cm

\end{center}

\begin{abstract}
 \vspace{-.2cm}Big bang/big crunch closed universes can be realized in AdS/CFT, even though they lack asymptotically AdS boundaries. With enough bulk entanglement, the bulk Hilbert space of a closed universe can be holographically encoded in the CFT. We clarify the relation of this encoding to observer-clone proposals and refute recent arguments about the breakdown of semiclassical physics in such spaces. In the limit of no bulk entanglement, the holographic encoding breaks down. The oft-cited one-dimensional nature of the closed universe Hilbert space represents the limitation of the external (CFT) Hilbert space to access the quantum information in the closed universe, similar to the limitations imposed on observers outside a perfectly isolated quantum lab. We advocate that the CFT nevertheless continues to determine the physical properties of the closed universe in this regime, showing how to interpret this relationship in terms of a final state projection in the closed universe. We provide a dictionary between the final state wavefunction and CFT data. We propose a model of the emergence of an arrow of time in the universe with a given initial or final state projection. Finally, we show that the conventional EFT in the closed universe, without any projection, can be recovered as a maximally ignorant description of the final state. This conventional EFT is encoded in CFT data, and it can be probed by computing coarse-grained observables. We provide an example of one such observable. Taken together, these results amount to a clean bill of health for baby universes born of AdS/CFT.
\end{abstract}

\vspace{-.1cm}

{\small  \;\;\;\,A video abstract is available at \url{https://youtu.be/s_9VqF-N8uQ}.}

\vspace{.2cm}

{
\small{\noindent\texttt{E-mail}:
 \href{mailto:santonini@berkeley.edu}{santonini@berkeley.edu}, \href{mailto:pratik_rath@berkeley.edu}{pratik\_rath@berkeley.edu}, \href{mailto:martinsasieta@brandeis.edu}{martinsasieta@brandeis.edu}, \\[.02cm]   \href{mailto:bswingle@brandeis.edu}{bswingle@brandeis.edu}, \href{mailto:alejandro.vilarlopez@ubc.ca}{alejandro.vilarlopez@ubc.ca}}}

\end{titlepage}

\setcounter{tocdepth}{2}
\tableofcontents
\noindent\makebox[\linewidth]{\rule{\textwidth}{0.4 pt}}

\cleardoublepage

%%%%%%%%%%%%%%%%%%%
\section{Introduction}
%%%%%%%%%%%%%%%%%%%%

Much of the recent theoretical progress in our understanding of quantum gravity is built upon the holographic principle \cite{Susskind:1994vu,tHooft:1993dmi}, which asserts in broad strokes that the physics of gravity is essentially lower-dimensional than it naively appears. Black holes provide the classic example, as they are predicted to have an entropy proportional to their area rather than their volume. In practice, the holographic principle is sharpest in the context of the Anti-de Sitter space/conformal field theory (AdS/CFT) correspondence \cite{Maldacena:1997re,Witten:1998qj,Aharony:1999ti}, and much of the progress has been made on questions about quantum gravity in asymptotically Anti-de Sitter spacetimes. However, we do not appear to live in Anti-de Sitter space, so it is important to understand quantum gravity in more general spacetimes. This paper addresses a simple class~\cite{Antonini:2023hdh}\footnote{We suggest the acronym AS$^2$ for \cite{Antonini:2023hdh}.} of cosmological spacetimes with initial and final singularities which can serve as toy models of more realistic Big Bang cosmologies. Using the construction in \cite{Antonini:2023hdh} as a simple theoretical laboratory, our overall goal is to move toward a framework for the holography of closed universes.

First, however, we will step back and set the stage. The study of quantum gravity in more general spacetimes is an independent subject, not a priori bound to AdS/CFT-style holography; nevertheless, in light of the aforementioned progress, it is tempting to leverage some version of holography to try to gain insight into quantum gravity in cosmological settings. Some of the approaches that can be broadly grouped under the umbrella of holographic cosmology include \cite{Strominger:2001pn,Alishahiha:2004md,dsds,Gorbenko:2018oov,Coleman:2021nor,Susskind:2021dfc,Susskind:2021omt,Susskind:2021esx,Shaghoulian:2022fop,McFadden:2009fg,Afshordi:2017ihr,Nastase:2018cbf,Nastase:2019rsn,Maldacena_2004,McInnes:2004nx,McInnes:2005sa,Freivogel:2005qh,Engelhardt:2014mea,Cooper_2019,Antonini:2019qkt,Chen:2020tes,VanRaamsdonk:2020tlr,VanRaamsdonk:2021qgv,antonini2022cosmologyvacuum,Antonini:2022fna,Fallows:2022ioc,Antonini:2022ptt,Ross:2022pde,VanRaamsdonk:2023ion,sahu2023bubblescosmologyadscft,Akers:2022max,Betzios:2024oli,Betzios:2024zhf,Barbon:2025vvh}.

While this general direction holds promise, there are also some immediate confusions.\footnote{Quite apart from the significant challenge of building realistic models that are consistent with actual data.} For example, if the cosmological universe is closed, as might be the case for our universe, one has the issue that there are no obvious boundaries with which to formulate some version of holography.\footnote{One approach is thus to find some non-obvious boundaries, e.g. by using a $T\overline{T}$-deformation~\cite{Coleman:2021nor} or going to imaginary time~\cite{antonini2022cosmologyvacuum,Maldacena_2004}.} Another confusion is that closed universes are similar to perfectly isolated quantum labs, thus bringing a whole host of foundational issues in quantum physics that further complicate matters. 

With these motivations and confusions in mind, we revisit the model studied in \cite{Antonini:2023hdh}, which is a simple example of a semiclassical closed universe built using the tools of AdS/CFT. This work builds on significant prior literature, including the wormholes of \cite{Maldacena_2004} and early work interpreting them cosmologically~\cite{McInnes_2004}, the later ``microstate cosmology'' approach of \cite{Cooper_2019}, subsequent generalizations and related approaches including~\cite{VanRaamsdonk:2021qgv,antonini2022cosmologyvacuum}, and the microstates introduced in \cite{Balasubramanian:2022gmo}.\footnote{This, at least, is how one of the authors (BS) recalls the developments leading to \cite{Antonini:2023hdh}.} And the field continues to move rapidly; for example, results similar to those of \cite{Antonini:2023hdh} were later obtained in a model based on JT gravity \cite{Usatyuk:2024mzs,Usatyuk:2024isz}, and there are many additional developments including \cite{sahu2023bubblescosmologyadscft,VanRaamsdonk:2024kxm,sahu2024holographicblackholecosmologies,Maloney:2025tnn,Maldacena:2024uhs,Ivo:2024ill,Maldacena:2024spf,chakravarty2024newobservableholographiccosmology}.

Our first purpose in revisiting~\cite{Antonini:2023hdh} is to address some points raised in the recent literature.
One of these is \cite{Harlow:2025pvj}, which proposed some rules to treat ``observers'' in the gravitational path integral in closed universes.\footnote{See the independent discussion in \cite{Abdalla:2025gzn} for a different set of rules; we have less to say about these rules here. See also \cite{Akers:2025ahe,Chen:2025fwp,Nomura:2025whc,Blommaert:2025bgd,engelhardt2025observercomplementarityblackholes,higginbotham2025testsbabyuniversesadscft} for recent work on related topics.} Similar rules had already been derived and used in a specific AdS/CFT setup in \cite{Antonini:2023hdh}, but \cite{Harlow:2025pvj} asserted that the physics was essentially different despite identical mathematics. While we agree it is important to understand how the physics of observers interacts with other issues in quantum gravity, we argue that the physical interpretation of \cite{Harlow:2025pvj} is already implicit in \cite{Antonini:2023hdh} and, indeed, in the structure of EFT more generally. We therefore regard \cite{Harlow:2025pvj} as a natural generalization and formalization of this framework to more general models of closed universes, beyond the specific setup of \cite{Antonini:2023hdh}.

The other is \cite{Engelhardt:2025vsp} which, building on a recently raised puzzle~\cite{Antonini:2024mci} based on the construction of \cite{Antonini:2023hdh}, claimed to develop a procedure for testing whether a bulk description including a closed universe is the correct bulk dual to the CFT state introduced in \cite{Antonini:2023hdh}. They claim that the result of the test implies that a semiclassical closed universe is not the correct bulk dual.
As we discuss below, the test boils down to a low-energy boundary swap test, and the result is that the Hilbert space of the closed universe, as seen from the boundary theory, is one-dimensional. This result is interpreted by the authors of \cite{Engelhardt:2025vsp} as a breakdown of a $G\hbar$ expansion, which is how they choose to define semiclassicality. 
However, we argue in detail below that this test does not imply anything about physics in the closed universe, hence it does not provide evidence against semiclassicality of physics in the closed universe.
A similar perspective was in fact adopted in \cite{engelhardt2025observercomplementarityblackholes}, building on the work of \cite{Harlow:2025pvj}.

Our second purpose in revisiting~\cite{Antonini:2023hdh} is to continue to develop a framework for gravitational effective field theory (EFT) in closed universes. We focus first on using AdS/CFT-based constructions as a simple theoretical sandbox, and later propose some principles that may have more general applicability.  
To start, we recall how the tools of AdS/CFT can be used to construct sensible semiclassical closed universes. In particular, there is a large class of CFT questions for which the answer is parsimoniously given in terms of conventional gravitational effective field theory on a large closed universe. Entanglement (in the bulk effective field theory) between the closed universe and a more conventional AdS region plays a crucial but subtle role, in part because the Hilbert space of the closed universe is a confusing concept with multiple useful definitions. When the entanglement is large, one gets a more-or-less conventional encoding of the closed universe in the CFT; when the entanglement is small, the conventional encoding breaks down, but the properties of the closed universe are still reflected in CFT data. Throughout, we argue that gravitational EFT is never breaking down as long as the closed universe remains large; rather, the kinds of CFT questions that the bulk EFT answers change with entanglement. Then, inspired by tensor network models, we formulate a novel dictionary relating the bulk EFT state to CFT data and initiate the study of its relationship to the gravitational path integral. 

A key unifying thread throughout the paper is a tensor network model developed in \cite{Antonini:2023hdh} and refined here (see Section \ref{sec:2} and Appendix \ref{app:MERA}). This model makes many features of the closed universe and its relation to the CFT manifest, and it suggests a framework for an intrinsic description of the closed universe that connects to CFT data. Another guiding theme is the idea that our prior for semiclassicality of a large closed universe should be high, and we should worry about the UV completion afterwards. 

It is also worth stating explicitly again that we are not claiming that AdS/CFT will, in the end, be essential or even useful for describing closed universes in general. However, it does provide a powerful set of tools, and, via constructions like \cite{Antonini:2023hdh}, a simple theoretical playground to search for more general principles.

\subsection{FAQ}

We begin with a conceptual guide to the paper and main results via a series of questions. The responses here and throughout the paper should be viewed as articulating a framework for the holography of closed universes, as exemplified by the construction in \cite{Antonini:2023hdh} and the associated tensor network model.

%%%
\noindent\textbf{Question 1: Can we construct good semiclassical closed universes in AdS/CFT?}

Yes. In Section~\ref{sec:2}, we review the setup of \cite{Antonini:2023hdh} and discuss the full parameter space of solutions of this type. We address in detail throughout the paper why it is reasonable to think that this closed universe has semiclassical physics. We emphasize that the closed universe can be large and can be highly entangled with other disconnected spacetime regions, so there is no a priori reason to doubt that a semiclassical description should work well.

%%%
\noindent\textbf{Question 1.1: What does semiclassical mean in this context?} 

We won't give a comprehensive definition here. At the least, if we imagine the Standard Model inside the closed universe, then the experience of observers in small labs in a large universe will be essentially the same as what we experience on Earth. It also means that there is a large class of CFT questions for which the answer is parsimoniously given in terms of EFT in a large spacetime with weak quantum gravity fluctuations.

%%%
\noindent\textbf{Question 1.2: How can we ``see'' the closed universe most directly from the CFT?}

As part of the construction reviewed in Section~\ref{sec:2}, the tensor network model provides direct evidence for the existence of the closed universe. There is literally a large structure directly related to the CFT state which is interpretable as the space of the closed universe, see { Figures} \ref{fig:setup} and \ref{fig:TNMERA}. This tensor network perspective provides a unifying thread through the paper. 

%%%
\noindent\textbf{Question 1.3: Isn't the tensor network model just a toy model?}

Yes and no. There is no denying that the tensor network leaves out many features/details that would be necessary in a full holographic CFT with an Einstein gravity dual. Nevertheless, such tensor networks have been used to accurately represent states of other simpler CFTs, suggesting that, if the details were properly included, the tensor network could capture a realistic holographic system\footnote{For example, see \cite{geng2025ethmultiintervalentanglementreplica} for some recent progress.}. It also captures some quantitative details of the gravity solution, as discussed in Appendix \ref{app:MERA}. Moreover, the particular thermal entangled construction we employ has been shown to accurately capture thermal physics~\cite{sewell_tmera_2022}.

%%%
\noindent\textbf{Question 2: Can we have observers in this closed universe?}

If the physics is rich enough, of course! Moreover, in the first approximation, the rules of these observers are not mysterious. They are our rules. The Hilbert space of these observers is not mysterious. It is our Hilbert space. On our view, the question is less whether semiclassical physics holds and more how to relate that physics to some UV-complete description like the CFT. We address in detail how this relationship works in Section~\ref{sec:entanglobs}.

%%%
\noindent\textbf{Question 2.1: Should we expect the experiences of the observers in the closed universe to be accessible from the CFT if we believe in AdS/CFT?}

Yes and no. A closed universe has important features in common with an isolated quantum lab. In quantum physics, we cannot know what another observer experiences unless we become dynamically entangled with the same environment which is decohering that observer and leading to effectively classical physics. Hence, we don't expect that the CFT can tell us whether an observer in a closed universe measured a spin to be up or down. What the CFT can calculate includes the potential observations of a bulk observer along with the probabilities of those observations, just as would be the case in ordinary quantum theory when dealing with an isolated lab.\footnote{For now, we are including the Born rule in the rules of quantum theory.} Our view is that the experience of a given observer in the closed universe is not a sensible CFT question, but this does not mean there is anything wrong with AdS/CFT.  In the special case in which the bulk EFT is highly entangled between the closed universe and the AdS spacetimes in \cite{Antonini:2023hdh}, an effectively isometric map between the closed universe EFT Hilbert space and the CFT Hilbert space emerges for a set of semiclassical states as discussed in Section \ref{sec:encodinguniverse}. In this case, the CFT inner product approximately reproduces the inner product relevant to describing the semiclassical physics of the observer and the CFT can be used to simulate such an observer.

%%%
\noindent\textbf{Question 2.2: How does the CFT encode these predictions?}

In the regime where the entanglement with the closed universe is large, the encoding can be obtained via more-or-less conventional tools of entanglement wedge reconstruction, including islands and the like. This is discussed in Section~\ref{sec:entanglobs}. The result is similar to the encoding of the black hole interior at late time. When the entanglement is small or zero, one can still extract the properties of the closed universe from CFT data via other methods as discussed in Sections~\ref{sec:intrinsic} and \ref{sec:EFT}.

%%%
\noindent\textbf{Question 2.3: But doesn't a closed universe have a one-dimensional Hilbert space?}

In a sense, yes, but the interpretation is not obvious from the point of view of observers living in the universe. In particular, we advocate for an interpretation in which the one-dimensional Hilbert space is indicative of the fact that the closed universe is indistinguishable from the outside, e.g. from the perspective of the CFT. This Hilbert space is not the one relevant for the description of the experience of a gravitating observer, who can (and should!) still experience rich physics.\footnote{This point of view is not dissimilar from that advocated for in \cite{Abdalla:2025gzn}. } We address the physics of observers in Section~\ref{sec:entanglobs}, the meaning of the one-dimensional Hilbert space in Section~\ref{subsec:swap}, and further articulate a more intrinsic description in Section~\ref{sec:intrinsic}.

%%%
\noindent\textbf{Question 2.4: But how can you have entanglement with a 1d Hilbert space?}

As above, this depends on what you mean. The entanglement we speak about is literally entanglement in a certain bulk description and an entanglement-like structure in the CFT description. We discuss this in more detail in Section \ref{subsec:swap}. In particular, we address the question of SWAPs as recently discussed in \cite{Engelhardt:2025vsp}.

%%%
\noindent\textbf{Question 3: What happens when the entanglement is zero?}

The CFT still describes the properties of the closed universe, which is all an external party could hope to know about the closed universe. As discussed in Section~\ref{sec:EFT}, one can in principle carry out simple CFT experiments to measure certain matrix elements which are directly encoding properties of the closed universe. These experiments typically involve some amount of statistics, but this does not imply that the closed universe is only a property of the average.

%%%
\noindent\textbf{Question 4: Can we associate a closed universe with a particular CFT state? }

We argue that, yes, each element of a suitable ensemble is associated with a closed universe. This is most clearly seen via the tensor network. But closed universes also arise in the statistical description obtained from averaging over the ensemble of heavy operators used to create the state. If we first average over the heavy operators, then the average state has a simple interpretation in terms of quantum field theory on the closed universe. For multi-replica observables, we need a more complicated procedure involving either wormhole contributions (to compute statistical moments) or a projection (to model individual microstates). This is discussed in detail in Sections \ref{sec:intrinsic} and \ref{sec:EFT} and in the next part of the Introduction.

%%%
\noindent\textbf{Question 5: How does the closed universe setup compare to an evaporating black hole after the Page time?}\footnote{We thank Geoff Penington for a discussion on this point.}

They share many features. Consider the setup in \cite{Antonini:2023hdh} in which bulk fields in a closed universe are entangled with two AdS spacetimes, AdS${}_\mathsf{l}$ and AdS${}_\mathsf{r}$, see Figure \ref{fig:setup}. Suppose the left entanglement entropy is much larger than the right entanglement entropy. The dual description is given by two entangled CFTs (living on the boundary of the two spacetimes), and the closed universe is an entanglement island for the left CFT \cite{Antonini:2023hdh}. This setup has strong parallels to evaporating black holes after the Page time. In particular, in the microscopic (i.e. boundary) description CFT${}_\mathsf{L}$ plays the role of Hawking radiation and CFT${}_\mathsf{R}$ the role of the CFT dual to the black hole spacetime. From the bulk EFT point of view, AdS${}_\mathsf{l}$ plays the role of Hawking radiation in the Hawking state, the closed universe plays the role of the island \cite{Penington:2019npb,Almheiri:2019psf,Penington:2019kki,Almheiri:2019qdq,Akers:2022qdl} for Hawking radiation, and AdS${}_\mathsf{r}$ accounts for the remaining black hole microstates.\footnote{If the bulk entropy between the closed universe and AdS${}_\mathsf{r}$ is $O(1)$, which corresponds to the very late stage of evaporation for the black hole which is now Planckian in size, one might in principle worry that the analogy breaks down, because we lose semiclassical control over the black hole spacetime. Notice, however, that if we have e.g. $S_{\mathsf{r}}=O(G^\alpha)$ with $-1<\alpha<0$ -- which is possible in our setup as we will see in Section \ref{sec:encodinguniverse} -- the corresponding black hole is still under semiclassical control.} The main difference between the two setups is in the nature of the bulk representation of entanglement between the two microscopic systems. In our setup, this is simply bulk entanglement entropy between the closed universe and AdS${}_\mathsf{r}$. In the evaporating black hole case, the entanglement is geometrical, i.e. it is given (at leading order) by the area of the boundary of the island. We believe this distinction to be irrelevant for our holographic discussion; thus, the two setups can be considered very similar.

%%%
\noindent\textbf{Question 5.1: So is the closed universe encoded in the CFT just like the interior is encoded in the Hawking radiation?}

It depends. In the regime in which the bulk entanglement between closed universe and AdS spacetimes is large, yes. In particular, if the entanglement with both AdS${}_\mathsf{l}$ and AdS${}_\mathsf{r}$ is large, this is similar to the interior encoding shortly after the Page time. If the entanglement with AdS${}_\mathsf{l}$ is large and the entanglement with AdS${}_\mathsf{r}$ is small, the situation is analogous to the black hole near the endpoint of evaporation. However, in the black hole setup there is no analog of the case where the entanglement with both AdS${}_\mathsf{l}$ and AdS${}_\mathsf{r}$ is small, because the naive bulk entanglement between the interior and Hawking radiation always stays large.

%%%
\noindent\textbf{Question 6: With all this in mind, how is the paper structured? }

In Section \ref{sec:2} we review and expand on the construction of \cite{Antonini:2023hdh}. Highlights include the observation that the bulk entanglement can be made large, scaling like $G^{-\alpha}$ for $\alpha \leq \alpha_{\max}$ where $\alpha_{\max}<1$ depends on the theory, and the tensor network model (further refined in Appendix \ref{app:MERA}).

In Section \ref{sec:holo_enc} we discuss the holographic encoding of the closed universe in the regime of large entanglement. Highlights include a demonstration of the equivalence between the rules of \cite{Antonini:2023hdh} and those discussed in \cite{Harlow:2025pvj}, a discussion of the physics of decoherence and pointer states in the context of EFT and the models in \cite{Antonini:2023hdh}, and a discussion of the recent claims about SWAP operators in \cite{Engelhardt:2025vsp}.

In Section \ref{sec:intrinsic} we propose a more intrinsic description of the closed universe which is inspired by the tensor network model and which is sensible even at low entanglement. Highlights include a discussion of the crucial role of projections and their relation to the final state proposal, a proposed dictionary relating bulk states and CFT data, and a toy model of the emergence of the arrow of time.

In Section \ref{sec:EFT} we show how the results of a more basic form of bulk EFT can be recovered by averaging over projections/microstates. Highlights include a discussion of how this averaging manifests in the gravitational path integral and the exhibiting of simple CFT experiments that reveal properties of the closed universe even at zero entanglement.

In Section \ref{sec:outlook} we summarize our main results and perspective, and outline some directions for further work. Appendix \ref{app:MERA} elaborates on the MERA-based model introduced in Section \ref{sec:2}. Appendix \ref{app:UniverseBottleneck} elaborates on the discussion in Section \ref{sec:entanglobs} via a model with a geometrical bottleneck. Appendix \ref{app:gravargument} elaborates on the discussion in Section \ref{sec:intrinsic} using tools from canonical quantum gravity.

\subsection{Comments on perspectives}
\label{sec:notation}

For later use, we also spell out several useful perspectives that represent both specific computational frameworks and approaches to interpreting the physical meaning of the closed universe. 

\textbf{\iqft:} We use \iqft\, to mean quantum field theory on curved spacetime, potentially including perturbative quantum gravity corrections. This is a well-defined physical theory, at least in perturbation theory, with a large Hilbert space even when defined on a compact space. In this way of thinking, we are not solving the full gravitational constraints, implementing any holographic encoding map, or anything else. This setting provides the computational ingredients for the other perspectives mentioned below, e.g. we use the classical action of the bulk quantum field theory to evaluate saddle point contributions to the gravitational path integral (GPI) and we compute one-loop determinants in the same quantum field theory to study corrections to the saddle point.

\textbf{\ieftc:} We use \ieftc\, (or just EFT) to mean gravitational effective field theory, potentially including non-perturbative quantum gravity corrections via subleading saddle points of the GPI. This perspective acknowledges that the bulk states considered in \iqft\, cannot be taken seriously as microscopic states, at least without invoking some holographic map. However, we emphasize that \iqft, viewed as an ingredient in evaluating the GPI in \ieftc, is still an essential part of the story. 

To illustrate these notions, consider the case of black hole evaporation. We view Hawking's calculation as an example of \iqft. The entropy of the bulk state of Hawking radiation in \iqft\, is interpreted as the entropy of a coarse-grained or averaged microscopic state, $S(\overline{\rho})$. We view the replica wormhole calculation as an example of \ieftc. The entropy obtained from this computation is interpreted as the average of the entropy of microscopic states, $\overline{S(\rho)}$ \cite{Bousso:2020kmy}. We emphasize that suitable notions of averaging or coarse-graining are crucial to the interpretation of both calculations and that neither perspective really purports to give access to the microscopic state. 

\textbf{\ieftp:} Finally, we use \ieftp\, to indicate the inclusion of an explicit final state projection (possibly associated with the future and past singularities in our case). This perspective incorporates greater access to the microscopic state in the semi-classical description. Depending on how we average, we recover either \iqft\, or \ieftc. One may speculate that this perspective is related to the further inclusion of ``half-wormholes''~\cite{saad2021wormholesaveraging} in the GPI.

These perspectives have particular bearing on how we interpret the closed universe and its relationship to microscopic states in the CFT. For example, \iqft\, has a simple notion of entanglement with a closed universe, but the bulk state in this perspective has nothing directly to do with any CFT state - in particular, it lacks the erratic features that should be present in the microscopic state. Just as in the black hole case, the entropy of the bulk state in \iqft\, turns out to be the entropy of a coarse-grained state. Similarly, the entropy computed using \ieftc\, can be interpreted as a microscopic entropy but averaged over microstates. Both \iqft\, and \ieftc\, leave open the question of whether the closed universe ``exists'' for a single microstate. This question is clarified by \ieftp, as we discuss in detail in the later parts of the paper.

%%%%%%%%%%%%%%%%
\section{Setup}
\label{sec:2}
%%%%%%%%%%%%%%%%%%

Here we set the stage by reviewing and expanding on the central aspects of the AS$^2$ model \cite{Antonini:2023hdh}.

\subsection{Semiclassical state}
\label{sec:semisetup}

\begin{figure}[h]
    \centering
    \includegraphics[width=.98\linewidth]{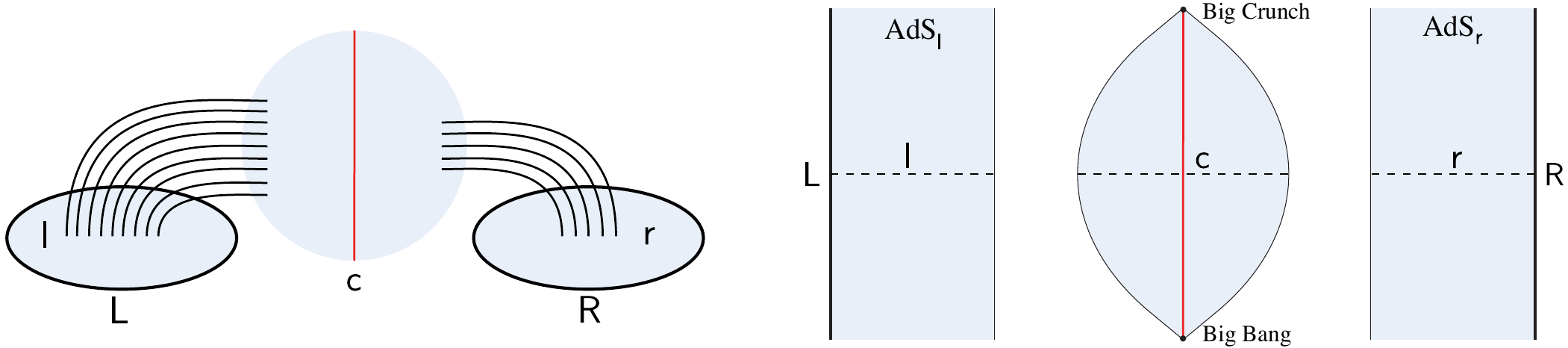} 
    \caption{On the left, representation of the semiclassical state on the time-symmetric slice. The lines connecting $\mathsf{l},\mathsf{r}$ and $\co$ correspond to entanglement lines of matter. On the right, Penrose diagram of the spacetime, where each point represents a sphere $\mathbf{S}^{d-1}$. The dashed line is the time-symmetric slice. The thin black lines are $r=0$ where the spheres cap off smoothly.}
    \label{fig:setup}
\end{figure}

The semiclassical state is defined on two asymptotically global AdS slices (with AdS scale $\ell$), denoted by left ($\mathsf{l}$) and right ($\mathsf{r}$), respectively. Additionally, the system includes a spatially closed universe  ($\co$). The initial data on the time-symmetric slice of the universe is taken spherically symmetric
\be\label{eq:solution1}
\text{d}s^2_{\co} = \ell^2 \text{d}\rho^2 + r(\rho)^2 \text{d}\Omega_{d-1}^2\,,
\ee
where $\rho \in [-\rho_0,\rho_0]$ is a { dimensionless radial coordinate, with proper radial distance measured by $\ell\,\mathrm{d}\rho$}. The geometry caps off smoothly at both ends $r(\pm\rho_{0}) =0$ and the universe has the spatial topology of a sphere $\mathbf{S}^{d}$. The function $r(\rho)$ is determined by the Hamiltonian constraint given some distribution of energy. 

In the simplest model, the closed universe is supported by a negative cosmological constant (of the same value as the one in the AdS regions) and a spherical thin shell of dust particles whose only parameter is its rest mass $m$. In this case, the profile of the geometry is 
\be\label{eq:solution2}
r(\rho) = \ell \sinh(\rho_0-|\rho|)\,.
\ee
The stress-energy of the localized shell at $\rho=0$ serves to glue the two vacuum initial data. The gluing determines the radius of the shell $R_* \equiv r(0) $ as a function of its rest mass (for details, see section $2$ of \cite{Antonini:2023hdh})
\be\label{eq:Rstar}
\dfrac{R_*}{\ell_P} \sim \begin{cases}
    (m\ell_P)^{\frac{1}{d-2}}\quad \quad \text{if }\; O(N^{\frac{2}{d-1}})\ll m\ell \ll O(N^2)\,,\\[.2cm]
    (m\ell)^{\frac{1}{d-1}}\quad \quad \hspace{.3cm} \text{if } \;O(N^2)\ll   m\ell\,,
\end{cases}
\ee
where $\ell_{ P} = (G_{ N})^{\frac{1}{d-1}}$ is the $(d+1)$-dimensional Planck length and, for future convenience, we have introduced the parametric scaling with the CFT central charge $N^2 \sim (\ell/\ell_P)^{d-1}$. The spatial volume of the closed universe is uniquely determined by the rest mass of the heavy shell
\be\label{eq:volume} 
\dfrac{\text{Vol}(\co)}{G_N \ell_P}\sim   \begin{cases}
    (m\ell_P)^{\frac{d}{d-2}}  \quad \quad\quad   \text{if }\; O(N^{\frac{2}{d-1}})\ll m\ell \ll O(N^2)\,,\\[.2cm]
      \dfrac{m \ell^2}{\ell_P}\quad \quad\quad  \hspace{.9cm} \text{if }\; O(N^2)\ll   m\ell\,.
\end{cases} 
\ee 
Thus, provided that the shell is heavy, $m\ell_P \gg 1$, or equivalently that $m\ell\gg O(N^{\frac{2}{d-1}})$, the universe contains a macroscopic volume. The mass density of the shell remains parametrically smaller than the Planck density. In fact, given that we are implicitly thinking about theories containing perturbative strings, such as the dual of $\mathcal{N}=4$ super Yang-Mills, we will also have a larger scale, the string scale $\ell_s$, which sets the limit of what we can call a weakly curved semiclassical universe. We will later determine the stricter lower bound that $m\ell$ must satisfy for the closed universe to remain semiclassical in string theory in AdS$_5\times \mathbf{S}^5$.

Under Lorentzian evolution, the spacetime includes the two AdS regions, while $\co$ evolves into a big bang/big crunch cosmological spacetime,\footnote{We remark this is not a typical cosmological spacetime with evolution driven by some homogeneous (and possibly isotropic) perfect fluid. However, it does contain a past and future curvature singularity, with its time-symmetric evolution driven by the dust shell's trajectory.} as shown on the right of Figure~\ref{fig:setup}. Moreover, on top of the classical geometry on the reflection-symmetric slice, there is a state of matter fields $\ket{\psi_{\mathsf{clr}}}$ in the \iqft\, description, which is generally entangled between the universe and the AdS regions.

\subsection{CFT microstate and tensor network model}
\label{sec:CFTmicroTN}

The conventional holographic description of this setup corresponds to a (pure) microstate $\ket{\Psi_{\mathsf{LR}}}$ of two holographic CFTs, denoted by left ($\mathsf{L}$) and right ($\mathsf{R}$), associated with the AdS asymptotic boundaries. Let us start from a Hilbert space way of constructing this microstate.

Let $\ket{\Psi_{AB}}$ be the CFT state corresponding to an entangled but geometrically disconnected semiclassical state $\ket{\psi_{ab}}$ between two AdS spaces. We consider the product of two such states, $\ket{\Psi_{AB}} \ket{\Psi_{CD}'}$, defined on four copies of the holographic CFT. Then we apply to the $C$ factor an operator ${ \Op^\dagger}$ that creates the shell, and finally we project into a maximally entangled state $\ket{\text{MAX}_{BC}}$, effectively gluing the $B$ and $C$ factors. Identifying $A = \mathsf{L}$ and $D = \mathsf{R}$ leads to the microstate
\be\label{eq:microgluing}
 \ket{\Psi_{\mathsf{LR}}} \,\propto\, \bra{\Op_{BC}}\left(\ket{\Psi_{\mathsf{L}B}} \ket{\Psi'_{C\mathsf{R}}}\right)\,,
\ee
where $\ket{\Op_{BC}} = \Op_C\ket{\text{MAX}_{BC}}  $.

\begin{figure}[h]
    \centering
    \includegraphics[width=.4\linewidth]{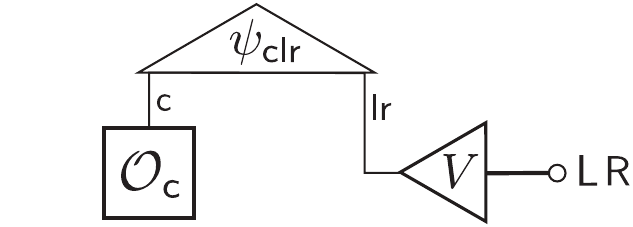} 
    \caption{Tensor network model in \cite{Antonini:2023hdh} for the microstate. }
    \label{fig:TN}
\end{figure}

The initial bulk states of two AdS spacetimes are mapped via holographic isometries $V$ to the boundary, $\ket{\Psi_{\mathsf{L}B}} = V_{\mathsf{L}B}\ket{\psi_{\mathsf{l}b}}$ and $\ket{\Psi'_{C\mathsf{R}}} = V_{C\mathsf{R}}\ket{\psi'_{c\mathsf{r}}}$. In each case, the isometry factorizes into the product of isometries for each AdS factor, $V_{\mathsf{L}B} = V_\mathsf{L} \otimes V_B$ and $V_{C\mathsf{R}} = V_C \otimes V_{\mathsf{R}}$. Using \eqref{eq:microgluing}, this allows us to write the microstate as
\be
\ket{\Psi_{\mathsf{LR}}} \; \propto \; V_{\mathsf{L}} V_{\mathsf{R}} \bra{\mathcal{O}_{BC}} V_B V_C \ket{\psi_{\mathsf{l}b}} { \ket{\psi'_{c\mathsf{r}}}} \, .
\ee
Assume for now the shell is taken to be heavy, $m \ell \gg N^2$, so that by \eqref{eq:Rstar} it is located in the asymptotic region $R_{\star} \gg \ell$. In this limit, the state of the closed universe will factorize between both sides of the shell, and in fact, the full bulk state is just $\ket{\psi_{\mathsf{clr}}} \approx \ket{\psi_{\mathsf{l}b}} { \ket{\psi'_{c\mathsf{r}}}}$, where the degrees of freedom in the closed universe $\co$ are identified with those of $bc$. We can then write the map from the bulk state to the two boundary CFTs as
\be\label{eq:bulktobdy} 
 \ket{\Psi_{\mathsf{LR}}} \;\propto\; V_\mathsf{L} V_\mathsf{R} \bra{\Op_{\co}} \ket{\psi_{\mathsf{clr}}}\,,
\ee
where $\ket{\Op_{\co}} = V_B^\dagger V_C^{\dagger} \ket{\Op_{BC}}$ is a state of the closed universe (determined by the shell operator) onto which we are projecting. A tensor network model of \eqref{eq:bulktobdy} is shown in Figure \ref{fig:TN}, with $V = V_{\mathsf{L}} V_{\mathsf{R}}$. In \cite{Antonini:2023hdh}, the state $\ket{\Op_{\co}}$ was drawn from a Gaussian random distribution; with this choice, this very simple random tensor network model captures many of the properties of the gravitational setup.

\begin{figure}[h]
    \centering
    \includegraphics[width=.4\linewidth]{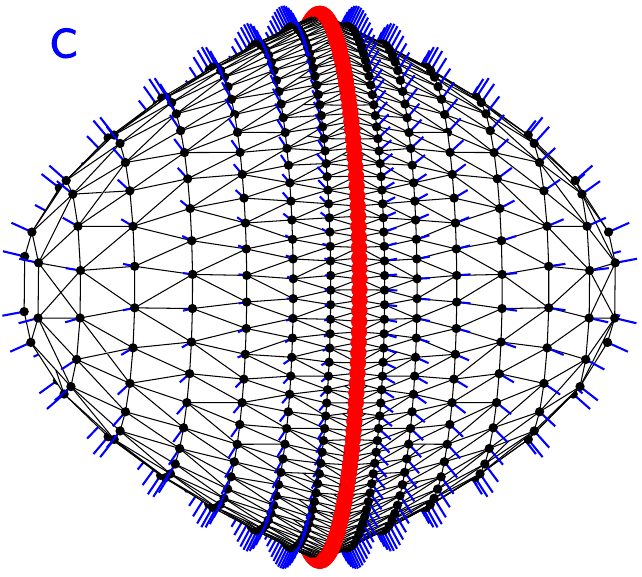} 
    \caption{AdS-local tensor network model of the closed universe state $\ket{\Op_{\co}}$ in the bulk Hilbert space. The tensor network is constructed from two MERA codes with large bond dimension, glued at a finite radius $R_*$ with the insertion of the shell operator (red). In section \ref{sec:intrinsic} we will interpret this state as the initial/final state of the closed universe.}
    \label{fig:TNMERA}
\end{figure}

In Appendix \ref{app:MERA}, we develop the model shown in Figure \ref{fig:TNMERA}, based on the MERA tensor network for the state $\ket{\Op_{\co}}$. Readers may think of this as an elaboration of the holographic map in Figure \ref{fig:TN} including the AdS-scale geometric structure of the closed universe. In this refined model, it becomes easier to understand what happens when we go away from the heavy shell limit. We leave the details to the Appendix, but the outcome is that \eqref{eq:bulktobdy} is essentially unmodified, provided we use an appropriate bulk state of the closed universe $\ket{\psi_{\mathsf{clr}}}$ (which is no longer a product state between both sides of the shell).

\subsection{Euclidean preparation}
\label{sec:Euclideanprep}

\begin{figure}[h]
    \centering
    \includegraphics[width=.98\linewidth]{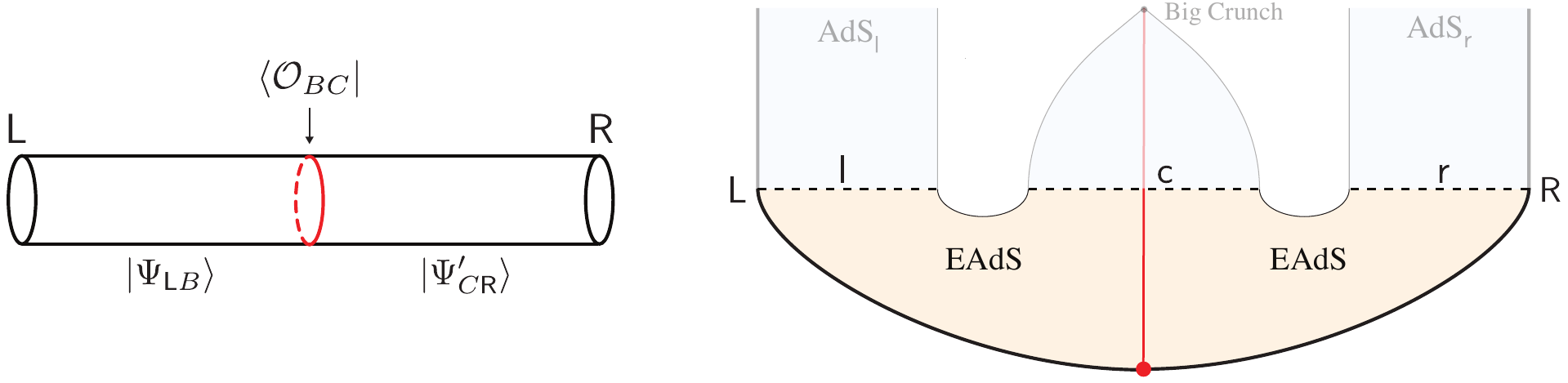} 
    \caption{On the left, Euclidean CFT path integral on $\mathbf{R}\times \mathbf{S}^{d-1}$ which prepares the microstate. On the right, the leading Euclidean gravitational solution which prepares the semiclassical state. Neglecting the backreaction of matter except for the heavy shell, the manifold is composed of two Euclidean AdS spaces (EAdS), cut and glued across the trajectory of the shell.}
    \label{fig:Euclideanprep}
\end{figure}

We now explain the Euclidean CFT path integral preparation of this microstate, as represented on the left of Figure \ref{fig:Euclideanprep}. We will assume that the states $\ket{\Psi_{AB}}$ and $\ket{\Psi_{CD}'}$ can be prepared by suitable Euclidean CFT path integrals on a cylinder, generally via a time-dependent Hamiltonian evolution.\footnote{Where necessary, we can appeal to a microcanonical path integral to change what saddles dominate~\cite{Marolf_2018}.} We can then connect both contours together through the $B$ and $C$ ends, inserting the shell operator $\Op^{\dagger}$ in between them. Identifying $A = \mathsf{L}$ and $D = \mathsf{R}$ leads to the microstate \eqref{eq:microgluing}.

In the simplest example of low-temperature TFD states $\ket{\text{TFD}_{AB}},\ket{\text{TFD}'_{CD}}$, at inverse temperature $\beta$, the Euclidean preparation is achieved by evolving with the CFT Hamiltonian $H$, $\exp(-\frac{\beta}{2} H)$. In this case, the state prepared by the CFT path integral including the shell operator insertion ``gluing'' { $B$ and $C$} is then given by a partially entangled thermal state (PETS) \cite{Goel:2018ubv,Antonini:2023hdh}
\begin{equation}
    |\text{PETS}\rangle=\frac{1}{\sqrt{Z}}\sum_{n,m}e^{-\frac{1}{2}\left(\tilde{\beta}_{\mathsf{L}} E_n+\tilde{\beta}_{\mathsf{R}}E_m\right)}\mathcal{O}_{nm}\ket{E_n}\ket{E_m},
    \label{eq:PETS}
\end{equation}
where the $\tilde{\beta}_{\mathsf{L}},\tilde{\beta}_{\mathsf{R}}$ determine the amount of Euclidean time evolution to the left and right of the shell insertion.\footnote{To keep with the conventions common in the literature on PETS, the operator $\mathcal{O}$ is the adjoint of the one used in \eqref{eq:microgluing}. Later sections using PETS (such as \ref{sec:coarseCFT}) are also written with this convention.} In Section \ref{sec:coarseCFT}, we will focus our attention on PETS states, but in the rest of the paper we will keep the discussion more general to include generic states prepared by time-dependent Hamiltonian evolution.

The semiclassical state is prepared by a Euclidean gravitational solution with boundary conditions fixed by the corresponding CFT path integral, as shown in Figure \ref{fig:Euclideanprep}. The Euclidean manifold is connected, which means that it generally prepares an entangled state of the bulk matter fields between the closed universe and the AdS regions. The actual bulk state for the matter is determined by the semiclassical sources preparing the states on the cylinder. If the mass of the shell is sufficiently large, its trajectory always remains in the asymptotic AdS region in the Euclidean diagram of Figure \ref{fig:Euclideanprep}, independent of other matter fields. Therefore, the Euclidean manifold effectively factorizes into two Euclidean manifolds. The presence of the shell does not modify that the dominant Euclidean saddle point is the one in Figure \ref{fig:Euclideanprep} provided that $\ket{\Psi_{\mathsf{L}B}}$ and ${ \ket{\Psi'_{C\mathsf{R}}}}$ correspond to entangled states of two gases of particles in two disconnected AdS factors. Also, in this case, the bulk entanglement becomes mostly bipartite between each AdS and each side of the universe, reflecting the fact that $\ket{\psi_{\mathsf{clr}}} \approx \ket{\psi_{\mathsf{l}b}} { \ket{\psi'_{c \mathsf{r}}}}$. 

An important property of the microstates is that the heavy CFT operator is ``screened'' by long Euclidean evolutions, which makes the energy of the state $\ket{\Psi_{\mathsf{LR}}}$ not scale with its (large) scaling dimension.
In the semiclassical picture, this corresponds to the fact that the heavy shell lives in the closed universe and thus it does not contribute to the ADM mass. For literature that helps better understand the background and details of this construction, we refer the reader to \cite{Antonini:2023hdh,Kourkoulou:2017zaj,Goel:2018ubv, Penington:2019kki,Chandra:2022fwi,Lin:2022zxd,Sasieta:2022ksu,Balasubramanian:2022gmo,Balasubramanian:2022lnw,Boruch:2023trc,Climent:2024trz,Barbon:2025bbh}.

\begin{figure}[h]
    \centering
    \includegraphics[width=.8\linewidth]{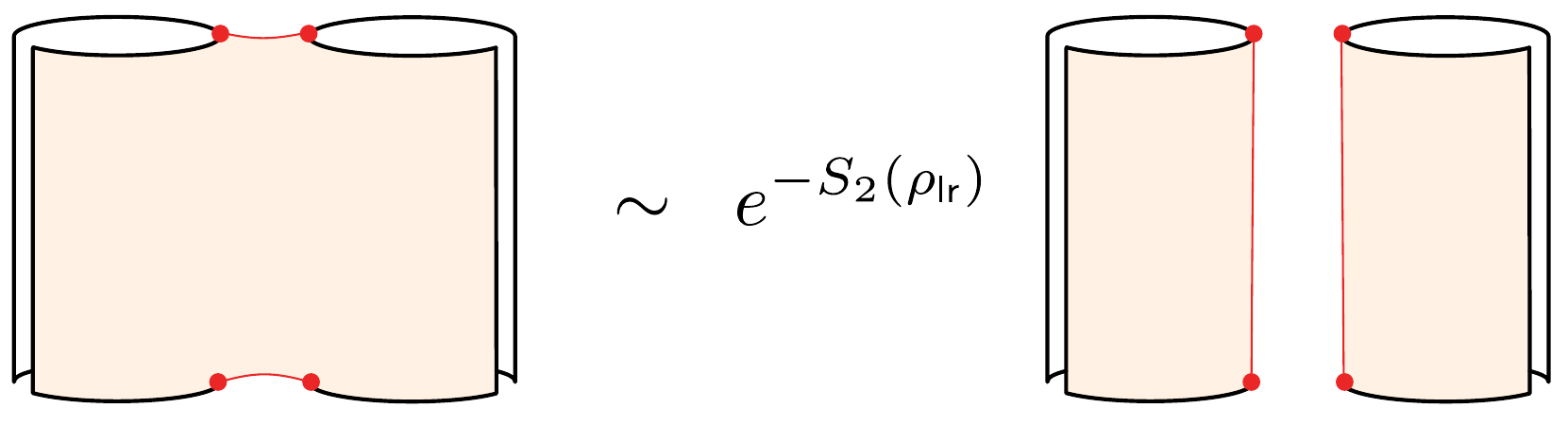} 
    \caption{The Euclidean wormhole reproducing the variance of the norm of the microstate over an ensemble of Gaussian microscopic thin shell operators. The boundary vertical black lines connect the bra and ket contours in the CFT Hilbert spaces, where all of the microscopic states live. For a detailed construction of this wormhole, see \cite{Antonini:2023hdh}.}
    \label{fig:variance}
\end{figure}

\subsubsection{On the role of operator statistics}

For the heavy shell, the gravitational path integral computes statistical moments of microscopic quantities over an ensemble of CFT operators that satisfy ETH-like microscopic ansatze \cite{Sasieta:2022ksu} (see also \cite{Saad:2019lba,Saad:2019pqd,Pollack:2020gfa,Belin:2020hea,Stanford:2020wkf,Chandra:2022bqq, Balasubramanian:2022gmo,deBoer:2023vsm,Chen:2024hqu,Chandra:2024vhm,Liu:2025ztq}). The most conservative interpretation of this is that there is no fundamental average; semiclassics is just an approximation to an actual microscopic operator. This is most useful in the conventional regime in which the statistical fluctuations of quantities are suppressed. As represented in Figure \ref{fig:variance}, the two-replica wormhole of \cite{Antonini:2023hdh} predicts a statistical variance to the norm of the state $Z = \|\bra{\Op_{BC}}\left(\ket{\Psi_{\mathsf{L}B}} \ket{\Psi'_{C\mathsf{R}}}\right)\|^2$ given by
\be\label{eq:whvariance} 
\dfrac{\overline{Z^2}-\overline{Z}^2 }{\overline{Z}^2} \sim  e^{-S_2(\rho_{\mathsf{lr}})}\,,
\ee 
where $S_2(\rho_{\mathsf{lr}})$ is the second R\'enyi entropy of the state of the AdS regions $\rho_{\mathsf{lr}} = \text{Tr}_{\co}(\ket{\psi_{\mathsf{clr}}} \bra{\psi_{\mathsf{clr}}})$. This expression agrees with the variance-to-norm squared ratio of $Z$ over an ensemble of microscopic thin shell operators. As long as the entanglement to the closed universe is large, the semiclassical description reliably captures properties of the CFT states, and thus it can be used to approximate the dual to individual microstates. If the entanglement is small, we cannot a priori conclude that the semiclassical description of the closed universe breaks down, only that it is not a correct picture for the CFT microstates. We will postpone to Sections \ref{sec:intrinsic} and \ref{sec:EFT} a detailed proposal for how to make sense of the regime of low entanglement.

The ETH ansatz is also the reason to select a random tensor $\ket{\Op_{\co}}$ in the tensor network of Figure \ref{fig:TN}; with this choice, the model reproduces \eqref{eq:whvariance} and other gravitational properties \cite{Antonini:2023hdh}.

\subsection{Parameter space in AdS\texorpdfstring{$_5\times \mathbf{S}^5$}{5xS5}}
\label{sec:ParameterSpace}

As emphasized in \cite{Antonini:2023hdh}, a feature which makes this setup particularly interesting is that it allows us to continuously vary the entanglement to the closed universe and reach a regime in which its holographic description should be fairly conventional. This is possible because the AdS spaces can accommodate a parametrically large entropy in the stable gas phase. 

We will be more quantitative in the top-down model of the type IIB string theory in AdS$_5\times \mathbf{S}^5$ dual to the holographic $\mathcal{N}=4\; SU(N)$ SYM theory with { 't Hooft} coupling $\lambda = g_s N$, where $g_s$ is the string coupling. In this case, we think of the dust particles that form the shell as coming from massive KK supergravity modes on the $\mathbf{S}^5$.\footnote{A dust particle is chosen to be a singlet state under the isometry group $SO(6)$ built from multiple KK modes. The corresponding CFT operator is uncharged under $R$-symmetry.}  Since the excitations are weakly interacting, the rest mass of the shell is essentially $m\ell  \approx n \Delta$, where $n$ is the number of particles and $\Delta\gg 1$ is the conformal dimension of the single-particle CFT primary. To build specific solutions, we can assume that the mass of each dust particle is small enough, $\Delta \ll N^2$, so that the backreaction on the $\mathbf{S}^5$ factor is negligible. The AdS part of the metric on the closed universe is then determined by \eqref{eq:solution1} and \eqref{eq:solution2}, again neglecting the backreaction of other matter fields except for the shell.

\begin{figure}[h]
    \centering
    \includegraphics[width=\linewidth]{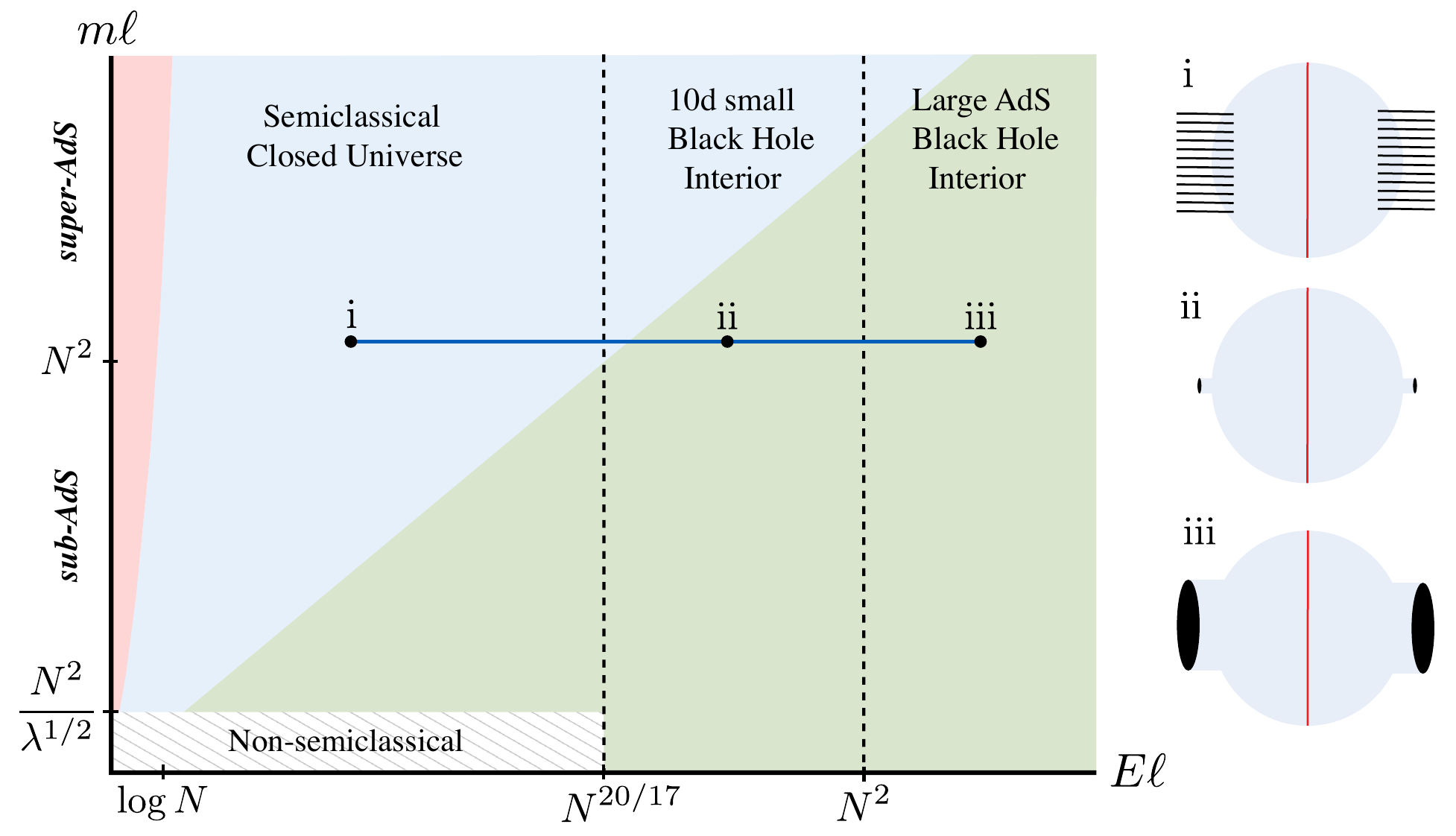} 
    \caption{Parameter space of the semiclassical states in AdS$_5 \times \mathbf{S}^5$.}
    \label{fig:phasediagram}
\end{figure}

Restricting to a symmetric situation between $\mathsf{l}$ and $\mathsf{r}$, there are two independent parameters that determine the features of the semiclassical state, as represented in Figure \ref{fig:phasediagram}:
\begin{enumerate}
    \item The rest mass $m\ell $ of the heavy operator. This determines the geometric volume of the closed universe. The AdS$_5$ part of this quantity is given by \eqref{eq:volume}.
    \item The ADM mass $E\ell $ of the gases of particles in the AdS spaces.
\end{enumerate}
The universe will be weakly curved and under semiclassical control if its { characteristic radius is larger than the string length}, $R_*\gg \ell_s$. Equivalently, using \eqref{eq:Rstar} and writing the parameters in terms of $\lambda$ and $N$ (recall that $\ell_P$ is the five-dimensional Planck length), this is true as long as $m \ell \gg O\left({N^2}/{\lambda^{1/2}}\right)$. Within this regime, the two parameters $m \ell$ and $E \ell$ individually determine the coarse-grained entropies of the closed universe and the AdS spaces, respectively.

Let us explain the three different phases separated by vertical dashed lines in Figure \ref{fig:phasediagram} (we leave the explanation of the coloring for Section \ref{sec:encodinguniverse}). At low energies, we have AdS spaces which can accommodate gases of particles with an entropy
\be\label{eq:gasentropy}
S(\mathsf{lr}) \sim (E\ell)^{9/10}\,,
\ee 
corresponding to the microcanonical entropy of the ten-dimensional gas in a box of size $\ell$. The entropy \eqref{eq:gasentropy} upper bounds the allowable entanglement between $\co$ and $\mathsf{lr}$. This is true up to energies $E\ell \lesssim O(N^{20/17})$, where the gas of particles ceases to be the dominant microcanonical phase \cite{Horowitz:1999uv}. We want to emphasize that close to this threshold, the entanglement to the universe can be made parametrically large in $N$, since at the transition point $S(\mathsf{lr}) \sim O(N^{18/17})$. Above the threshold, the dominant phase is a ten-dimensional black hole localized in the $\mathbf{S}^5$. In this regime, the universe becomes geometrically connected to the AdS spaces as part of the black hole interior (ii).\footnote{We assume that there is a tunable parameter on the states $\ket{\Psi_{\mathsf{L}B}}$ and $\ket{\Psi_{C\mathsf{R}}}$ in \eqref{eq:microgluing} that changes their energy. Above the threshold, these states correspond to high-energy equilibrium states of two-sided small black holes with connected interiors. We will not contemplate how these states are prepared by { a} Euclidean path integral like Figure \ref{fig:Euclideanprep} in this regime.} We are taking the string coupling to be sufficiently large, $g_s \gg N^{-9/17}$, so that there is no stringy transient between these two phases \cite{Horowitz:1996nw,Horowitz:1997jc,Barbon:2004dd}. As the energy is increased above $E\ell \gtrsim O(N^2)$, the black hole becomes large, of positive specific heat (iii).

The other relevant notion of entropy is that of the matter in the universe. Here we do not have a restriction on the ADM energy, as the universe is closed, but we nevertheless require the state at the time-symmetric slice to be particle gas-like. From the considerations above, we will assume that the universe can accommodate $N^{18/17}$ matter degrees of freedom per AdS scale volume; this defines our code subspace within the closed universe. For the purposes of the green coloring explained in the next section, it is important that we will not consider the heavy shell as a part of this code subspace, but rather as part of a fixed background which holds the universe together. This gives an estimate for the entropy 
\be 
S(\co)  \sim \begin{cases}
     \dfrac{(m\ell)^2}{N^{50/17}}\quad \quad\quad  \hspace{.4cm} \text{if }\; O(N^{2}/\lambda^{1/2})\ll m\ell \ll O(N^2)\,,\\[.2cm]
      \dfrac{m \ell}{N^{16/17}}\quad \quad\quad  \hspace{.6cm} \text{if }\; O(N^2)\ll   m\ell\,,
\end{cases}
\ee
where, in the first line, we are using the extensivity of entropy at sub-AdS scales.

\subsection{Regimes of holographic encoding}
\label{sec:encodinguniverse}

\begin{figure}[h]
    \centering
    \includegraphics[width=.44\linewidth]{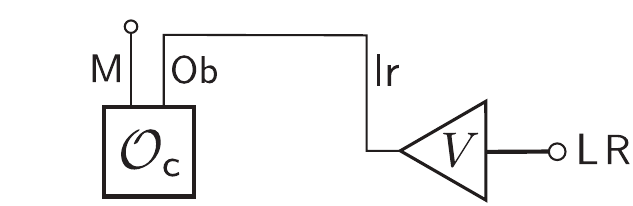} 
    \caption{The holographic encoding of the closed universe $W$ is defined through the fixed $\mathsf{Ob}$-$\mathsf{lr}$ bulk entanglement. }
    \label{fig:code}
\end{figure}

We now explain the different regimes of holographic encoding of the closed universe, corresponding to the coloring of Figure \ref{fig:phasediagram}. In order to do this, we will use the tensor network model of Figure \ref{fig:TN} as a guide. We will use the notation $|\cdot| = { e^{S(\cdot)}}$ for the bond dimension in this qubit model{ , with entropies measured using natural logarithms}. To define an encoding map, we fix the entanglement of the state $\ket{\psi_{\mathsf{clr}}}$ to the universe by distilling it in terms of a collection of EPR pairs $\ket{\text{MAX}_{\mathsf{Oblr}}} \propto \sum_{i=1}^{|\mathsf{lr}|} \ket{i_{\mathsf{Ob}}^*}\ket{i_{\mathsf{lr}}}$ between the $\mathsf{lr}$ AdS spaces and the corresponding degrees of freedom in the universe, which we shall call $\mathsf{Ob}$ for reasons that will become apparent in the next section. This amounts to separating the rest of the unentangled degrees of freedom in the universe, which we will denote $\mathsf{M}$, so that $\co = \mathsf{MOb}$. Allowing for general states in $\mathsf{M}$, this defines the holographic encoding map $W: \mathcal{H}_{\mathsf{M}}\rightarrow \mathcal{H}_\mathsf{LR}$, given by
\be 
W \,=\,\dfrac{1}{\sqrt{|\mathsf{lr}|}} V \bra{\Op_{\co}} \ket{\text{MAX}_{\mathsf{Oblr}}}\,,
\ee 
shown in Figure \ref{fig:code}. Recall that in this model $\ket{\Op_{\co}}$ is a Gaussian random state and $V$ is the product of two conventional holographic isometries of two separate AdS spaces. 

In what follows, we will use colors for the headings associated with the coloring of Figure \ref{fig:phasediagram}.

{\color{nicegreen}\subsubsection{ Approximately isometric regime}}

The green region in Figure \ref{fig:phasediagram} corresponds to a parameter regime in which the holographic encoding is expected to be approximately isometric. This can only be the case if there is more entropy in the AdS spaces $\mathsf{lr}$ than in the rest of the universe $\mathsf{M}$.\footnote{In the black hole phase, this can only happen if the Bekenstein-Hawking entropy of the black holes exceeds the EFT entropy of the black hole interior, in which case there is no ``bag of gold'' (see e.g. \cite{Antonini:2024yif}).} In our setup, the stability of the gas of particles in AdS implies that the universe cannot be too large, $O(N^2/\lambda^{1/2}) \ll  m\ell \lesssim O(N^{2}) $, where the lower bound guarantees that the universe is semiclassical.

The quantitative transition between isometric and non-isometric regimes depends on specific details of the encoding map. We will now quantify this for the random tensor network model of Figure \ref{fig:code}. On average over the Gaussian random state $\ket{\Op_{\co}}$, the map $W$ is isometric, $\overline{W^\dagger W} = \mathbf{1}_{\mathsf{M}}$ { for any ratio of the bond dimensions}. However, this does not mean that individual instances $W$ are likely to be isometric. To quantify this, we will compute the average distance to isometry $\big\|W^\dagger W- \mathbf{1}_{\mathsf{M}}\big\|_{1}$, for the one-norm ${\big\|X\big\|_{1}} = \text{Tr}\sqrt{X^\dagger X}$. In this random tensor model, $W^\dagger W$ corresponds to a Wishart random matrix. It will be useful to define the ratio of the bond dimensions
\be 
r \equiv \dfrac{|\mathsf{M}|}{|\mathsf{lr}|}\leq 1\,, 
\ee 
 In the large bond dimension limit, the eigenvalues $\lambda$ of $W^\dagger W$ are distributed according to the Marchenko-Pastur distribution \cite{marchenko1967distribution}
\be\label{eq:MPdistribution} 
p(\lambda) = \dfrac{1}{2\pi r \lambda} \sqrt{(\lambda_+ - \lambda)(\lambda - \lambda_-)}\,,
\ee 
where $\lambda \in [\lambda_-,\lambda_+]$ and the two cuts of the distribution are
\be 
\lambda_{\pm} = {(1\pm \sqrt{r})^2}\,.
\ee 
The mean of this distribution is $\overline{\lambda} =1$ and the variance is $\overline{\lambda^2} -1 =r$.

On average, the distance to isometry is determined by the spread of the eigenvalues of $W^\dagger W$,
\be 
\overline{\big\|W^\dagger W- \mathbf{1}_{\mathsf{M}}\big\|}_{1}  = |\mathsf{M}| \overline{|\lambda-1|}\,.
\ee

Using the Marchenko-Pastur distribution \eqref{eq:MPdistribution} this yields
\be 
\overline{|\lambda-1|}  = { \dfrac{(r+2)\sqrt{r(4-r)}-4(1-r)\arccos\!\left(1-\frac{r}{2}\right)}{2\pi r}}
\ee 
{ for $0<r\leq 1$.} As $r\rightarrow 0$ we have
\be 
\overline{|\lambda-1|} = { \dfrac{8}{3\pi}\sqrt{r} - \dfrac{r^{3/2}}{15\pi} + O(r^{5/2})}\,,
\ee
and thus 
\be\label{eq:avgonenormR} 
\overline{\big\|W^\dagger W- \mathbf{1}_{\mathsf{M}}\big\|}_{1}  \approx { \dfrac{8}{3\pi}\sqrt{r}\,|\mathsf{M}|  = \dfrac{8}{3\pi}\dfrac{|\mathsf{M}|^{3/2}}{|\mathsf{lr}|^{1/2}}}\,.
\ee 

{ The unnormalized trace norm is parametrically small when $S(\mathsf{lr})-3S(\mathsf{M})\gg 1$, or equivalently when $4S(\mathsf{lr})-3S(\co)\gg 1$. Thus the AdS entropy must exceed three-quarters of the entropy of the closed universe, including the entangled degrees of freedom, by a parametrically large margin. This trace-norm criterion is stronger than requiring small operator-norm deviation from isometry.} This regime can only be reached if the universe is of sub-AdS size, but still weakly curved.\\[.2cm]

{\color{niceblue} \subsubsection{Effectively isometric regime}}

Under some mild assumptions about the holographic map, the non-isometric region can be divided into ``effectively isometric'' and ``highly non-isometric'', as explained in \cite{Akers:2022qdl}. These correspond to the blue and red regions in Figure \ref{fig:phasediagram}, respectively. In the effectively isometric regime (blue), the map acts as an approximate isometry for a large collection of bulk states.\footnote{This is much weaker than having a fixed isometry approximating the holographic map for any state \cite{Akers:2021fut,Akers:2022qdl,Antonini:2024yif}.} Again, the details of this transition and the degree of effective isometricity are model-dependent. For the random tensor model of Figure \ref{fig:code}, the overlap between two states $\ket{\psi_{\mathsf{M}}},\ket{\phi_{\mathsf{M}}}$ is preserved on average, up to exponentially small errors in the entropy,
\begin{gather}
\overline{\left| \bra{\psi_{\mathsf{M}}}W^\dagger W\ket{\phi_{\mathsf{M}}} - \bra{\psi_{\mathsf{M}}} \ket{\phi_{\mathsf{M}}}\right|^2} = e^{-S_2(\rho_{\mathsf{lr}})} \;, \label{eq:varianceisom}
\end{gather}
where we are writing the result in terms of the second R\'enyi entropy $S_2(\rho) = - \log \text{Tr} \,\rho^2$ as this result is valid if we allow $\mathsf{Ob}$ and $\mathsf{lr}$ to be in a more general fixed entangled state.

This model satisfies a much stronger effective-isometricity property, due to the strong measure concentration of Gaussian random states \cite{Akers:2022qdl,Kar:2022qkf}.\footnote{Note that this model is very similar to the model in \cite{Akers:2022qdl}, with the different interpretation that part of the entanglement is non-geometric. The mathematical difference is that the state $\ket{\Op_{\co}}$ is Gaussian random, so Gaussian isoperimetric measure concentration theorems replace those of the Haar measure in \cite{Akers:2022qdl}.} In particular, given two states on the universe which are independent of the map, these states are mapped isometrically up to exponentially small corrections with exponentially large probability \cite{Antonini:2023hdh}
\be\label{eq:gaussianconcentration} 
\text{Pr}\left[\left|\bra{\psi_{\mathsf{M}}}W^\dagger W { \ket{\phi_{\mathsf{M}}}} - \bra{\psi_{\mathsf{M}}} \ket{\phi_{\mathsf{M}}}\right|\geq \sqrt{18}|\mathsf{lr}|^{-\gamma}\right] \leq 12  \exp\left(-\dfrac{|\mathsf{lr}|^{1-2\gamma}}{2}\right)\;.
\ee
for any fixed $0<\gamma < \frac{1}{2}$ and $|\mathsf{lr}|>4$.

Within the circuit model of quantum computation, we can define the set of poly$(S(\mathsf{M}))$-complex states in the closed universe. From a counting argument, it then follows from \eqref{eq:gaussianconcentration} that with extremely high probability, the map acts effectively isometrically for all of them as long as the entropy in the closed universe remains sub-exponential in the entropy of the AdS spaces, $S(\mathsf{lr}) \gg  \log S(\co)$ (see \cite{Akers:2022qdl,Antonini:2023hdh}). Although this particular fact relies on the measure concentration of the model, the weaker form of effective isometricity \eqref{eq:varianceisom}, corresponding to the average case scenario, will much more generally be true, e.g., if the state $\ket{\Op_{\co}}$ is drawn randomly from a quantum state $2$-design.\\[0cm]

{\color{nicered}\subsubsection{Highly non-isometric regime}}

In the red region, the map is highly non-isometric, and even ``simple'' states on the closed universe are mapped non-isometrically to the CFT. Many such states that look perfectly distinguishable in $\mathsf{M}$ look almost parallel in the CFT, and thus become indistinguishable in the CFT. Again, the details of the transition to this regime are dependent on the model and on the required degree of non-isometricity. For the tensor network model of Figure \ref{fig:code} the transition happens when the entropy of the universe becomes exponential in the entropy of the AdS gases, $S(\mathsf{lr}) \lesssim \log { S}(\co)$. In this regime the family of poly$(S(\mathsf{M}))$-complex states in $\mathsf{M}$ is mapped non-isometrically with $O(1)$ probability.

In any case, the most extreme regime of non-isometricity is always reached when there is no bulk entanglement at all with the closed universe. In this case, all of the states of the closed universe are mapped to the same CFT state. As pointed out in \cite{Antonini:2023hdh}, this signals a clear limitation of the conventional holographic encoding of the closed universe as the map $W$ breaks down. We will consider this regime in Sections \ref{sec:intrinsic} and \ref{sec:EFT}.\\[0cm]

\section{Aspects of the holographic encoding}
\label{sec:holo_enc}

In this section, we will work in the regime in which the entanglement between the closed universe and the AdS spaces is fixed by $S(\mathsf{lr})$ and is large enough to be in the approximately isometric or effectively isometric regimes explained in Section \ref{sec:encodinguniverse} (corresponding to the green and blue regions in Figure \ref{fig:phasediagram}). Our discussion in the previous section shows that the holographic encoding $W$ into the $\mathsf{LR}$ CFT Hilbert spaces is rather conventional in this situation, sharing several features with the holographic encoding of the black hole interior. In particular, as shown in \cite{Antonini:2023hdh}, the universe belongs to the entanglement wedge of the CFTs, and it can be understood as an entanglement island. The goal of this section is twofold. In the first part, we will explain how our setup relates to the recent proposal of \cite{Harlow:2025pvj} in which observers in closed universes play a central role. In the second part, we will call into question a recently proposed CFT test in \cite{Engelhardt:2025vsp} claimed to prove that the closed universe cannot be part of the bulk description in this setup.

\subsection{Entanglement and observers in closed universes}
\label{sec:entanglobs}

We will use Figure \ref{fig:code} as a reference of the holographic encoding $W$. Recall that the quantum information flows from the closed universe to the CFT because there is a fixed set of entangled pairs between $\mathsf{Ob}$ in the closed universe and the AdS gases $\mathsf{lr}$. We will now denote the entangled half-pairs in the AdS regions by $\mathsf{Ob}' \subseteq \mathsf{lr}$ (Figure \ref{fig:code} represents the equality $\mathsf{Ob}' = \mathsf{lr}$, but more generally the entanglement could be sub-maximal). The conventional holographic isometry $V$ maps $\mathsf{lr}$ to the CFT Hilbert space $\mathsf{LR}$. This defines the map $W$ from states in the closed universe $\mathsf{M}$ to the CFT $\mathsf{LR}$.

The encoding map $W$ is almost equivalent to the one recently discussed in \cite{Harlow:2025pvj}. In \cite{Harlow:2025pvj}, the degrees of freedom $\mathsf{Ob}$ are interpreted as an ``observer'' living in the closed universe, while $\mathsf{Ob}'$ is an external ``clone'' of the observer (thus explaining the labels). The need to include such a clone arises in \cite{Harlow:2025pvj} from demanding the observer to be a classical system, as a consequence of the decoherence of the $\mathsf{M}\mathsf{Ob}$ wavefunction. We shall, for the moment, ignore the different origin of the degrees of freedom $\mathsf{Ob}'$ in both setups (we will comment on this at the end of this section) and focus on the resulting encoding $W$, which has the form presented in Figure \ref{fig:code} in both cases. Recall that the $\mathsf{Ob}$-$\mathsf{Ob}'$ entanglement is what permits us to describe the highly non-trivial Hilbert space in $\mathsf{M}$ from the Hilbert space of the CFT. We will now explicitly show how the dimension of the Hilbert space describing the closed universe is set by the dimension of $\mathsf{Ob}'$ \cite{Antonini:2023hdh}.

Before moving on, it is important to notice a key difference between our setup and \cite{Harlow:2025pvj}. In our construction, the clone $\mathsf{Ob}'$ also gravitates, and it is holographically encoded in the CFT. Therefore, overlaps and similar objects in the Hilbert space of $\mathsf{Ob}'$ can be computed using the standard rules of the gravitational path integral.

\begin{figure}[h]
    \centering
    \includegraphics[width=.9\linewidth]{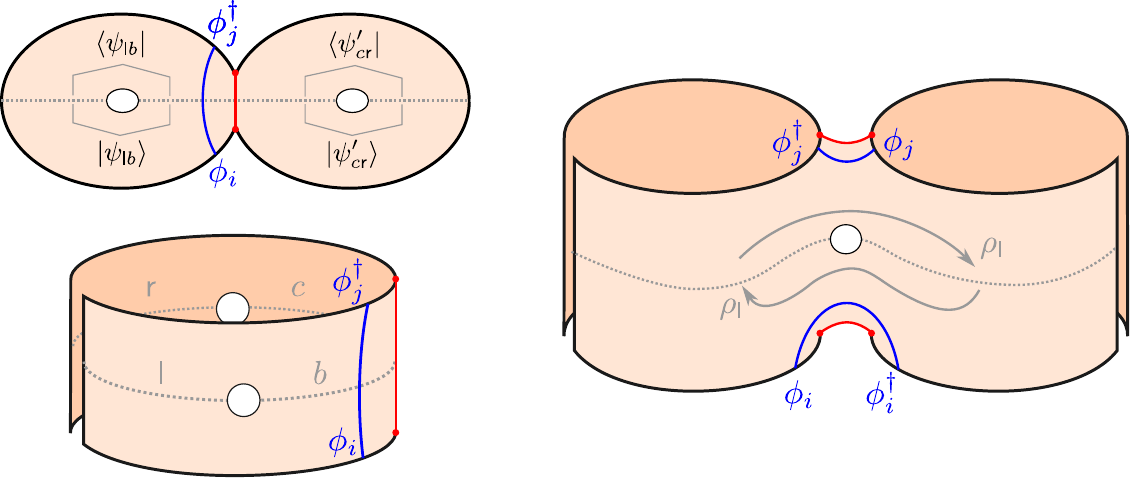} 
    \caption{\textbf{Left}: Two representations of the bulk saddle computing $\overline{\bra{\Psi_j} \ket{\Psi_i}}$. The blue line is the (dominant) matter geodesic and the red line pinches off in the heavy shell limit. The bottom diagram is more like those used to discuss closed cosmologies. Notice that the ``past'' and ``future'' boundaries (or ``bra'' and ``ket'') are not disconnected due to the external regions connecting to the CFT: this plays a similar role to the observer lines in \cite{Harlow:2025pvj}. \textbf{Right}: Two-boundary wormhole computing the square of the overlap. The front face gives ${\rm Tr}(\rho_{\mathsf{l}}^2)$, the back computes ${\rm Tr}(\rho_{\mathsf{r}}^2)$. This reflects how the external observer-like lines connect between copies through the bulk.}
    \label{fig:OverlapAndSquare}
\end{figure}

To see how this works, we will consider CFT states prepared as explained in Sections \ref{sec:CFTmicroTN} and \ref{sec:Euclideanprep}, and illustrated in Figure \ref{fig:Euclideanprep},
\begin{equation}\label{eq:statei}
\ket{\Psi_i} \equiv \ket{\Psi_i}_{\mathsf{LR}} = \bra{\mathcal{O}_{BC}} ( \ket{\Psi_{i,\mathsf{L}B}} \ket{\Psi'_{C \mathsf{R}}} ) \,,
\end{equation}
where the index $i$ labels the insertion of a local primary with $1 \ll \Delta_i \ll N^2$ in the preparation of $\ket{\Psi_{i,\mathsf{L}B}}$. The Euclidean preparation includes a localized particle following a geodesic that intersects the cosmological component of the time-symmetric slice. Accordingly, the semiclassical dual to \eqref{eq:statei} contains a point particle in the left part of the closed universe. This provides a concrete situation where different insertions generate different states in $\mathsf{M}$.\footnote{Choosing the particles to be in the left part is of course arbitrary; more general states are also possible, but they do not add much to this discussion.} As a completely specific case shown in the left of Figure \ref{fig:OverlapAndSquare}, we can choose the states $\ket{\Psi_{i,\mathsf{L}B}}$ and $\ket{\Psi'_{C \mathsf{R}}}$ to be thermal-like states below the Hawking-Page temperature
\begin{equation}
\label{eq:ThermofieldStates}
\ket{\Psi_{i,\mathsf{L}B}} = \left|e^{- (\beta_\mathsf{L} - \beta_{\mathsf{M}}) H /2} \phi_i(x) e^{- \beta_{\mathsf{M}} H/2}\right\rangle  \, , \qquad \ket{\Psi'_{C \mathsf{R}}} = \left|e^{- \beta_\mathsf{R} H /2}\right\rangle \, ,
\end{equation}
where $\beta_{\mathsf{M}}$ is some fixed parameter, and the condition $\beta_{\mathsf{M}}< \beta_L/2$ determines that the matter geodesic in the Euclidean section crosses the closed universe part of the time-symmetric slice.\footnote{In this case, the CFT state dual to the bulk setup is simply given by a slight modification (due to the insertion of the matter particle) of the PETS \eqref{eq:PETS}. } We however remark that the construction can be made much more general, where $\ket{\Psi_{i,\mathsf{L}B}}$ and $\ket{\Psi'_{C \mathsf{R}}}$ are chosen to be more general Euclidean-prepared states dual to disconnected AdS regions.

The states in \eqref{eq:statei} are the CFT images of the corresponding bulk states, $\ket{\Psi_i} = W \ket{\psi_i}$, under the holographic map $W$. We choose our matter insertions so that the $\ket{\psi_i}$ are orthogonal in the bulk EFT, so $\bra{\psi_i} \ket{\psi_j} = \delta_{ij}$. We want to gravitationally compute the CFT overlaps $G_{ij} = \bra{\Psi_i} \ket{\Psi_j}$ and their moments. In order to completely simplify the gravitational calculations, it is convenient to take the shell to be very heavy, $m\ell/N^2 \gg 1$, so that all of the Euclidean geometries pinch off at the shell's trajectory. 

Using this, we can easily compute the gravitational contribution to the overlap, represented by an overline $\overline{G_{ij}}$. The relevant saddle point is represented on the left of Figure \ref{fig:OverlapAndSquare}. Its contribution to the overlap is 
\begin{equation}
\label{eq:OverlapsSingleShell}
\overline{G_{ij}} = \delta_{ij} \bra{\psi_{\mathsf{l} b}} \ket{\psi_{\mathsf{l}b}} \bra{\psi'_{c \mathsf{r}}} \ket{\psi'_{c \mathsf{r}}} Z_0 e^{- m_i L_i} \, .
\end{equation}
Here, $\bra{\psi_{\mathsf{l} b}} \ket{\psi_{\mathsf{l}b}}$ and $\bra{\psi'_{c \mathsf{r}}} \ket{\psi'_{c \mathsf{r}}}$ are the overlaps of the bulk states dual to $\ket{\Psi_{\mathsf{L}B}}$ and $\ket{\Psi'_{C \mathsf{R}}}$, $Z_0$ is a universal factor coming from the heavy shell, and $e^{-m_i L_i}$ is the contribution from the particle geodesic of length $L_i$ (more details about this and similar computations can be found in \cite{Antonini:2023hdh}).\footnote{In periodically identified Euclidean AdS, there are infinitely many topologically distinct geodesics joining two boundary points. We keep only the leading contribution for $m_i \ell \sim \Delta_i \gg 1$, both in this and subsequent computations (in particular, the alternative matter contraction in the wormhole of Figure \ref{fig:OverlapAndSquare} is subdominant).} Note that the states have not been normalized, and that the overall factor multiplying the Kronecker delta $\delta_{ij}$ cancels with the normalization. 

The square of the overlap follows from a similar computation, with two asymptotic boundaries instead. In addition to a disconnected piece formed by two copies of the overlap, there is a connected contribution represented on the right of Figure \ref{fig:OverlapAndSquare}. Both gravitational contributions produce
\begin{equation}
\label{eq:OverlapsSquareSingleShell}
\overline{G_{ij}G_{ji}} = \delta_{ij} \left( \bra{\psi_{\mathsf{l} b}} \ket{\psi_{\mathsf{l}b}} \bra{\psi'_{c \mathsf{r}}} \ket{\psi'_{c \mathsf{r}}} Z_0 e^{- m_i L_i} \right)^2 + {\rm Tr} ( \rho_{\mathsf{l}}^2) {\rm Tr} ( \rho_{\mathsf{r}}^2) Z_0^2 e^{-m_i L_i - m_j L_j} \, ,
\end{equation}
where we have introduced the (unnormalized) bulk reduced states $\rho_{\mathsf{l}} = {\rm Tr}_b (\ket{\psi_{\mathsf{l}b}} \bra{\psi_{\mathsf{l}b}})$ and $\rho_{\mathsf{r}} = {\rm Tr}_c (\ket{\psi'_{c \mathsf{r}}} \bra{\psi'_{c\mathsf{r}}})$ (the traces are taken over the corresponding AdS component). Since the geodesic length in Euclidean AdS only depends on the separation of the insertions, in the heavy shell limit, the geodesic lengths $L_i$, $L_j$ in the wormhole become the same as those in the disconnected saddle. Dividing by the normalization, the connected contribution reduces to
\begin{equation}\label{eq:secondmomentwormhole}
\frac{\overline{|G_{ij} - \overline{G_{ij}} |^2}}{\overline{G_{ii}} \,\overline{G_{jj}}} = \frac{{\rm Tr} ( \rho_{\mathsf{l}}^2) {\rm Tr} ( \rho_{\mathsf{r}}^2)}{({\rm Tr} \rho_{\mathsf{l}})^2 ({\rm Tr}  \rho_{\mathsf{r}})^2} = e^{-S_2(\rho_{\mathsf{lr}})} \, ,
\end{equation}
which is determined by the second R\'enyi entropy of the state of the AdS gases $S_2(\rho_{\mathsf{lr}}) =S_2(\rho_{\mathsf{l}}) + S_2(\rho_{\mathsf{r}})$, as the state is factorized (recall that the second R\'enyi entropy is $S_2(\rho) = -\log { \frac{\text{Tr}(\rho^2)}{(\text{Tr}\rho)^2}}$). We are implicitly choosing a maximally entangled state, so $S_2(\rho_{\mathsf{lr}}) = S({\mathsf{lr}})$. Since, as we explained above, the $\mathsf{lr}$ gases are the clones $\mathsf{Ob}'$ in our model, \eqref{eq:secondmomentwormhole} translates to the result of \cite{Harlow:2025pvj}.\footnote{A minor difference is that the model of \cite{Harlow:2025pvj} includes the assumption of gauging the discrete {$\mathsf{CRT}$}-symmetry under which the thin shell operator $\Op$ is charged. This feature does not modify the main conclusion of the paper, and in particular in this computation it just adds another suppressed piece to the variance of the overlap.} 

From \eqref{eq:secondmomentwormhole} and the higher moments, a conventional resolvent calculation outlined in \cite{Antonini:2023hdh} yields that the states $\ket{\Psi_i}$ span the microcanonical Hilbert space dimension of the clone,
\be 
{\text{dim}(\mathcal{H}_{\mathsf{Ob}'})}= \overline{\text{rank}(G_{ij})} = e^{S(\mathsf{lr})}\,.
\ee 
The variance of the dimension, computed gravitationally, vanishes, up to possible off-shell contributions. We reach the common conclusion in both papers:\footnote{Here we { assume that the observer's} Hilbert space $\mathcal{H}_{\mathsf{Ob}}$ has smaller dimension than the rest of the universe $\mathcal{H}_{\mathsf{M}}$, namely, that we are in the effectively isometric encoding regime discussed in Section \ref{sec:encodinguniverse}.}
\begin{tcolorbox}[colframe=black, colback=white, boxrule=0.5pt, sharp corners]
     \it When the closed universe is entangled via an ``observer'' $\mathsf{Ob}$ with an external ``clone'' $\mathsf{Ob}'$, the Hilbert space that describes it is that of the clone, and its dimension is $e^{S_{\mathsf{Ob}}}$ for $S_{\mathsf{Ob}}$ the entanglement entropy between the observer and the clone.
\end{tcolorbox}
Further matching with the results of \cite{Harlow:2025pvj} can be found in Appendix \ref{app:UniverseBottleneck}, where we also introduce a more elaborate model in which the closed universe has a geometric bottleneck.

\subsubsection{Comments on observers, pointer variables, and all that}

We now comment on how decoherence, pointer variables, and similar notions were implicit in the model of ~\cite{Antonini:2023hdh}. We will then return to the discussion of the physical status of the ``observer'' $\mathsf{Ob}$ or ``clone'' $\mathsf{Ob}'$ in both models (since, as we have already seen, once we assume \emph{a} clone is present, both our model and that of \cite{Harlow:2025pvj} are identical).

First, consider the dynamics within the closed universe as defined by the background. As we reviewed in Section~\ref{sec:2}, the closed universe is filled with a weakly interacting thermal gas. There is also the possibility of including additional ``probe'' particles as insertions; these particles are light enough to not cause significant backreaction but heavy enough to be well described by a classical trajectory. This setup---a thermal gas with weak interactions---is a classic dynamical system which exhibits decoherence in the position basis for heavier ``probe'' particles.\footnote{In more detail, we get the emergence of classical trajectories, indicating decoherence to a pointer basis of wavepackets with well-localized position and momentum.} In the present context, we have an additional length scale, the AdS scale, which complicates the kinetic theory when it is comparable to the mean free path. However, while it would be fun to analyze the details of this setup with the extra AdS scale, we do not expect the essential conclusions to change.\footnote{Depending on the details of the bulk theory, we can presumably arrange for the mean free path to be much less than the AdS radius by dialing up the density and entropy of the bulk gas.} Modulo this proviso, we thus expect the internal dynamics of the closed universe to exhibit decoherence in a standard fashion. 

In fact, the special role of position variables is arguably built into the basic structure of EFT. A classic example from particle physics is the puzzle of $\alpha$-ray tracks in a cloud chamber, solved by Mott almost 100 years ago~\cite{Mott1929}. He showed that, despite the spherical scattering wavefunction, the tracks were overwhelmingly likely to form straight lines (the classical trajectory). Although his analysis predates the idea of decoherence and pointer variables, it is compatible with them and illustrates the crucial role of position variables as dictated by the locality of interactions. Since the Standard Model is built on data derived from such tracks in ever more sophisticated detectors, there is a real sense in which EFT has position as pointer variables built in.

More generally, any particle that couples to the electromagnetic field is going to have its position rendered effectively classical due to scattering with the background radiation. We emphasize that 
this is precisely the situation envisioned in \cite{Antonini:2023hdh}: particles, moving on geodesics, in the presence of a background thermal gas. This physics is well known in the classic references on the subject, including ~\cite{Joos:1984uk,PhysRevD.24.1516}. It has also been argued in the context of cosmology that local field variables form a natural class of pointer variables~\cite{burgess2014efthorizonstochasticinflation}. Taking this into account, we now argue that the rules used in \cite{Antonini:2023hdh} and extended to generic settings by \cite{Harlow:2025pvj} are naturally compatible with an observer interpretation in a standard way, even though \cite{Antonini:2023hdh} did not use the language of observers. 

How is this realized in the present setup? The clone degrees of freedom are in the AdS regions, $\mathsf{Ob}' \subseteq \mathsf{lr}$, and together with their partners in the closed universe they form entangled pairs. If these entangled pairs are introduced by inserting particles in the Euclidean preparation, in such a way that they follow geodesics and have well determined positions, we can write the entangled bulk state as
\begin{equation}
\label{eq:ObserverCloneEntanglement}
\ket{\omega} = \sum_{a} C_{a} \ket{a}_{\mathsf{Ob'}} \ket{a}_{\mathsf{Ob}} \, ,
\end{equation}
where the states $\ket{a}_{\mathsf{Ob'}}$ and $\ket{a}_{\mathsf{Ob}}$ have a collection of particles localized in different positions. This was discussed in \cite{Antonini:2023hdh}. Tracing out the clone, we get a state for $\mathsf{Ob}$ which is diagonal in the position (pointer) basis. The clone can be seen in this way as a tool to guarantee this. 

We can now make contact with observers and the construction of \cite{Harlow:2025pvj}. In \cite{Harlow:2025pvj}, the clone $\mathsf{Ob}'$ was introduced as a tool to force the observer to be in a classical state, i.e., diagonal in a pointer basis (without committing to any particular pointer basis). It was argued that this is needed because any observer will experience interactions with the environment in the closed universe which decohere its state, and the description of the closed universe has to assume a priori the existence of such a classical observer. The connection with our setup is clear, and we thus advocate that \cite{Antonini:2023hdh} is already implicitly incorporating a rich physical context related to decoherence and pointer variables. In fact, \cite{Antonini:2023hdh} has an additional feature relative to \cite{Harlow:2025pvj} insofar as the purifiers explicitly live in a gravitating region (as would be the case for any in-universe environment which is decohering the local state). We emphasize that this does not mean that the rules introduced in \cite{Harlow:2025pvj} are entirely equivalent to \cite{Antonini:2023hdh}. In fact, one main novelty in \cite{Harlow:2025pvj} is that regardless of the specific bulk setup, an external ``observer clone'' can be introduced, leading to a non-trivial Hilbert space for the closed universe. In this sense, the setup of \cite{Antonini:2023hdh} and discussed here is equivalent to considering the setup of \cite{Harlow:2025pvj} after the cloning procedure has been carried out.

There is one additional distinction that is worth emphasizing. The observer-clone setup is at best a ``frozen'' simulation of what is naturally occurring inside the closed universe. Indeed, we argued at the beginning of this subsection that the native in-universe dynamics is going to exhibit a fairly conventional kind of decoherence as occurs with a wide variety of background gases and radiation. Of course, the overall universe state is going to be a superposition of all the ways this could happen, so that an external party will still not know which way a probe particle was observed to move. What the observer-clone setup does is to take a snapshot of this dynamical process, a frozen moment in time, which preserves the natural role of position variables (at least in \cite{Antonini:2023hdh}).

Finally, it is worth reflecting also on the different status of the clone degrees of freedom in both models. These constitute, in the end, the physical system by which the cosmology is being described. For us and \cite{Antonini:2023hdh}, the clones are just AdS gas degrees of freedom, holographically encoded in the CFTs. They thus have a very clear physical existence, and we could identify the fundamental description with the CFT. On the contrary, in \cite{Harlow:2025pvj} the clone $\mathsf{Ob}'$ is an auxiliary system introduced to force the observer to be in a classical state (a requirement which does not come out naturally of a standard state preparation for the closed universe with an observer). It is thus somewhat less clear to what extent the Hilbert space of $\mathsf{Ob}'$ has significance as a fundamental object. This is to be contrasted with similar encodings in, e.g., black hole evaporation \cite{Akers:2022qdl}, where the interior degrees of freedom $\mathsf{i}$, entangled with early Hawking radiation $\mathsf{R}$, are encoded via $V \otimes I_\mathsf{R}: \mathcal{H}_{\mathsf{i}} \otimes \mathcal{H}_{\mathsf{R}} \rightarrow \mathcal{H}_{\mathsf{B}} \otimes \mathcal{H}_{\mathsf{R}}$ into two systems with clear physical significance: the fundamental black hole Hilbert space $\mathcal{H}_{\mathsf{B}}$ (e.g., a subspace of the CFT in AdS/CFT) and the radiation $\mathcal{H}_{\mathsf{R}}$. At late times, it makes sense to say that the interior degrees of freedom are in the entanglement wedge of the radiation $\mathsf{R}$. Similarly, in our model, it is natural to say that the closed universe is in the entanglement wedge of the CFTs. It is more difficult to imagine in what sense the closed universe is in the entanglement wedge of the clone $\mathsf{Ob}'$ when this is just a tool to make the state of the observer classical.

\textit{Note added: in the final stage of the preparation of this paper, the independent work \cite{engelhardt2025observercomplementarityblackholes} appeared. The results for a closed universe presented there are reproduced in our setup using the language of observers as described in this section. In particular, an ``interior observer $\beta$'' can be modeled as, e.g., the degrees of freedom in the cosmology entangled with the $\mathsf{l}$ AdS space. Then $\mathsf{l}$ is viewed as the clone, which arises by forcing the observer to be classical, while the entanglement of the universe with $\mathsf{r}$ is viewed as actual entanglement with an external gas. The expectation value of the partial SWAP operator for the external gas (the $\mathsf{R}$ CFT SWAP) is the maximum between $e^{-S_2(\mathsf{r})}$ and $e^{-S_2(\mathsf{l})}$, just like in \cite{engelhardt2025observercomplementarityblackholes}. An ``external observer $\alpha$'' measures the global SWAP operator of the gases $\mathsf{l r}$ (the full $\mathsf{LR}$ SWAP), with expectation value $1$. A further mention { of} this partial SWAP can be found in Section \ref{subsec:swap}.} 

\subsection{On SWAP tests}
\label{subsec:swap}

Here we address recent claims made in \cite{Engelhardt:2025vsp} (referred to in this subsection as EG) and show that their argument against semiclassicality fails. For simplicity and clarity, the analysis is broken into several small parts. First is a review of the puzzle raised in \cite{Antonini:2024mci} (referred to in this subsection as AR). Second is a direct refutation of the argument in EG. A key point is that the SWAP operator discussed in EG, when properly defined in the CFT, is just the boundary SWAP operator (in a low energy subspace) and, accordingly, it will also swap the closed universes. Third is a discussion of various approaches to effectively not swapping the closed universes and detecting the entanglement with it. Fourth is a collection of further comments that clarify the previous discussions and further address the relationship to AR. 

\subsubsection{Reviewing AR and the EG argument}

\begin{figure}[h]
    \centering
    \includegraphics[width=.85\linewidth]{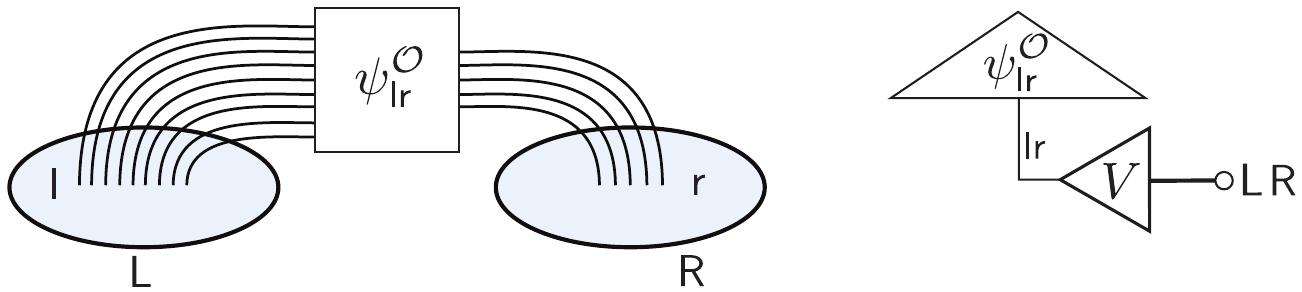} 
    \caption{On the left, Description 2 consists of a pure state on the $\mathsf{lr}$ AdS spaces. On the right, tensor network model of the CFT state. The closed universe tensor has been absorbed into the state of the quantum fields.}
    \label{fig:setupaltern}
\end{figure}

In \cite{Antonini:2024mci} a subset of us pointed out that there exists an alternative semiclassical description of the CFT state of interest. The idea is to absorb the universe into the state of the matter fields in the two AdS spaces, which according to \eqref{eq:bulktobdy} is
\be 
\ket{\psi^{\Op}_{\mathsf{lr}}} \,\propto\, \bra{\Op_{\co}} \ket{\psi_{\mathsf{clr}}} \,.
\ee 
This is possible because the CFT state $\ket{\Psi_{\mathsf{LR}}}$ is low-energy and can be expanded in a basis built by acting with left and right light primaries on the vacuum of the two CFTs. Each state in this basis is dual to some simple bulk state of matter in two copies of pure AdS spacetime. Once the exact coefficients (given by light-light-heavy OPE coefficients) defining the state $\ket{\Psi_{\mathsf{LR}}}$ in this low-energy basis are known, the alternative description can be built simply using the HKLL dictionary and does not manifestly involve a closed universe. Thus, there is this alternative dual description to $\ket{\Psi_{\mathsf{LR}}}$ in terms of two entangled AdS gases in this particular state, as shown in Figure \ref{fig:setupaltern}. This was labeled ``Description 2'' in AR, while ``Description 1'' is the saddle of the gravitational path integral including the closed universe and reviewed in Section \ref{sec:2}. Both Description 1 and Description 2 are indistinguishable from the CFT, as they correspond to the same CFT state $\ket{\Psi_{\mathsf{LR}}}$.\footnote{In the black hole evaporation case after the Page time, one can consider a similar puzzle, with Description 1 given by the usual semiclassical saddle with the interior highly entangled with radiation in the Hawking state, and Description 2 built by considering the exact microscopic state of radiation in the bulk [QFT], which would lead to a non-smooth interior, as it was discussed in \cite{Antonini:2024mci}.}

This creates a puzzle because it seems to indicate that in the Description 2, without the closed universe, the AdS spaces are in a pure state, while in the description with the closed universe, the AdS spaces are in a mixed state if we follow the \iqft\, interpretation above. However, there isn't an obvious discrepancy for an EFT of quantum gravity, even at the semiclassical level. An important difference is that in semiclassical gravity, wormholes swapping closed universes can be important. To distinguish this from \iqft, we will call this semiclassical gravity description including wormholes as \ieftc, as we explained in Section \ref{sec:notation}. Note that \ieftc\, is still only an effective description and not a UV-complete description of quantum gravity, although it includes non-perturbative effects captured by the gravitational path integral. 

If one insists on using \iqft\, instead of \ieftc\, or \ieftp, then indeed these two descriptions are at odds with each other. AR suggested that this tension means we need to give up on something, and they suggested various alternatives. One option is to prefer Description 2 since it was constructed using the extrapolate dictionary, which is a basic tenet of AdS/CFT. One is then forced to conclude that the description with the closed universe does not exist, or at the very least the EFT in the closed universe has large corrections. Other possibilities included the presence of an ensemble of CFTs and the CFT not being a complete description of the bulk theory.

Another option that we will discuss more below is that the two descriptions are both dual to the same CFT state and there is in fact no unique bulk dual for a CFT state. This option was called gauge-equivalence in v1 of \cite{Antonini:2024mci}.\footnote{Not all the authors of the present paper agree that this is the best term to describe situations of this sort.} However, such an option only works if one uses \ieftc\, instead of \iqft. It is worth mentioning that the alternatives proposed by AR may not be exhaustive.

Recently, \cite{Engelhardt:2025vsp} claimed that the breakdown of semiclassicality in the closed universe description was indeed the correct resolution to the puzzle. Essentially, the claim of that paper is that there is a CFT operator in two replicas that is able to distinguish the two descriptions, and that it singles out the description without the universe as the true description. This is presented as concrete evidence against a semiclassical closed universe in AdS/CFT, though it is unclear why the effective description of a large semiclassical closed universe breaks down. Below, we explain why their analysis is not conclusive.

\subsubsection{Addressing the EG argument}

The microscopic state $\ket{\Psi_{\mathsf{LR}}}$ from the CFT perspective is pure. The semiclassical bulk state built with the gravitational path integral in the saddle point approximation is supported on a spacetime that includes the $\mathsf{lr}$ AdS region and a disconnected component, $\co$, which is the closed universe. In the \iqft\, description, the quantum fields in the closed universe are entangled with the fields in the $\mathsf{lr}$ AdS regions. At the outset, we recall that this entanglement can be made large, even scaling with $N^\alpha$ to some power $\alpha<2$, as explained in Section \ref{sec:2}. In contrast, the authors of \cite{Engelhardt:2025vsp} assumed that this entanglement would always be $O(N^0)$. Our arguments in this section are not sensitive to the precise amount of entanglement.

EG proposes to construct a ``causal wedge'' SWAP operator which swaps the state of the bulk fields in two copies of the $\mathsf{lr}$ AdS regions. At the level of QFT on a given background, the action of this operator is straightforward: it simply swaps the bulk state in the two copies of the $\mathsf{lr}$ AdS regions in some basis, say the Fock basis. Using the standard holographic map $V$, this bulk operator maps to the boundary SWAP operator, or more precisely to the boundary SWAP operator restricted to a low-energy subspace (although the full SWAP does the job just fine).\footnote{The authors of \cite{Engelhardt:2025vsp} claim this SWAP operator is not the same as the boundary SWAP operator (see Footnote 9 in \cite{Engelhardt:2025vsp}). However, following the ordinary holographic dictionary, it seems clear it indeed is (or at least a boundary SWAP restricted to the low-energy sector of the CFT).}

\begin{figure}
    \centering
    \includegraphics[width=0.7\linewidth]{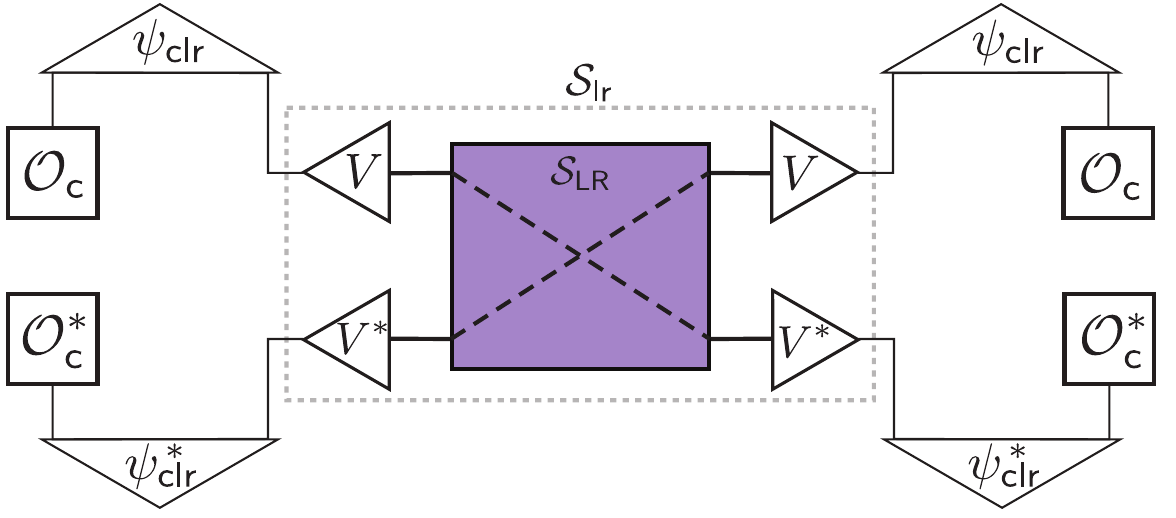}
    \caption{The causal wedge SWAP operator $\mathcal{S}_{\mathsf{LR}}$ as defined in \cite{Engelhardt:2025vsp} also swaps the closed universes and has unit expectation value.}
    \label{fig:swap}
\end{figure}

If we now carry out a SWAP test using this operator and multiple copies of the same CFT state (meaning the same microstate), what will we find? From the CFT perspective, the answer is clear: the state is pure and the SWAP test will give unity.\footnote{Each run of the experiment involves a controlled application of the SWAP to two copies of the CFT state conditioned on an ancilla qubit. Running the standard SWAP trick circuit, a measurement of the ancilla qubit in the computational basis gives $0$ with probability $P(0) = \frac{1+\tr(\text{SWAP}\rho \otimes \rho)}{2}$ and $1$ with probability $P(1) = \frac{1-\tr(\text{SWAP}\rho \otimes \rho)}{2}$. When $\rho$ is pure, every measurement will return $0$ (assuming an ideal quantum computer) and the experimenter will quickly conclude that $P(0)\approx 1$ and $P(1)\approx 0$.} Moreover, from the bulk perspective, the answer is also clear. For example, using the tensor network model (see Figure \ref{fig:swap}), we see immediately that the closed universes are also ``swapped'' (although their position in the diagram is meaningless); indeed, they are just preparing a state of the bulk fields. From another point of view, the closed universe is in the entanglement wedge of the bulk fields in the $\mathsf{lr}$ AdS regions and hence it is swapped when they are. 

\begin{tcolorbox}[colframe=black, colback=white, boxrule=0.5pt, sharp corners]
\emph{Because the causal wedge SWAP, when mapped to the CFT using the holographic map, is just the boundary SWAP, it also swaps the closed universes. Thus, in both Descriptions we obtain the correct value of unity for the SWAP expectation value.}
\end{tcolorbox}

We emphasize that this situation is conceptually quite similar to the situation with a mostly evaporated black hole. For example, imagine a small black hole in AdS which is well past the Page time, so that it has mostly evaporated into radiation (radiation that is still contained in the asymptotically AdS region). At late times, the interior of the black hole is located within the entanglement wedge of the radiation and the causal wedge SWAP operator, which at the bulk QFT level does not swap the interior; in fact, it does so after applying the holographic map. 

Therefore, the SWAP operator defined in EG cannot detect whether a closed universe is present in the bulk QFT description. Regardless of whether we consider a bulk description with or without a closed universe (Descriptions 1 and 2 in AR, respectively), the expectation value of the SWAP operator will always give unity. In Description 1, this is obtained by swapping the closed universes together with the state of bulk fields in the AdS spacetimes; in Description 2, simply by swapping the state of bulk fields in the AdS spacetimes, which is pure in the first place. 

Before proceeding, we make one other comment. After their main argument, EG concludes that the only way Description 1 can make sense is if the closed universe has a one-dimensional Hilbert space, and interpret this fact as a breakdown of the $G\hbar$ expansion defining semiclassical EFT in the closed universe. However, it is not clear that the large fluctuations arising when computing CFT inner products using the bulk EFT have an impact on ordinary observations within the closed universe. The closed universe Hilbert space is one-dimensional when ``viewed from the outside'', namely from the CFT point of view. Nonetheless, it is true that the holographic map acts as a projector in the universe onto the state $\ket{\mathcal{O}_{\co}}$, and in this sense, the entanglement that we are describing is not conventional. In the remainder of this section, we will describe how to make sense of this entanglement.

\subsubsection{How to not swap the closed universes}

Here we discuss various senses in which one can in fact swap only part of the state, keeping the closed universe fixed, and detect the $\mathsf{lr}$ entanglement with $\co$. We speak about ``senses'' because these different approaches don't always correspond to operators acting on a two-copy Hilbert space. Some of them require taking a more general perspective on the task.

%%%%

\paragraph{Coarse-grained SWAP.} The simplest way to not swap the closed universe is to not use fine-grained information about the state. Consider an ensemble of microstates, corresponding to different microscopic operators $\mathcal{O}$ (and thus states $\ket{\mathcal{O}_{\co}}$) consistent with a semiclassical heavy shell of dust, as in \cite{Sasieta:2022ksu}. Imagine an agent who is not able to probe the detailed microstate in this ensemble. Operationally, this means the agent cannot tell apart different states in the ensemble. For all they know, the ensemble consists of a single mixed state.

Now suppose they want to measure the purity of the state. They obtain a bunch of ``copies'' of the state from some device, unaware that these ``copies'' can all be different states. They carry out the swap test many times and build statistics. What they ultimately estimate for the purity is in fact the purity of the ensemble-averaged state. This is given by the entropy of the bulk state in the \iqft\, interpretation, which we interpret as a coarse-grained version of the \ieftc\, or quantum gravity state.\footnote{To reinforce the analogy with the evaporating black hole, the coarse-grained state discussed here is equivalent to the coarse-grained state of Hawking radiation, which can in fact be interpreted as an ensemble-averaged state \cite{Bousso:2020kmy}. The result of the SWAP test in that case yields the coarse-grained entropy of Hawking radiation, which follows the Hawking curve. }

Crucially, this will work even when the entanglement is very low. This coarse-grained SWAP test will still generate a purity that is consistent with the thermal R\'enyi entropy of a very weakly excited gas of particles in a large closed universe. This is an example of a semi-classical computation giving the right answer to a properly designed CFT question.

A different way to phrase the above discussion is that the coarse-grained state can be purified using an auxiliary reference system to represent our ignorance of the precise microstate. In this language, the closed universe is, in fact, in the entanglement wedge of the reference system. Here, we aren't claiming that such a reference system physically exists; instead, it is just a trick to capture our ignorance. A similar trick was used by \cite{Almheiri:2021jwq} to capture the sensitive dependence of the interior of the black hole on the precise couplings of the theory.

\paragraph{Partial SWAP.} 

\begin{figure}
    \centering
    \includegraphics[width=0.78\linewidth]{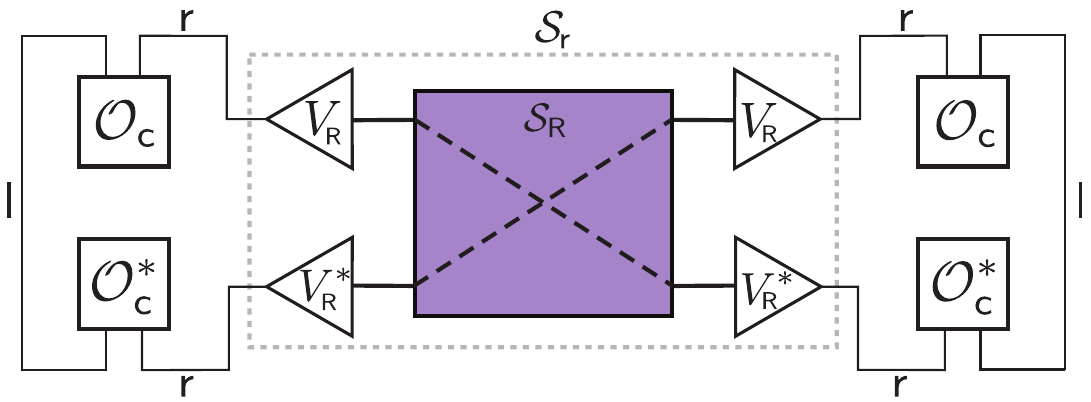}
    \caption{If the $\mathsf{cl}$ entanglement is sufficiently large, the partial CFT SWAP $\mathcal{S}_{\mathsf{R}}$ does not swap the closed universe and detects the bulk entanglement to the $\mathsf{r}$ AdS space.}
    \label{fig:swapentangle}
\end{figure}

Related to the above coarse-grained approach, we can also split the $\mathsf{LR}$ system and use the $\mathsf{R}$ CFT as a physical version of a reference which keeps track of the microstate. In the regime in which the bulk entanglement is largest between the closed universe and the $\mathsf{r}$ bulk fields, the closed universe is in the entanglement wedge of the $\mathsf{r}$ bulk fields (and the $\mathsf{R}$ CFT) and not the $\mathsf{L}$ CFT. Now the same bulk QFT SWAP on the $\mathsf{l}$ bulk fields, which still maps to the $\mathsf{L}$ CFT SWAP, does not SWAP the closed universes, as shown in Figure \ref{fig:swapentangle} { with $\mathsf{L}$ and $\mathsf{R}$ interchanged}. In this case, the result of the SWAP is controlled by the entanglement entropy between the two CFTs, which in the bulk is captured by the bulk entanglement entropy between fields in the left AdS spacetime and fields in the closed universe.

This relationship between the closed universe and the entanglement wedge of various CFTs was already discussed in~\cite{Antonini:2023hdh}. Moreover, as made clear in Sec.~\ref{sec:entanglobs}, one can also view this construction from the point of view of introducing explicit bulk observers. It is certainly true that~\cite{Antonini:2023hdh} did not explicitly discuss observers in this language, but as argued in Sec.~\ref{sec:entanglobs}, the mathematical rules of \cite{Harlow:2025pvj} are explicit in~\cite{Antonini:2023hdh} (where they were derived from the rules of AdS/CFT) and the physical content was implicit as well.

The situation in which we re-introduce the $\mathsf{R}$ CFT is also directly related to the coarse-graining approach via measurements of the $\mathsf{R}$ CFT. By post-selecting specific measurement outcomes that single out a consistent microstate and conveying this information to the $\mathsf{L}$ CFT, one can select a corresponding subset of the SWAP test outcomes for which the two states in the SWAP test happened to agree, and that subset of the data will give an estimate of unit purity.

\paragraph{Holographic map approach.} It is also possible to implement versions of the causal wedge SWAP without explicit coarse-graining or the introduction of a reference. Using the general encoding map in Figure \ref{fig:TN}, we can generate a set of states $|i \alpha\rangle$ where $i$ indexes operators acting on $\mathsf{clr}$ (the fields of the bulk QFT) and $\alpha$ indexes the heavy shell operator. As a simple example, if the encoding maps are all random, then these states can have exponentially small overlaps,
\begin{equation}
    \langle i \alpha | j \beta \rangle = \delta_{ij} \delta_{\alpha \beta} + O(e^{-S/2}),
\end{equation}
where $S = S(\mathsf{lr})$. We can then map bulk QFT operators to the CFT Hilbert space by replicating matrix elements. The reader may look back at Section \ref{sec:2} for more quantitative statements.

For example, we could consider four states defined by fixing a heavy shell operator and considering $i \in \{0,1\}^2$ to label the absence or presence of a particle in the closed universe and the absence or presence of a particle in one of the two AdS spacetimes. Denote these states $|n_{\text{AdS}},n_{\text{CU}}\rangle$. These are all states in the Hilbert space of the two CFTs. Now taking two copies of such a Hilbert space, we can define a version of the causal wedge SWAP as
\begin{equation}
\mathcal{S}_{\text{CW}}\, | n_{1,\text{AdS}} n_{2,\text{AdS}}, n_{1,\text{CU}} n_{2,\text{CU}} \rangle = | n_{2,\text{AdS}} n_{1,\text{AdS}}, n_{1,\text{CU}} n_{2,\text{CU}} \rangle.
\end{equation}
This is an operator that effectively SWAPs the state of the particle in AdS but not the state of the particle in the closed universe. One could then apply this version of the causal wedge SWAP to an ``entangled'' state like ${ \frac{1}{\sqrt{2}}\bigl(|0,1\rangle + |1,0\rangle\bigr)}$ to obtain an expectation value close to $1/2$ up to $e^{-\# S}$-type corrections.

Now, it is not straightforward to extend this to a SWAP of the entire closed universe because the typical matrix elements of the total SWAP are of the same size as the corrections to the overlap. This can be gotten around in various ways, but it also misses the main point.

A more instructive thought experiment is as follows. With sufficiently large entropy $S$ (and sufficiently rich bulk physics), we can plausibly hide a large quantum computer inside the closed universe and replicate any desired quantum computation. We say plausibly only because we do not fill in the details, not because we know of any serious obstacle to this working. 

Thus imagine a quantum physics lab inside the closed universe built from the bulk fields. This lab contains a quantum computer operating on $K$ qubits. From these qubits we get $2^K$ computational basis states, each of which is mapped to a not-quite-orthogonal state in the CFT Hilbert space. We can also construct a set of elementary unitary operations, gates, that are universal for computations on the $K$ qubits. These gates can also be mapped to operators acting on the CFT Hilbert space, operators which can be approximated as unitary. 

This setup already suffices to implement a set of approximate quantum computations, where approximate means up to $e^{-\# S}$-type errors. Based on \cite{Antonini:2023hdh}, we expect these states and operations to look random from the CFT perspective, but from the closed universe perspective, they can be ordered structures. Moreover, when asked where the quantum computation is occurring, one reasonable answer would be that it is occurring in the closed universe.

The overall point of this thought experiment is simply to show how very complex quantum processes in the closed universe can be seen to play out from the CFT perspective, provided there is enough ``entanglement''. This is exactly what we would expect for an ordinary quantum physics lab in a large semiclassical universe with weak gravity.

\subsubsection{Interpreting the bulk entanglement}

One of the central lessons of recent years is that, in gravitating systems, fine-grained entropies of ``microscopic'' systems can be captured in terms of coarse-grained entropies of dynamically determined regions of the geometry. To be clear, these entropies are coarse-grained in that they are entropies computed in states that don't directly know about the detailed microstructure of the microscopic state. For example, we compute Page curves using bulk states that don't encode any of the detailed erratic amplitudes in the microscopic state. The situation with the closed universe is essentially the same, right down to the erratic/random nature of the states \cite{Antonini:2023hdh}.

Consider again the state { of the} $\mathsf{LR}$ pair of CFTs, which has the form
\begin{equation}
    |\psi \rangle = \sum_{\text{Fock basis}} \sqrt{\text{thermal factor}} \;|\text{Fock state}\rangle \times (\text{erratic amplitude}).
    \label{eq:erratic}
\end{equation}
Here the ``erratic amplitude'' is the closed universe part. Microscopically, it is just some light-light-heavy matrix element in the CFT (which is understood to behave erratically \cite{Belin:2020hea}), but this matrix element can also be interpreted as a ``microstate'' of the closed universe, as made precise via the tensor network model.

We emphasize that the ``one-dimensional Hilbert space of the closed universe'' is no obstacle. The form of the state in \eqref{eq:erratic} can be thought of as ``entanglement'' between the CFT and a one-dimensional Hilbert space of numbers; indeed, it is explicitly in the form of a Schmidt decomposition. Of course, we know this can also be viewed simply as a superposition of states in the CFT, but the analogy is powerful because the amplitudes are erratic and for any suitably coarse-grained or averaged property, these erratic amplitudes behave as if they are orthogonal vectors in some larger Hilbert space.

We can also as above split the $\mathsf{LR}$ pair and use the $\mathsf{R}$ CFT to replace the erratic amplitudes in \eqref{eq:erratic} with erratic states in the Hilbert space of the $\mathsf{R}$ CFT. One can think of this as an auxiliary system that prepares particular microstates of the $\mathsf{L}$ CFT conditioned on the state of an ancilla system (the $\mathsf{R}$ CFT). One can also think of it in terms of observers along the lines of \cite{Harlow:2025pvj}. The tensor network makes it clear that the same underlying structure is present in all these cases. The tensor network is suggesting that the closed universe remains semiclassical when the entanglement is low, but the ability of the CFT to ``access'' the closed universe is becoming more tenuous. 

Nevertheless, even in the limit of zero entanglement, the CFT knows about the closed universe in a more general sense. For example, considering the statistical properties of the erratic amplitudes in the CFT, one would eventually find that they have a parsimonious description in terms of correlators of bulk fields in a semiclassical closed universe as discussed in Section~\ref{sec:coarseCFT}. Within the tensor network model, each such amplitude is built from the same underlying structure, leading to the conjecture that the closed universe exists amplitude by amplitude. We discuss this further in Section \ref{sec:intrinsic}.

\subsubsection{Closing comments}

Reflecting back on the analysis in this section, we argued that the causal wedge SWAP described by EG is simply the boundary SWAP and the result of such a SWAP test on the CFT state is unit purity. This would imply a one-dimensional Hilbert space for the closed universe, inconsistent with \iqft. EG then claims that one-dimensionality of closed universe Hilbert space in this perspective implies that the closed universe is not semiclassical. 

However, as we noted above, the one-dimensionality of the closed universe Hilbert space, suitably defined, is a prediction of \ieftc. Moreover, this is presumably consistent with the experience of an observer in the closed universe being well described by \iqft\, even though it gives the wrong value for the SWAP. The postselection implemented by the holographic map from the bulk QFT to the two CFTs is similar to the one considered in \cite{Akers:2022qdl} for { an} evaporating black hole, and can also be interpreted in terms of a final state projection \cite{Horowitz:2003he} and leads to the \ieftp\, perspective as discussed in Section \ref{sec:intrinsic}.\footnote{For more discussion on the final state proposal, see \cite{Gottesman:2003up,Bousso:2013uka,Almheiri:2025ugo}.} While there are certainly universe-level effects that are inconsistent with QFT in the closed universe, it is still plausible that observers limited to small subsets of the closed universe experience standard QFT and that their experience is perfectly semiclassical.\footnote{Indeed, even a large network of such observers spread over the whole universe and given ample time to communicate and compute would probably be unable to find evidence of any projection.} Similar ideas have been emphasized recently by \cite{Nomura:2025whc}. 

Zooming out, we want to emphasize that it is natural to have a high prior that the EG argument had to fail for a simple but compelling meta-reason. We clearly experience semiclassical physics in our own weakly gravitating corner of the universe. Yet it is certainly possible that the universe as a whole is closed. To claim that our experience of semiclassical physics is somehow evidence for the universe as a whole being open or having asymptotic boundaries or anything else is simply too great an extrapolation. 
To be clear, this is a meta-argument---we are not saying the literature contains exactly this claim---rather, we are claiming that it is clear that the semiclassicality of the closed universe must be compatible with it having, from the CFT perspective, a one-dimensional Hilbert space.

From the fact that the argument in EG fails, we only learn that a semiclassical closed universe is not ruled out by the analysis therein. Put differently, all the possible resolutions listed in AR to the puzzle raised there are back on the table. However, in this paper we are going further, arguing in favor of the semiclassicality of the closed universe, as we will discuss further in Sections \ref{sec:intrinsic} and \ref{sec:EFT}. 

So where do we stand in terms of the possible resolution of the puzzle raised in AR? As we reviewed above, the proposed categories of resolutions were ensemble averaging, non-semiclassicality of the closed universe, gauge equivalence, and AdS$\neq$CFT. As we have emphasized, we have a high prior for a semiclassical closed universe in the absence of a mechanism leading to its breakdown. Ensemble averaging actually does play a role; as we emphasize, this need not correspond to ensembles of theories as ensembles of heavy operators are sufficient for many purposes.\footnote{In the context of holographic models, ensembles of theories are best understood in low-dimensional toy models and it is unclear how they generalize to higher dimensions due to the paucity of CFTs. For an interesting idea to fix this problem, see \cite{Marolf:2024jze}. We also note that the condensed matter literature contains a huge variety of ensembles of higher-dimensional theories which may be useful for constructing examples.} In the absence of a concrete generalization of AdS/CFT, we are then led to the option of gauge equivalence, i.e., both descriptions are dual to the same CFT state. A somewhat different take on the idea of gauge equivalence is to say that the different descriptions are best suited to certain computations, for example, the naive bulk state in \iqft\, is suitable for computing averages of single replica observables, the wormholes in \ieftc\, are suitable for computing averages of multi-replica observables, and the projected bulk state in \ieftp\, is suitable for modeling specific microstates within a minimal extension of the semiclassical framework. In this way, different interpretations of the state of the closed universe are not right or wrong so much as they are tailored to different questions. We will discuss these ideas in further detail in Section \ref{sec:EFT}.

\textit{Note added: in the final stage of the preparation of this paper, the independent work \cite{higginbotham2025testsbabyuniversesadscft} appeared. It reaches a similar conclusion as to the inability of the EG proposal to distinguish the two descriptions in AR. }

%%%%%%%%%%%%%%%%%%%%%%%%%%%%%%%%%%%%
\section{Towards an intrinsic description of the closed universe}
\label{sec:intrinsic}
%%%%%%%%%%%%%%%%%%%%%%%%%%%%%%%%%%%%%

We now turn our attention to formulating a more intrinsic description of the closed universe. As we will see, CFT data still permeates the discussion, but the conceptual focus is different. We begin by discussing the meaning of the one-dimensionality of the closed universe Hilbert space and the associated notion of a closed universe final state. We then propose a dictionary relating intrinsic closed universe properties to CFT data. We also comment on the role of singularities and quantum chaos and discuss a model of the arrow of time. The reader can think of this section as fleshing out the idea of \ieftp\, discussed in the introduction. 

\subsection{The meaning of the one-dimensional Hilbert space}
\label{sec:meaning}

\begin{figure}[h]
    \centering
    \includegraphics[width=.98\linewidth]{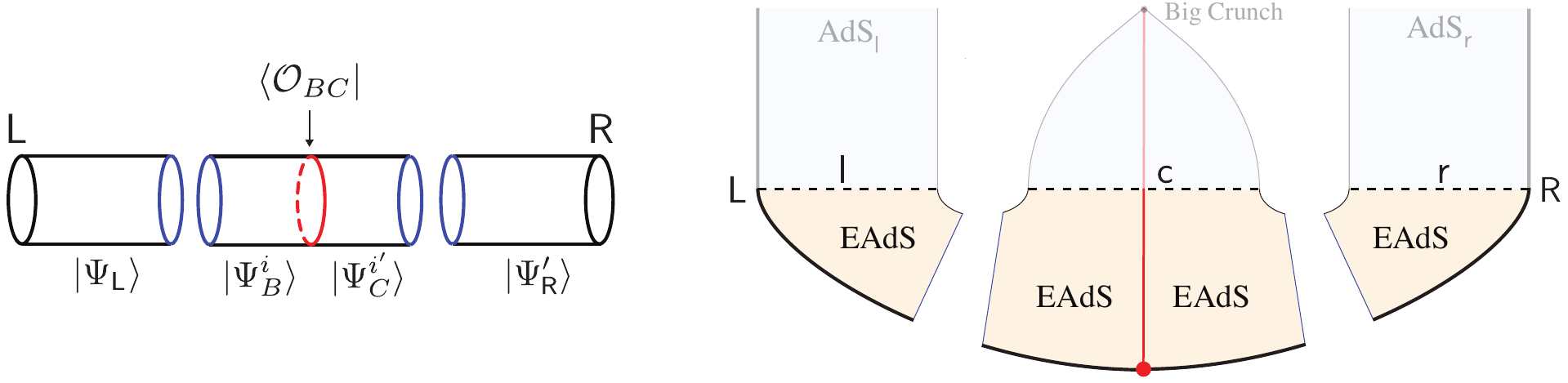} 
    \caption{On the left, the Euclidean CFT path integral that prepares this state. The blue boundaries represent a fixed low-energy input state, which could e.g. be the CFT ground state.  On the right, the bulk saddle point preparing the semiclassical unentangled state, represented in a way to make contact with Figure \ref{fig:Euclideanprep}.}
    \label{fig:Euclideanprepnoent}
\end{figure}

Let us take the regime of no entanglement. In this regime, we start from a factorized state $\ket{\Psi_{\mathsf{L}}}\ket{\Psi^i_{B}}\ket{\Psi^{i'}_{C}}\ket{\Psi'_{\mathsf{R}}}$ and project into $\bra{\Op_{BC}}$ as shown in Figure ~\ref{fig:Euclideanprepnoent}. The unnormalized CFT microstate corresponds to
\be\label{eq:microgluing0ent}
 |\tilde{\Psi}^I_{\mathsf{LR}}\rangle = \Op_{I}^* \ket{\Psi_{\mathsf{L}}} \ket{\Psi'_{\mathsf{R}}}\,, \quad \quad \Op_{I} = \bra*{\Psi^{i'}} \Op \ket*{\Psi^{i}}\,,
\ee
where we are suppressing the $B$ and $C$ labels as these are auxiliary Hilbert spaces. The Euclidean path integral that prepares this state is shown in Figure \ref{fig:Euclideanprepnoent}. A particular way to factorize the CFT path integral is to take infinitely long thermal evolutions after inserting the shell operator, as in \cite{Antonini:2023hdh}, but we will not restrict to this case here. In this regime, the information of the universe is just encoded in the matrix element $\Op_{I}$ where we are using the shorthand notation for the indices $I = i'i$. 

From the Hilbert space of the CFT, two microstates representing different states of the universe become indistinguishable, as they are exactly parallel quantum states. An alternative way to present this is to say that the product of these matrix elements defines the microscopic CFT Gram matrix of unnormalized overlaps between $\Omega$ different closed universe microstates
\be 
\widetilde{G}_{IJ}\equiv  \langle \tilde{\Psi}^I_{\mathsf{LR}}|\tilde{\Psi}^J_{\mathsf{LR}} \rangle \;\propto \;\Op_I \Op^*_J \quad\quad\text{for } I,J = 1,...,\Omega\,.
\ee
Thus $\widetilde{G}$ has rank $1$ for any $\Omega\geq 1$ and this implies that the dimension of the ``closed universe Hilbert space'' is therefore one, in accord with the often-cited result in the literature \cite{McNamara:2019rup,McNamara:2020uza,Marolf:2020xie,Antonini:2023hdh,Usatyuk:2024mzs,Harlow:2025pvj,Abdalla:2025gzn}.

We advocate that the result above is nothing specially deep about quantum gravity, but can be interpreted simply as a straightforward fact about any isolated quantum system:\footnote{Here we mean an eternally isolated quantum system which is in a pure state. One important difference between the closed universe and an isolated quantum lab is that, as hinted by the external CFT description, the closed universe is not a separate system because it is defined by a fixed global pure state. Therefore, the closed universe cannot be truly entangled to an external system, as opposed to an isolated quantum lab, which could be fundamentally entangled. A perspective to see this difference will be presented in Section \ref{sec:finalstate}.}

\begin{tcolorbox}[colframe=black, colback=white, boxrule=0.5pt, sharp corners]
\emph{The one-dimensionality of closed universes represents that from the external observer's Hilbert space (CFT in this case) every closed universe is indistinguishable.}
\end{tcolorbox}

Above we have kept the heavy operator $\Op$ fixed and defined the holographic map by considering its matrix elements between different light states. This amounts to having a fixed background in the closed universe. We could also consider modifying the physical properties of the heavy operator, like its rest mass, or including several of them,  which would lead to topologically and geometrically distinct closed universe backgrounds (see Appendix \ref{app:UniverseBottleneck} for an example). From the external observer (the CFT) the conclusion would still be the same: in the regime of no entanglement, all of the universes are indistinguishable and thus the external Hilbert space describing them is one-dimensional.

\subsection{Closed universe initial/final state}
\label{sec:finalstate}

For a quantum system in eternal isolation, the same one-dimensionality trivially follows if one tries to describe the system from an external Hilbert space. There is, however, something which works differently for the closed universe than for an isolated lab, as signaled by the holographic map. The difference is that the bulk state of the quantum fields gets always projected into the fixed state $\ket{\Op_{\co}}$ in the universe, instead of traced over states in $\co$.

 From the point of view of the holographic map, the projection in this regime is the extreme non-isometric case discussed in Section \ref{sec:2}. The specific state $\ket{\Op_{\co}}$ the map projects the closed universe onto was defined after \eqref{eq:bulktobdy} and it depends on the microscopics of the heavy CFT operator used to create the shell. In Figure \ref{fig:TNMERA} a tensor network model of the state was presented using an AdS-scale local tensor network of the closed universe developed in Appendix \ref{app:MERA}.

\begin{figure}[h]
    \centering
    \includegraphics[width=.75\linewidth]{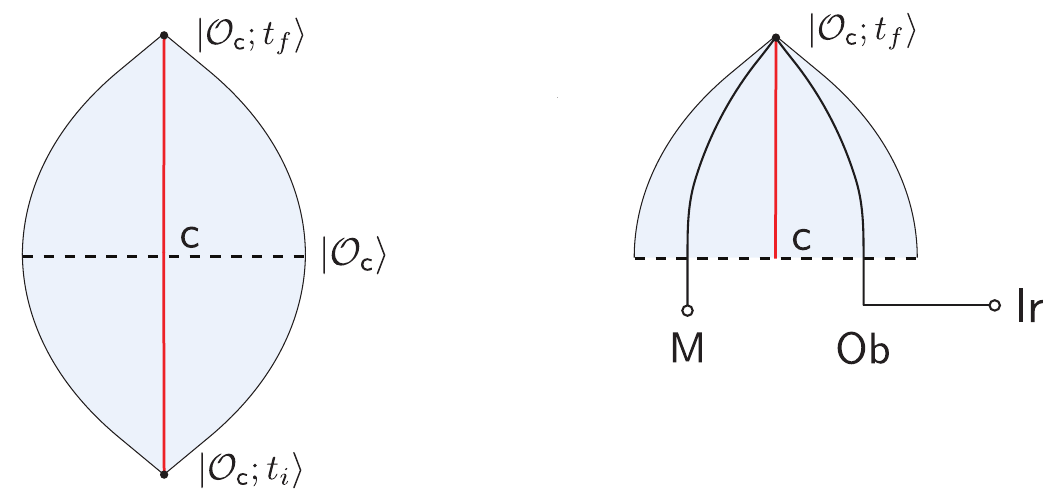} 
    \caption{On the left, the big bang / big crunch singularity implements an initial / final state projection compatible with each other. Both states propagate to $\ket{\Op_{\co}}$ at the time-symmetric slice.  On the right, the holographic map. The universe is prepared { in} an entangled state with $\mathsf{lr}$. The final state projection generates a backflow of quantum information from the singularity to $\mathsf{Ob}$, which finally reaches $\mathsf{lr}$ through the bulk entanglement. }
    \label{fig:gravinifinal}
\end{figure}

One possibility is that this projection physically happens in the closed universe, in analogy with the black hole final state projection \cite{Horowitz:2003he}. In this case, this can either be thought of as an initial/final state projection, happening because of the presence of a past/future singularity.\footnote{In closed universes without past/future singularities, such as those which eternally inflate, this projection can be specified at the asymptotic past/future boundaries.} 
Accordingly, the initial and final state projections are compatible with each other. The holographic map is then a Heisenberg picture model of such a projection, where the final state has been suitably propagated to the time-symmetric slice of the universe using the Hamiltonian defined by the background, as shown in Figure \ref{fig:gravinifinal}. From now on, we will refer to this as a final state projection, and comment on the initial state and the arrow of time in Section \ref{sec:arrowoftime}.\footnote{As we will see in Section \ref{sec:EFT}, the final state can be interpreted as a specific $\alpha$-state in the baby universe literature \cite{COLEMAN1988867,GIDDINGS1988854,Marolf:2020xie,Marolf:2020rpm,Saad:2021uzi}. We will give more details about the precise relation in Section \ref{subsec:AverageFinalStates}.}

Taking this perspective, the final state projection modifies the global structure of the quantum mechanics in the closed universe, irrespective of whether there is entanglement with the exterior or not. Some basic properties, such as causality, unitarity, and no-cloning, can be violated at the scale of the universe. Still, restricting to sufficiently local subsystems, one expects to recover conventional quantum mechanics without any projection \cite{Horowitz:2003he,Lloyd:2013bza}. This permits describing conventional effects such as decoherence and local observers in this framework. We will discuss in Section \ref{sec:outlook} possible subtleties in recovering local physics from a global decoherence functional when the initial/final states of the universe are pure (i.e., in the absence of entanglement with an external system). In what follows, we will refer to $\ket{\mathcal{O}_{\co}}$ as the final state of the closed universe in the sense that it is at least the projection implemented by the holographic map, without committing too much about its role in a theory for observers within the closed universe.

%, in the same manner as for an isolated quantum lab.   

\subsection{The dictionary}

We will propose that the CFT uniquely determines the microscopic structure of the final state of the closed universe. In \cite{Horowitz:2003he} the final state was conjectured to be sufficiently random.\footnote{For the evaporating black hole made from collapse, exact unitarity of the S-matrix requires some fine-tuning of the final state \cite{Gottesman:2003up}. In our case it is not clear that any fine-tuning is required.} Here, we will see that this is a consequence of the pseudo-random microscopics of heavy CFT data.

We will mainly motivate our proposal with the help of tensor network models, and leave a formal gravity argument for Appendix \ref{app:gravargument}. For simplicity, we consider the limit in which $m\ell/N^2 \gg 1$ so that the closed universe contains a large spatial volume (in AdS units) and the shell is in the asymptotic region. In this limit, we can parametrically distinguish the classical background generated by the thin shell from the perturbative excitations on top of this background.

\begin{figure}[h]
    \centering
    \includegraphics[width=.8\linewidth]{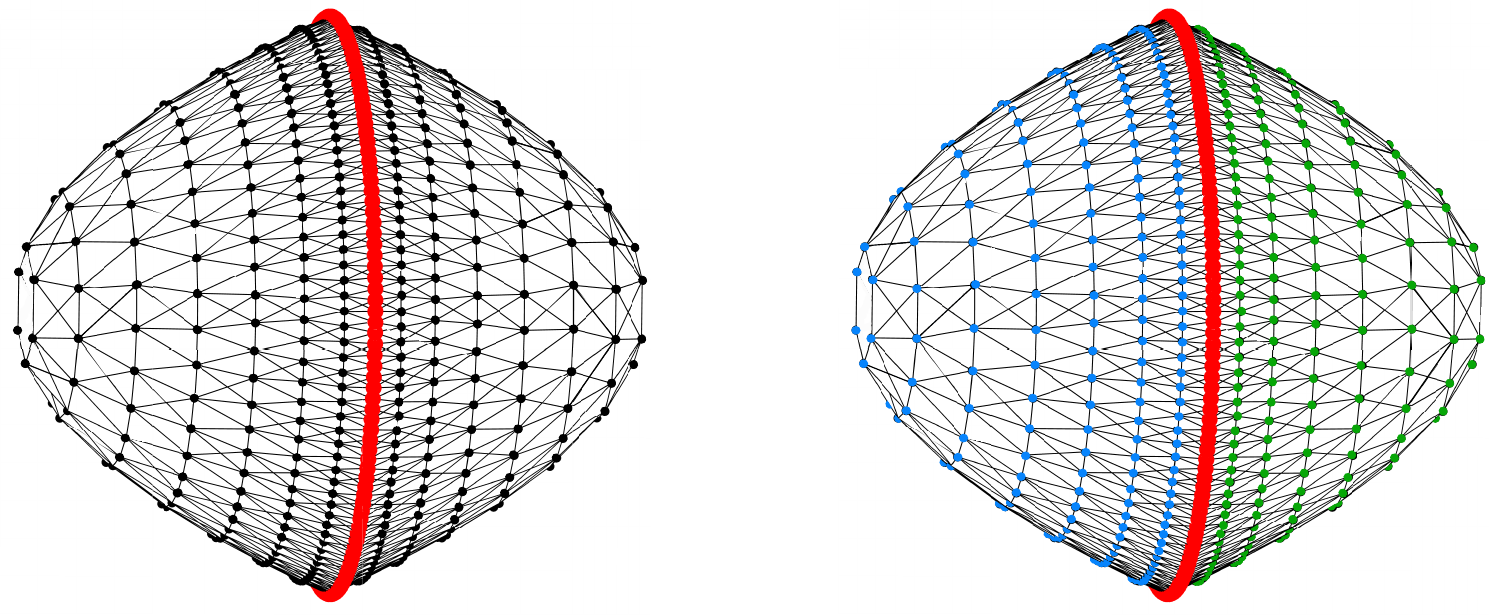} 
    \caption{The matrix elements define closed universe tensor networks. On the left, the closed universe tensor network computing the vacuum expectation value $\Op_{00} = \bra{0}\Op\ket{0}$ of the shell operator. On the right, the closed universe tensor network computing the matrix element $\Op_I = \bra*{\Psi^{i'}}\Op\ket*{\Psi^{i}}$ between two excited gas states  $\ket{\Psi^i}$ (blue) and $|\Psi^{i'}\rangle$ (green). These numbers determine the wavefunction of the closed universe final state $\ket{\Op_{\co}}$ represented in Figure \ref{fig:TNMERA}.}
    \label{fig:numbersuniverse}
\end{figure}

First, consider the matrix elements of the shell operator in a basis of light states, $\mathcal{O}_I = \bra*{\Psi^{i'}} \Op \ket*{\Psi^{i}}$. These matrix elements are microscopic, and in AdS/CFT they can be computed by suitable Euclidean CFT path integrals, as presented in Figure \ref{fig:Euclideanprepnoent}. Furthermore, the light states belong to the low-energy code subspace, so we can replace them by their bulk representation using the bulk-to-boundary map $V$, and write the light-light thin shell matrix element as
\be\label{eq:dictionaryTN} 
\mathcal{O}_I =\bra{\psi_I}\ket{\mathcal{O}_{\co}}\,,
\ee
where $\ket{\psi_I}$ is the corresponding state of the bulk fields, related to the CFT state by $V\ket{\psi^I } = \ket{\Psi^i}\ket{\Psi^{i'}}$ (recall that $V = V_\mathsf{L} V_\mathsf{R}$ is the product of two AdS isometries), and $\ket{\Op_{\co}}$ is the bulk state of the closed universe
\be\label{eq:statecloseduniverse} 
\ket{\Op_{\co}} = V^\dagger |\mathcal{O} \rangle \,.
\ee 
This leads to natural closed universe tensor network representations of the matrix elements \eqref{eq:dictionaryTN}, as shown in Figure \ref{fig:numbersuniverse}.\footnote{Although we are interpreting the tensor network in terms of the bulk geometry at the time-symmetric slice, it is also illuminating for the purposes of Appendix \ref{app:gravargument} to note that the tensor network can be interpreted as a suitable discretization of the CFT path integral in some conformal frame, as in \cite{Miyaji:2016mxg,Chandra:2023dgq,Geng:2025efs}.} 

Consider a basis of bulk states $\lbrace \ket{\psi_I}: I=1,...,d_{\co}\rbrace $. The closed universe final state can be expanded in this basis as
\be\label{eq:finalstateTN}
\ket{\Op_{\co}} = \sum_{I=1}^{d_{\co}} \Psi_{\Op}^I \ket{\psi_I}\,.
\ee 
The tensor network model of $\ket{\Op_{\co}}$ was presented in Figure \ref{fig:TNMERA}, constructed by adding bulk dangling legs to the MERA closed universe tensor network, which by construction implements \eqref{eq:statecloseduniverse}.

The intrinsic closed universe-to-CFT dictionary suggested by \eqref{eq:dictionaryTN} is thus:
\begin{tcolorbox}[colframe=black, colback=white, boxrule=0.5pt, sharp corners]
\vspace{-.7cm}
\begin{align}
 \textit{final state wavefunction  }&  \leftrightarrow \textit{   CFT light-light matrix elements of $\Op$}   \nonumber\\[.2cm]
 \Psi_{\Op}^I &\;=\;\Op_I\,. \label{eq:dictionary}
\end{align} 
\end{tcolorbox}

The final state wavefunction happens to be very simply related to the microscopic properties of the heavy shell operator. We believe that this could be a consequence of the point-like nature of the cosmological singularity in these setups. It is possible that the final state wavefunction is more involved for extended cosmological singularities, or in the black hole interior, but we will not study that here.

\subsubsection{Other closed universes, other final states}

Even if $\ket{\Op_{\co}}$ is the natural final state for the closed universe defined throughout this paper, this state is by no means the unique final state we could have defined for the same classical background. Recall that the state $\ket{\Op_{\co}}$ was defined via the holographic map, by keeping $\mathcal{O}$ fixed while studying its matrix elements between light states.

One can naturally think of semiclassically distinguishable final states of the same classical background by slightly modifying the semiclassical details of the heavy operator, adding a few dust particles, or by adding a few matter quanta anywhere else in the universe. One might think of these states as different universes really, as nothing inside of a closed universe can change its global state. Thus, a given universe, with a given semiclassical singularity, is {\it defined} by a global final state. 

In fact, one may also consider a mixed final state of the closed universe, where the CFT state is prepared by conditioning on an auxiliary system and then acting with different operators. This will lead to a non-trivial entanglement entropy for the CFTs, which in the boundary description can be understood as entanglement with the auxiliary system.

\subsubsection{Singularities and quantum chaos} 

An interesting fact is that this concrete proposal connects the quantum structure of singularities to the chaotic features of the CFT. It follows from \eqref{eq:dictionary} that the final state of the closed universe is quasi-random. This was important in the black hole final state \cite{Horowitz:2003he,Lloyd:2013bza} to guarantee that the sufficiently local quantum mechanics is unaffected by the projection.

Here, this property arises as a consequence of the quantum chaotic features of the CFT. The heavy operator $\Op$ is expected to have erratic matrix elements between light states of the CFT. The specific structure of these matrix elements can be viewed as a generalized version of the eigenstate thermalization hypothesis (ETH) \cite{PhysRevA.43.2046,PhysRevE.50.888,Belin:2020hea}. More precisely, the matrix elements $\Op_I$ are divided into a smooth and an erratic part. The erratic parts behave as i.i.d. random complex Gaussian variables, which leads to
\be\label{eq:ETHop} 
\overline{\Op_I} = 0\,,\quad\quad \overline{\Op_I \Op_J^*} = f_I\delta_{IJ}\,,
\ee
where $f_I$ is a smooth function over the $I$ index that can be found in \cite{Sasieta:2022ksu}. Here the overline represents the statistical coarse-graining implemented by the ETH ansatz. 

Such a model for light-light matrix elements of a heavy operator lies beyond the scope of conventional ETH, given that light bands of a holographic CFT are weakly chaotic. Still, evidence that supports this assumption comes mainly from two-dimensional CFTs \cite{Collier:2019weq} and from gravity \cite{Belin:2020hea,Stanford:2020wkf,Chandra:2022bqq,Sasieta:2022ksu}. In particular, higher moments, such as,
\be\label{eq:gaussianmoment} 
\overline{\Op_I \Op_J^*\Op_K \Op_L^*} =f_If_K\left( \delta_{IJ}\delta_{KL} + \delta_{IL}\delta_{KJ}\right)\,,
\ee
are reproduced by spacetime wormhole contributions to moments of the matrix elements of the shell operator \cite{Sasieta:2022ksu}.\footnote{We are ignoring non-Gaussianities, which can be incorporated and correspond to bulk diagrams where two shells cross. This kind of contribution is essential to reproduce higher-point correlators \cite{Foini:2018sdb} (see e.g. \cite{Belin:2021ryy,deBoer:2024mqg} for such corrections in the CFT data of two-dimensional CFTs reproduced by gravity).} The contribution \eqref{eq:gaussianmoment} is reproduced by the two-boundary wormholes connecting the four boundaries preparing the matrix elements.

As a consequence of this quasi-randomness, the final state appears locally thermalized, and the global projection becomes irrelevant to describe the universe locally.\footnote{This property of random states is often referred to as Page's theorem \cite{Page:1993df} or ``canonical typicality'' \cite{Goldstein:2005aib,Popescu_2006}.} In the thermodynamic sense, the final state of the closed universe is a highly entropic state. The initial state being highly entropic is only a consequence of the symmetry of the setup under time reflection.

\subsection{Arrow of time}
\label{sec:arrowoftime}

In this section, we draw inspiration from the idea of associating projections with singularities in the spacetime. In the context of the Page-Wootters formalism~\cite{PhysRevD.27.2885}, we show that the presence of both initial and final state projections has an interesting implication for the arrow of time. We exhibit a toy model in which the ground state of a Wheeler-DeWitt-like Hamiltonian spontaneously breaks time-reversal symmetry by taking the form of a Schrodinger's cat like state. The state is transformed into itself by time-reversal, but the terms in the superposition that accomplish this are macroscopically distinct. This phenomenon was anticipated by Page in \cite{PhysRevD.32.2496}. The considerations discussed here are also related to the quantum reference frame program~\cite{Bartlett_2007} and its application to problems in quantum gravity, e.g.~\cite{H_hn_2021}.

Consider a composite system consisting of a system $S$ and a clock $C$. In analogy with the Wheeler-DeWitt equation, $H_{\text{WDW}} | \psi_{\text{WDW}} \rangle =0$, we will introduce a system-clock Hamiltonian $H_{SC}$ and look for the ground state of the system-clock Hamiltonian.\footnote{In the notation introduced below, we could also enforce $H_{\text{sync}}=0$ on physical states as an analog of the Wheeler-DeWitt equation and look at the spectrum of $H_{SC}$ in this subspace.}

The system has a Hilbert space of dimension $D_S$ and a Hamiltonian $H$ which we will take to be random and of GOE form. Such a GOE Hamiltonian is a simple model of a $\mathsf{CRT}$-invariant bulk theory where we take the $\mathsf{CR}$ action to be trivial. Such a random matrix choice is of course unrealistic in detail but captures the basic intuition that the dynamics of the system should be chaotic.\footnote{As an example of a quantum chaotic system which presumably has random-matrix-like level statistics, the reader may have in mind the highly excited plasma in the early universe, prior to recombination.} What happens with a non-chaotic system dynamics is an interesting question that we do not address here. 

The clock has a Hilbert space with $2T+1=D_C$ states; here $T$ will play the role of the total time the system evolves for. The clock basis is labeled $|t\rangle_C$ for $t=-T,\cdots,0,\cdots,T$. See \cite{10.5555/863293} for an early use of such a construction in the theory of Hamiltonian complexity.

We also introduce a time-reversal symmetry implemented by an anti-unitary operator $\Theta$. After choosing a fixed reference basis $|a\rangle$ in $S$, this operator acts as complex conjugation combined with a basis transformation 
\begin{equation}
    \Theta |a \rangle_S |t \rangle_C = |a \rangle_S |-t\rangle_C.
\end{equation}
It will also be convenient to use $\Theta_S$, the restriction of this operator to act just on $S$; it is simply complex conjugation in the $|a\rangle$ basis. The Hamiltonian is time-reversal symmetric, $\Theta_S H \Theta_S^{-1}=H$.

The system-clock Hamiltonian consists of several pieces,
\begin{equation}
    H_{SC}=H_{\text{sync}} + H_{\text{init}} + H_{\text{final}},
\end{equation}
where
\begin{equation}
    \begin{aligned}
    H_{\text{sync}} &={ \sum_{t=-T}^{T-1}} \Bigl( I_S \otimes { \bigl(|t \rangle \langle t|_C+|t+1 \rangle \langle t+1|_C\bigr)} \\
    &\qquad{}- e^{-\iw H} \otimes |t+1\rangle \langle t|_C - e^{+\iw H} \otimes |t\rangle \langle t+1|_C \Bigr),
    \end{aligned}
\end{equation}
\begin{equation}
    H_{\text{init}} = g_{\text{init}} (I_S - P_0) \otimes |-T\rangle \langle -T|_C,
\end{equation}
and
\begin{equation}
    { H_{\text{final}} = g_{\text{final}} \Theta_S (I_S - P_0) \Theta_S^{-1} \otimes |T\rangle \langle T|_C.}
\end{equation}
{ When $g_{\text{init}}=g_{\text{final}}$, $H_{\text{final}}=\Theta H_{\text{init}}\Theta^{-1}$.} So far, $P_0$ is an arbitrary projector; we will make a particular rank one choice below. The first of these terms forces the system and clock to entangle, so that zero energy states are of the form
\begin{equation} 
    \sum_t e^{- i (t+T) H} |\psi\rangle_S  |t \rangle_C
    \label{eq:hist_state_general}
\end{equation}
for some system state $|\psi\rangle$. The second and third terms are zero energy when the system state is annihilated by { $I_S-P_0$ and $\Theta_S(I_S-P_0)\Theta_S^{-1}$} conditioned on the clock being in $|-T\rangle$ and $|+T\rangle$, respectively. We will speak of terms in $H_{SC}$ as being ``satisfied'' when they annihilate a state.

We can also check that the system-clock Hamiltonian is invariant under time reversal,
\begin{equation}
    \Theta H_{SC} \Theta^{-1} = H_{SC},
\end{equation}
so long as the Hamiltonian $H$ is real in the standard basis { and $g_{\text{init}}=g_{\text{final}}$}.

Now, when both $g_{\text{init}}$ and $g_{\text{final}}$ are zero, then any state of the form \eqref{eq:hist_state_general}
for any $|\psi\rangle$ is a zero energy state of $H_{SC}$. When $g_{\text{init}}>0$ and $g_{\text{final}}=0$, the zero energy states are of the form \eqref{eq:hist_state_general} with the additional constraint that $P_0 |\psi\rangle = |\psi\rangle$. Similarly for the case with $g_{\text{init}}=0$ and $g_{\text{final}}>0$. 

However, when both $g_{\text{init}}>0$ and $g_{\text{final}}>0$, then it is not clear if $H_{SC}$ has zero energy states or what the ground state looks like. It is possible that, for special choices of $|\psi_0\rangle$ and $H$, both $H_{\text{init}}$ and $H_{\text{final}}$ can be satisfied, but this is not generic.

So then what is the ground state of $H_{SC}$ in the generic case? There are two obvious candidates which satisfy either $H_{\text{init}}$ or $H_{\text{final}}$. Assuming for simplicity that $P_0 = |\psi_0\rangle \langle \psi_0|$ is rank one, the two states of interest are
\begin{equation}
    |\Psi_-\rangle = \frac{1}{\sqrt{2T+1}}\sum_t e^{- \iw (t+T) H} |\psi_0\rangle \otimes |t \rangle_C
\end{equation}
and
\begin{equation}
    |\Psi_+\rangle = \frac{1}{\sqrt{2T+1}}\sum_t e^{- \iw (t-T) H} \Theta_S |\psi_0\rangle \otimes |t \rangle_C.
\end{equation}
These two states correspond to a system state which either starts in $|\psi_0\rangle$ in the past ($-$) or ends up in state $\Theta_S |\psi_0\rangle$ in the future ($+$). These two states are mapped into each other under time-reversal,
\begin{equation}
    \Theta | \Psi_- \rangle = \frac{1}{\sqrt{2T+1}} \sum_t e^{\iw (t+T) H} \Theta_S |\psi_0 \rangle \otimes |-t\rangle_C = |\Psi_+\rangle.
\end{equation}

{ For equal boundary couplings,} the matrix elements of the system-clock Hamiltonian are
\begin{equation}
    (H_{SC})_{--} = { \frac{g_{\text{final}}}{2T+1}} \langle \psi_0 | e^{\iw (2T)H} (I_S - \Theta_S P_0 \Theta_S^{-1}) e^{-\iw (2T)H} | \psi_0 \rangle \equiv E_{--},
\end{equation}
\begin{equation}
    (H_{SC})_{++} = { \frac{g_{\text{init}}}{2T+1}} { \langle \Theta_S\psi_0 |} e^{-\iw (2T)H} (I_S -  P_0 ) e^{\iw (2T)H} { | \Theta_S\psi_0 \rangle} \equiv E_{++} = E_{--},
\end{equation}
and $(H_{SC})_{-+} = (H_{SC})_{+-}=0$. However, $|\Psi_-\rangle$ and $|\Psi_+\rangle$ are not orthogonal states, so we first construct an orthonormal basis to find the eigenvalues of $H_{SC}$ in the subspace. The overlap is
\begin{equation}
    \langle  \Psi_+ | \Psi_- \rangle = \frac{1}{2T+1} \sum_t { \langle \Theta_S\psi_0 |} e^{\iw (t-T)H} e^{-\iw(t+T)H} | \psi_0\rangle = { \langle \Theta_S\psi_0 |} e^{- i (2T) H} | \psi_0 \rangle \equiv \zeta.
\end{equation}

One orthonormal basis is spanned by $|u_1\rangle = |\Psi_-\rangle$ and 
\begin{equation}
    |u_2\rangle = \frac{|\Psi_+\rangle - \zeta^* |\Psi_-\rangle }{\sqrt{1-|\zeta|^2}}.
\end{equation}
In this orthonormal basis, the system-clock Hamiltonian reads
\begin{equation}
    H_{SC} \rightarrow E_{--} \begin{bmatrix}
        1 & \frac{- \zeta^*}{\sqrt{1-|\zeta|^2}} \\ \frac{- \zeta}{\sqrt{1-|\zeta|^2}} & \frac{1+|\zeta|^2}{1- |\zeta|^2}
    \end{bmatrix}.
\end{equation}

We focus on the physical limit of ${ |\zeta|} \ll 1$, which simplifies the expressions. The eigenvalues are $E_{--} (1 \pm |\zeta| + \cdots)$ and the eigenvectors are superpositions of $|\Psi_-\rangle$ and $|\Psi_+\rangle$. The ground state thus amounts to a Schrodinger's cat-like state for time-reversal symmetry. For example, if $|\Psi_-\rangle$ has some density perturbations which grow into the future (see e.g. \cite{Hawking_1993}), then $|\Psi_+\rangle$ has the same density perturbations growing into the past and the true ground state of $H_{SC}$ is a superposition of these two branches.

To verify that the restriction to the two-dimensional space spanned by $|\Psi_{\pm}\rangle$ accurately captures the ground state of $H_{SC}$, we implemented a simple exact calculation as follows. We fix $T$ and the dimension $D_S$ of the system, generate a random instance of a GOE Hamiltonian $H$, construct $H_{SC}$ with $g_{\text{init}}=g_{\text{final}}=1$, and numerically determine the entire spectrum. We verified numerically (1) that the ground state of $H_{SC}$ has the claimed cat state form, (2) that there is a small gap between the ground state and the first excited and a larger gap to the rest of the spectrum, and (3) that the ground state is primarily supported on the span of $|\Psi_{\pm}\rangle$.

We close this subsection with some comments on the main result. First, the system and the clock are both physical and should be viewed as different parts of the overall closed universe, e.g. the clock might be the scale factor or some co-evolving scalar field and the system would be the rest of the bulk degrees of freedom. Although we have idealized the system and clock as not interacting, this is not strictly necessary: the Schrodinger's cat like ground state, being a consequence of discrete symmetry breaking, should be stable to any small perturbation of $H_{SC}$ that respects the symmetry.\footnote{Of course, such interactions may be constrained by other considerations as well, such as desiring a unitary or nearly unitary effective evolution of the system with respect to the clock.}

Second, we can see the arrow of time emerge for a given branch by looking at correlations. For example, given a system operator at time $t_0$, its system-clock version is
\begin{equation}
    O_{t_0} = \sum_t e^{-\iw (t-t_0)H} O e^{\iw (t-t_0)H} \otimes |t \rangle \langle t|.
\end{equation}
The expectation value of this operator in the state $|\Psi_-\rangle$ is 
\begin{eqnarray}
    \langle \Psi_- | O_{t_0} | \Psi_- \rangle & = \frac{1}{(2T+1)} \sum_{t} \langle \psi_0 | e^{\iw (t +T) H} e^{-\iw(t-t_0)H} O e^{+\iw(t-t_0)H} e^{-\iw (t +T) H} |\psi_0\rangle \\
    & = \langle \psi_0 | e^{\iw (t_0 + T) H} O e^{-\iw (t_0 + T) H} | \psi_0\rangle,
\end{eqnarray}
which is the conventional expectation value at time $t_0$. A similar formula holds for the $|\Psi_+\rangle$ expectation value, with a suitably adjusted initial state at $-T$. Crucially, however, the cross matrix element is small for a generic $O$,
\begin{eqnarray}
    \langle \Psi_+ | O_{t_0} | \Psi_-\rangle & = \frac{1}{2T+1} \sum_t { \langle \Theta_S\psi_0 |} e^{\iw (t-T)H} e^{-\iw(t-t_0) H} O e^{\iw (t-t_0) H} e^{-\iw (t+T){ H}} | \psi_0 \rangle \\
    & = { \langle \Theta_S\psi_0 |} e^{\iw (t_0-T)H} O e^{-\iw(t_0 + T) H} | \psi_0 \rangle \sim  \langle \Psi_+ | \Psi_- \rangle \approx 0.
\end{eqnarray}
This indicates that correlation functions with multiple $O$s inserted will decompose into two decoherent branches, one in which the arrow of time is to the future and one in which it is to the past.

Third, although we motivated this section with the idea of initial and final state projections, in retrospect, we can distinguish two kinds of projections or preferred states. On one hand, we have the special system state which is singled out by the $H_{\text{init}}$ and $H_{\text{final}}$ terms in $H_{SC}$. These terms can be viewed as trying to enforce some version of initial and final state projections in a very direct way. In other words, it makes sense to say that this sort of projection is occurring at a particular moment in time (relative to the background geometry or clock). On the other hand, we have the overall state of the system plus clock. Projecting onto the ground state of $H_{SC}$ is in a sense ``timeless'' because it includes the clock degree of freedom and cannot be associated with any one moment in time. One could view this overall state as an analog of the 1d Hilbert space of the closed universe, although we did allow ourselves a larger Hilbert space in its construction. It is worth delving more into how these two classes of preferred states relate to the final state proposal in the black hole context.

%%%%%%%%%%%%%%%%%%%%%%%%%%%%%
\section{Recovering QFT in the closed universe via coarse-graining}
\label{sec:EFT}
%%%%%%%%%%%%%%%%%%%%%%%%%%%

In this section, we will show that the conventional global EFT of the closed universe, without any final state projection, can still be described from the CFT if one chooses to implement an average over the microscopics of the final state. From the bulk point of view, this can be interpreted as saying that conventional EFT, ignoring final state projections at terminal spacelike singularities, is equivalent to the maximally ignorant description of the final state. We will also comment on the mathematical relation to a statistical description over baby universe $\alpha$-states. The reader can think of this section as discussing the relationship between \iqft, \ieftc\, and \ieftp\, as discussed in the introduction and how [QFT] can be probed from the point of view of the CFT.

\subsection{ETH ensemble}
\label{subsec:ETHensemble}

The way to consider statistics in this context is to promote the ETH ansatz of the heavy operator $\mathcal{O}$ in \eqref{eq:ETHop} to an actual ensemble of microscopic CFT operators $\{\mathcal{O}^{(i)}\}$ compatible with the thin shell, as in \cite{Sasieta:2022ksu}. Here $i$ labels the operator drawn from the ETH ensemble. 

The heavy shell in the bulk can be approximated by any of the typical draws of the ensemble. Let us consider how to see this in terms of the light matrix elements of the heavy shell operator. Since the operator is heavy, applying each ${ \mathcal{O}^{(i)}}$ to the CFT ground state (or any other light state), the corresponding states $\mathcal{O}^{(i)}|gs\rangle$ will have approximately the same energy $E_i$ within some very narrow microcanonical band. Projecting this heavy state to the ground state produces the matrix element $\langle gs|\mathcal{O}^{(i)}|gs\rangle$. This matrix element is expected to behave erratically as a function of the particular instance of the heavy operator $\mathcal{O}^{(i)}$. From the CFT side, this erraticity is similar to how light-light-heavy OPE coefficients are expected to behave \cite{Collier:2019weq,Belin:2020hea}, as discussed in Section \ref{subsec:AverageFinalStates}. 

From the gravitational path integral point of view, in the semiclassical approximation, the insertion of $\mathcal{O}^{(i)}$ corresponds to the insertion of a spherically symmetric shell. The matrix element $\langle gs|\mathcal{O}^{(i)}|gs\rangle$, computed in gravity, vanishes. The interpretation is that the semiclassical gravity description of the shell does not have access to the microstructure of its matrix elements, but only to coarse-grained properties over these matrix elements. In particular, when computing the variance of the matrix element $|\langle gs|\mathcal{O}^{(i)}|gs\rangle |^2$ gravitationally, a connected wormhole contribution exists in which the shell propagates through the wormhole. In this case, the wormhole can be precisely constructed as the zero temperature limit of the Euclidean saddle computing the norm of the PETS discussed in Section \ref{sec:2} (see also \cite{Antonini:2023hdh}). This shows how the gravitational path integral, and thus the bulk EFT, does not compute the exact microscopic value of the matrix elements, but only { their} characteristic size over the ensemble. This structure is consistent with a pseudorandom behavior for $\langle gs|\mathcal{O}^{(i)}|gs\rangle$. Thus, the bulk description is compatible with the ETH ensemble. 
We can then approximate $\langle gs|\mathcal{O}^{(i)}|gs\rangle$, and similar one-point correlators on other light states, with random variables with zero average and non-zero variance, which leads to the generalized ETH ensemble \eqref{eq:ETHop}.

An important comment is that generic operators of the ETH ensemble will not be constructed as products of local primaries of low conformal dimension, and they will be completely non-local. Their semiclassical description will nevertheless be compatible with a pressureless domain wall of rest mass $m$, and thus with the same closed universe background. It is important to note that: (1) the ensemble of operators lives in a single CFT, so there is no need to invoke an averaging over theories, and (2) the use of the ensemble is purely practical here, we just want to find a CFT way to represent the conventional global [QFT] in the closed universe.

\subsection{Statistical average over final states}
\label{subsec:AverageFinalStates}

We can think of each operator of the ensemble as having its own microscopic final state $\ket{\Op_{\co}}$ given by the dictionary \eqref{eq:dictionary}. The statistical average over final states is given by
\be
\overline{\rho} = \overline{\ket{\Op_{\co}}\bra{\Op_{\co}}} = \dfrac{1}{d_{\co}}\sum_{I=1}^{d_\co}  \ket{\psi_I}\bra{\psi_I}\,.
\ee
This follows from \eqref{eq:ETHop} using the fact that in the large mass limit the microcanonical two-point function of the heavy shell becomes independent of the energy, and that, moreover, the entropy-suppression characteristic of the conventional ETH ansatz vanishes (see \cite{Sasieta:2022ksu}), yielding $f_I \,\propto\, 1$. Thus, we find that the average state is maximally mixed
\be 
\label{eq:MaximallyMixedFinalState}
\overline{\rho}  = \dfrac{1}{d_{\co}}\mathbf{1}_{\co}\,.
\ee
We can think of this average as a coarse-grained description of the final state, implemented by the completely depolarizing quantum channel. The average state \eqref{eq:MaximallyMixedFinalState}, when inserted as a final state, is equivalent to having no projection at the singularity, and thus it recovers the conventional EFT description in the closed universe. 

\begin{figure}[h]
    \centering
    \includegraphics[width=.9\linewidth]{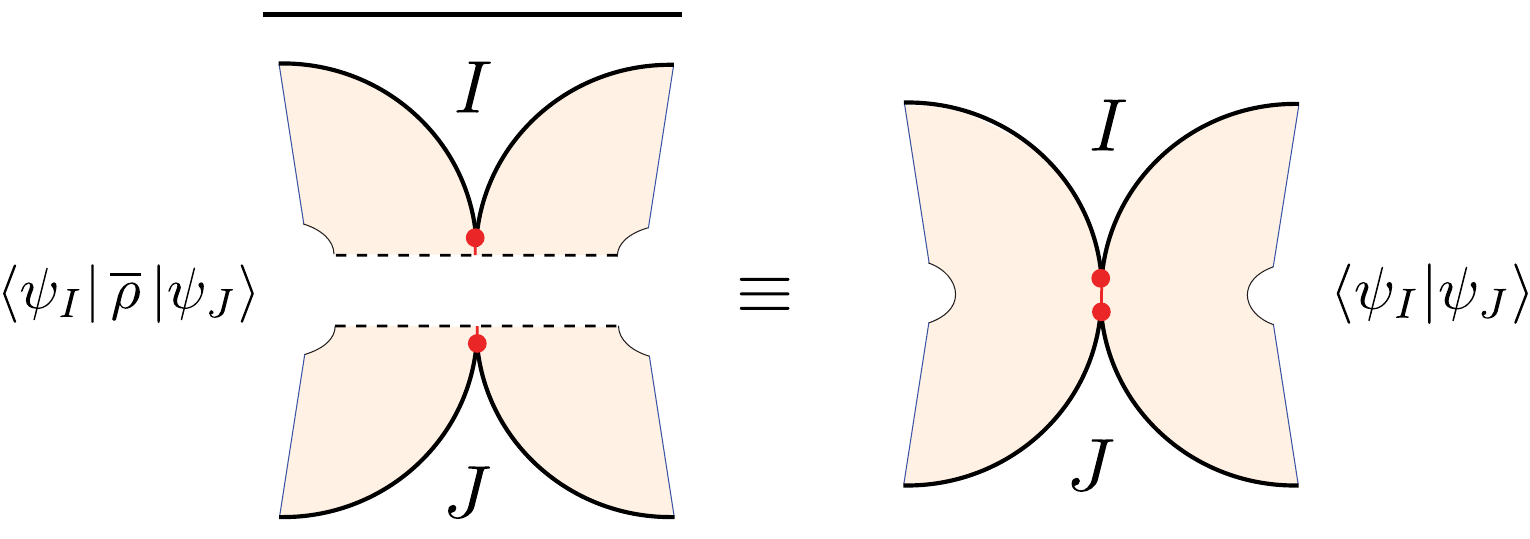} 
    \caption{The statistical average over final states generates the wormhole and prepares the bulk state $\ket{\psi_I}$ at the time-symmetric slice when $I=J$. The dashed line on the left represents the final state projection.}
    \label{fig:coarsegrainedstategpi}
\end{figure}

As an example, say we are preparing a state $\ket{\psi_{\co}}$ in the closed universe as in Figure \ref{fig:Euclideanprepnoent}, obtaining numbers in the CFT corresponding to the overlap of this state with the final state $\bra{\Op_{\co}}\ket{\psi_{\co}}$. We can repeat the experiment multiple times taking independent draws of the CFT operator in the ETH ensemble, and correspondingly of different final states, while fixing the light state we want to prepare. The statistical average of { products of an overlap and a conjugate overlap} will generate the wormhole, and this will prepare the state $\ket{\psi_{\co}}$ in the closed universe, with no projection associated with it. This is represented in Figure \ref{fig:coarsegrainedstategpi}.

\subsubsection{Replicas and dimensions}

When there are replicas of the closed universe, we have a choice to make about how the coarse-graining is implemented: independently on each replica or a single time for all of the replicas. If we average independently on each replica, this is equivalent to applying the maximally depolarizing channel in the replicated Hilbert space, which yields
\begin{equation}\label{eq:replicascg}
\overline{\rho} \otimes \dots \otimes \overline{\rho} = \dfrac{1}{d_{\co}}\mathbf{1}_{\co} \otimes \dots \otimes \dfrac{1}{d_{\co}}\mathbf{1}_{\co} \, .
\end{equation}
This reproduces the [QFT] results in multiple copies of the closed universe background. Each bra is correlated with its corresponding ket, so the gravitational path integral preparation corresponds to multiple copies of the right diagram in Figure \ref{fig:coarsegrainedstategpi}. This essentially agrees with the rules of \cite{Abdalla:2025gzn} at the perturbative level. From the CFT, such a coarse-graining is attained by picking an independent draw of the microscopic operator on each of the replicas, and then taking the average. Using these rules, the by now conventional gravitational resolvent calculation yields
\be 
\text{rank}(\overline{\rho}) = d_{\mathsf{c}}\,,
\ee
to be the dimension of the [QFT] Hilbert space. This is another example of the fact that using $\overline{\rho}$ is equivalent to having no projection, and recovering a conventional EFT.

\begin{figure}[h]
     \centering
     \includegraphics[width=.9\linewidth]{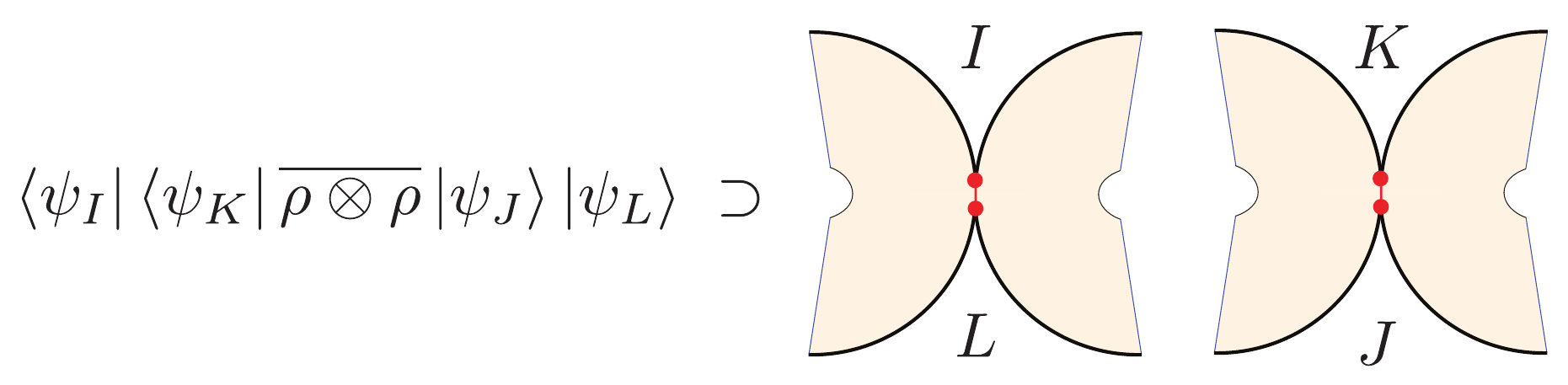} 
     \caption{Inter-replica correlated saddle point of the gravitational path integral.}
     \label{fig:lb}
\end{figure}

If instead we perform a correlated average for all the copies, we can access the average value of non-linear quantities depending on the final state $\rho$. This is more fine-grained than using $\overline{\rho}$, in that it provides information about the fact that there is a  pure final state, even though it does not know about the precise final state. From the CFT, the way to access this information is to pick the same microscopic operator in all of the replicas, and then to take the average. This produces extra contractions as a consequence of the Gaussian statistics, much like in \eqref{eq:gaussianmoment}. As an example, for two replica quantities, the average state is
\begin{equation}
\label{eq:TwoCopyRhoCoarseGrained}
\overline{\rho \otimes \rho} \; \propto \;\overline{\rho} \otimes \overline{\rho} + (\overline{\rho} \otimes \overline{\rho}) \mathcal{S} \, ,
\end{equation}
with $\mathcal{S}$ the swap operator interchanging the bras between copies. This state is invariant under left/right multiplication by $\mathcal{S}$. From the perspective of the gravitational path integral, the matrix elements of the first term are prepared by two copies of the diagram in Figure \ref{fig:coarsegrainedstategpi}. The second term is sketched in Figure \ref{fig:lb}. The generalization to multiple replicas is straightforward; by implementing all of the Wick contractions, we get an average state $\overline{\rho \otimes ... \otimes \rho}$ which is fully symmetrized between replicas. Accordingly, the gravitational path integral computation of this quantity contains one term for each possible pairing of bras and kets. Using a resolvent calculation, this allows { us} to access
\be\label{eq:unitrank} 
\overline{\text{rank}(\rho)} = 1\,,
\ee
consistent with the fact that the final state is pure.

To sum up, the two rules are different ways to coarse-grain the fact that there is a final state. Both of them coincide in single-replica quantities depending on $\overline{\rho}$. However, the former \eqref{eq:replicascg} is a self-consistent [QFT] description in many replicas, without any replica wormholes involved. The latter, where we include the effects of replica wormholes via the standard rules of the gravitational path integral, yields a more fine-grained [EFT], but one that is not fully self-consistent, because it considers $\overline{\rho}$ for single replica quantities, \eqref{eq:TwoCopyRhoCoarseGrained} for two replica quantities, and so on. Both of the rules can be explicitly understood from the CFT doing different experiments with the ensemble of microscopic heavy operators, the difference being whether we correlate draws of the ETH ensemble of operators between different replicas or not.\footnote{This discussion is clearly reminiscent of black hole evaporation, see e.g. \cite{Penington:2019kki,Almheiri:2019qdq,Bousso:2020kmy}, in which the growing entropy of the Hawking state (the equivalent of [QFT] in our setting) emerges from averaging the state in each replica separately, whereas the Page curve emerges when including spacetime wormholes, which implement correlated averages among replicas (the equivalent of [EFT] in our discussion).}

\subsubsection{Connection to $\alpha$-states in a many-universe Hilbert space}

To make contact with some of the ideas that have been developed in the literature \cite{COLEMAN1988867,GIDDINGS1988854,Marolf:2020rpm,Marolf:2020xie,Saad:2021uzi}, we can rephrase our discussion in terms of $\alpha$-states and a Hilbert space of baby universes. Consider the components of the final state in a basis of bulk states. For a microscopic instance of the heavy operator,
\begin{equation}
\bra{\psi_J} \rho \ket{\psi_I} = \mathcal{O}_J \mathcal{O}_I^{*} \, .
\end{equation}
The coarse-grained final state is then, using the explicit form of the Gaussian measure
\begin{equation}
\overline{\bra{\psi_J} \rho \ket{\psi_I}} = \mathcal{N}^{-1} \int \prod_K ({\rm d}\mathcal{O}_K {\rm d}\mathcal{O}^{*}_K) e^{-\sum_K  |\mathcal{O}_K|^2}  \mathcal{O}_J \mathcal{O}_I^{*} \; \propto \; \delta_{IJ} \, ,
\end{equation}
where $\mathcal{N}$ is the appropriate normalization factor. The generalization to $n$ copies is immediate
\begin{equation}
\overline{\bra{\psi_{J_1}, \ldots, \psi_{J_n}} \rho_1 \otimes \dots \otimes \rho_n \ket{\psi_{I_1}, \dots, \psi_{I_n}}} = \mathcal{N}^{-1} \int \prod_K ({\rm d}\mathcal{O}_K {\rm d}\mathcal{O}^{*}_K) e^{-\sum_K  |\mathcal{O}_K|^2}  \mathcal{O}_{J_1} \mathcal{O}_{I_1}^{*} \dots \mathcal{O}_{J_n} \mathcal{O}_{I_n}^{*} \, .
\end{equation}

The Gaussian integral over microscopic matrix elements $\mathcal{O}_I$ can be reinterpreted in the language of a Hilbert space of bosonic excitations. Given its form, we end up with a set of complex harmonic oscillators, with $\mathcal{O}_I$ acting as position variables. Associated to them, we introduce Hilbert space operators $\hat{\alpha}_I$ such that the above integrals compute vacuum expectation values,
\begin{equation}
\overline{\bra{\psi_{J_1}, \ldots, \psi_{J_n}} \rho_1 \otimes \dots \otimes \rho_n \ket{\psi_{I_1}, \dots, \psi_{I_n}}} = \bra{\Omega} \hat{\alpha}_{J_1} \hat{\alpha}^{\dagger}_{I_1} \dots \hat{\alpha}_{J_n} \hat{\alpha}^{\dagger}_{I_n} \ket{\Omega} \, .
\end{equation}
All the $\hat{\alpha}_I$ and $\hat{\alpha}_{I}^{\dagger}$ commute, and thus we can introduce simultaneous eigenstates for all of them ($\alpha$-states):
\begin{equation}
\hat{\alpha}_I \ket{\mathcal{O}} = \mathcal{O}_I \ket{\mathcal{O}} \, , \qquad \hat{\alpha}^{\dagger}_I \ket{\mathcal{O}} = \mathcal{O}^{*}_I \ket{\mathcal{O}} \, .
\end{equation}
Using this notation, if we compute expectation values in an $\alpha$-state instead of in the oscillator vacuum,\footnote{This state is sometimes called the Hartle-Hawking state in the baby universes literature, although it is not the same as the no-boundary state construction.} we would get the ($n$-copied) final state projection associated with a specific microscopic operator $\mathcal{O}$,
\begin{equation}
\bra{\mathcal{O}} \hat{\alpha}_{J_1} \hat{\alpha}^{\dagger}_{I_1} \dots \hat{\alpha}_{J_n} \hat{\alpha}^{\dagger}_{I_n} \ket{\mathcal{O}} = \mathcal{O}_{J_1} \mathcal{O}_{I_1}^{*} \dots \mathcal{O}_{J_n} \mathcal{O}_{I_n}^{*}
\end{equation}

To summarize, the different final state projections we have been discussing can be understood in the language of $\alpha$-states. A microscopic choice of operator $\mathcal{O}$ determines an $\alpha$-state, which just means that we have a definite final state projection. The main novelty introduced in Section \ref{sec:intrinsic} was to explain how such { a} final state can be determined from CFT data. If we coarse-grain in the way the gravitational path integral does, we lose information about the specific final state. This can be implemented by an appropriate Gaussian average or, equivalently, by computing expectation values in the oscillator vacuum $\ket{\Omega}$ (i.e. ``Hartle-Hawking state'') of the baby-universe Hilbert space.

\subsection{Reconstructing the closed universe from CFT data}

\label{sec:coarseCFT}

In this section, we explicitly demonstrate how simple coarse-grained CFT experiments, combined with sufficient operator statistics, reveal properties of [QFT] in the closed universe, such as its geometry, independent of the amount of entanglement in the bulk QFT description.

For concreteness, we consider the PETS state $\ket{\text{PETS}_i}$ presented in \eqref{eq:PETS}, where the $i$ index labels a specific shell operator $\mathcal{O}^{(i)}$, which can be thought of as one instance from the microscopic ETH ensemble of heavy shell operators discussed in Section { \ref{subsec:ETHensemble}}. In the limit of low temperatures, $\beta_{\mathsf{L},\mathsf{R}}\rightarrow \infty$, the state reduces to
\begin{equation}
      \frac{1}{\sqrt{Z_i}}\sum_{j,k}e^{-\frac{\beta_\mathsf{L} E_j+\beta_\mathsf{R} E_k}{2}}\mathcal{O}^{(i)}_{jk}|E_j\rangle|E_k\rangle \to \frac{1}{\sqrt{Z_i}} e^{{ -\frac{(\beta_\mathsf{L}+\beta_\mathsf{R})E_{gs}}{2}}}
\langle gs | \mathcal{O}^{(i)} | gs\rangle |gs_\mathsf{L} \rangle |gs_\mathsf{R} \rangle.
\label{eq:ground}
\end{equation}
The unnormalized state is then equal to the ground state, up to an overall constant proportional to $\langle gs| \mathcal{O}^{(i)} |gs\rangle$, where $\mathcal{O}^{(i)}$ is a heavy operator, such that the average energy of the state $\mathcal{O}^{(i)}|gs\rangle$ is $E_i=O(N^2)$. Taking the normalization $1/\sqrt{Z_i}$ into account, the state \eqref{eq:ground} is the vacuum up to a phase $e^{\iw\theta_i}=\langle gs| \mathcal{O}^{(i)} |gs\rangle/\sqrt{|\langle gs| \mathcal{O}^{(i)} |gs\rangle |^2}$. The specific phase depends on the specific heavy shell operator $\mathcal{O}^{(i)}$. We are of course quite used to discarding such constants (or overall phases) in quantum theory. However, this constant -- or more precisely the statistical properties of this constant as we vary the heavy operator $\mathcal{O}^{(i)}$ -- has interesting physics in it, as we will discuss.

The expectation value $\langle gs| \mathcal{O}^{(i)} |gs\rangle$ can be probed by a boundary CFT experiment using the following protocol. { On a finite-energy subspace of one copy of the CFT, define the controlled operator $CO=\mathbf{1}\otimes|0\rangle\langle0|+\mathcal{O}^{(i)}\otimes|1\rangle\langle1|$. Since $\mathcal{O}^{(i)}$ need not be unitary, choose a known $\kappa_i\geq\max\{1,\|\mathcal{O}^{(i)}\|\}$ and implement $K_i=CO/\kappa_i$ as a heralded Kraus operator. Defining $\nu_i=\langle gs|\mathcal{O}^{(i)\dagger}\mathcal{O}^{(i)}|gs\rangle$, the normalized output conditioned on success, starting from $|gs\rangle \frac{|0\rangle+|1\rangle}{\sqrt{2}}$, is}
\begin{equation}
   { |\chi_i\rangle} = \frac{|gs\rangle |0\rangle + \mathcal{O}^{(i)} |gs\rangle |1\rangle}{{ \sqrt{1+\nu_i}}}. 
\end{equation}
{ The success probability is $p_i=(1+\nu_i)/(2\kappa_i^2)$, so the measured success rate and known $\kappa_i$ determine $\nu_i$.} If we now prepare many copies of this state and repeatedly measure the ancilla in the $\sigma_x$ and $\sigma_y $ bases, then we can determine the matrix element via
\begin{equation}
    { \frac{1+\nu_i}{2}\left(\langle \sigma_x\rangle + \iw\langle \sigma_y\rangle\right)} = \langle gs |\mathcal{O}^{(i)} | gs\rangle
    \label{eq:phase}
\end{equation}
and therefore the overall phase of the state in equation \eqref{eq:ground} after normalization. This result can also be generalized to compute expectation values on { normalized} light states $|\mathcal{L}\rangle = { \frac{O_{\mathcal{L}}|gs\rangle}{\|O_{\mathcal{L}}|gs\rangle\|}}$ obtained by acting with a light operator (with scaling dimension $\Delta=O(1)$) on the ground state. Moreover, by employing another { heralded controlled operation} $CO_{\mathcal{L}}$ which conditionally applies a light operator depending on an ancilla, { and accounting for the corresponding success probabilities and normalization factors,} we can further obtain light-light matrix elements such as $\langle \mathcal{L}_1|\mathcal{O}^{(i)}| \mathcal{L}_2\rangle $.

Within the ETH ensemble of operators, the matrix element $\langle gs|\mathcal{O}^{(i)}|gs\rangle$ is a random variable with zero mean and non-zero variance. The expectation values of two different operators $\mathcal{O}^{(i)}, \mathcal{O}^{(j)}$ are uncorrelated in this approximation, i.e. $\overline{\langle gs|\mathcal{O}^{(i)}|gs\rangle { \langle gs|\mathcal{O}^{(j)}|gs\rangle^{*}}}=\sigma^2\delta_{ij}$.\footnote{{ In the circular complex Gaussian ETH ensemble, the phases $\theta_i$ defined after equation \eqref{eq:ground} are uniformly distributed.} This discussion should be understood in terms of the separate averaging for multiple replica quantities discussed in Section \ref{subsec:AverageFinalStates}, yielding [QFT].} From the bulk perspective, this simply means that, although the geometries and semiclassical states corresponding to two different microscopic operators are indistinguishable, their overlap vanishes at the disk level because of the microscopic ``internal'' degrees of freedom of the shell. This can be modeled by giving an internal global index to each shell created by an operator drawn from the ETH ensemble.
We remark that the statistical properties of this ensemble can be deduced directly from CFT experiments: we could compute, using the protocol defined above, $\langle gs|\mathcal{O}^{(i)}|gs\rangle$ for the heavy shell operators of interest $\mathcal{O}^{(i)}$ and deduce the mean, variance, and higher moments of the ensemble $\left\{\mathcal{O}^{(i)}\right\}$.

The picture emerging from this discussion is that semiclassical physics in the closed universe (i.e. [QFT]) can be used to answer specific questions in the CFT, namely coarse-grained questions -- where we do not keep track of the exact heavy operator in the ensemble $\{\mathcal{O}^{(i)}\}$ -- whose answer is well-approximated by an average over the ensemble. The statistics of the results of experiments in the CFT can in turn be used to deduce properties of the closed universe.

\begin{figure}[h]
    \centering
    \includegraphics[width=0.4\linewidth]{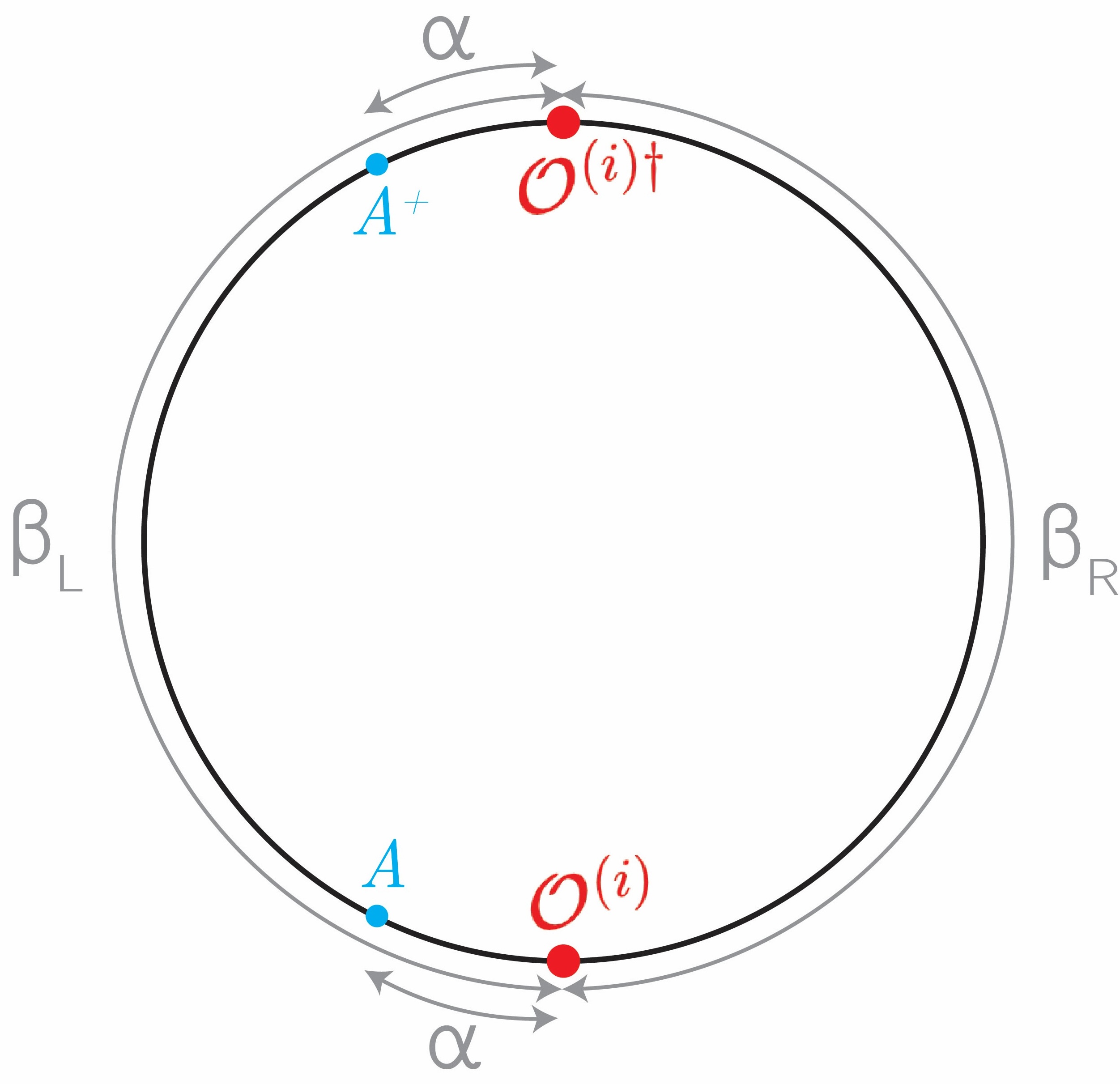}
    \caption{Boundary correlation function $G(\alpha,\alpha)$, with $\alpha$ fixed and $\beta_\mathsf{L},\beta_\mathsf{R}\to\infty$.}
    \label{fig:twoptfunction}
\end{figure}

As a practical example, consider a CFT operator $A$ with a vanishing one-point function in the ground state and heavy enough to be treated in the geodesic approximation (see e.g. \cite{Faulkner:2018faa}), but not so heavy { as} to backreact on the bulk geometry. This could be e.g. a primary operator with scaling dimension $1\ll \Delta_A\ll N^2$. 
We are interested in computing the two-point function 
\begin{equation}
   G(\alpha_1,\alpha_2)= \langle \text{PETS}_i| e^{\frac{(\beta_L-2\alpha_2) H}{2}} A^\dagger e^{-\frac{(\beta_\mathsf{L}-2\alpha_2) H}{2}} e^{-\frac{(\beta_\mathsf{L}-2\alpha_1) H}{2}}Ae^{\frac{(\beta_\mathsf{L}-2\alpha_1) H}{2}}|\text{PETS}_i\rangle\,.
\end{equation}
For simplicity, the primaries here are inserted at the same angular location in the CFT, so that the associated bulk geodesic will sit on the $r-\tau$ plane. Let us also consider $\alpha_1=\alpha_2\equiv \alpha$, although our analysis can be straightforwardly extended to $\alpha_1\neq\alpha_2$. We are interested in computing this correlation function for $\beta_\mathsf{L},\beta_\mathsf{R}\to\infty$ keeping $\alpha$ fixed. We represent the CFT correlation function in Figure \ref{fig:twoptfunction}.

\begin{figure}[h]
    \centering
    \includegraphics[width=0.5\linewidth]{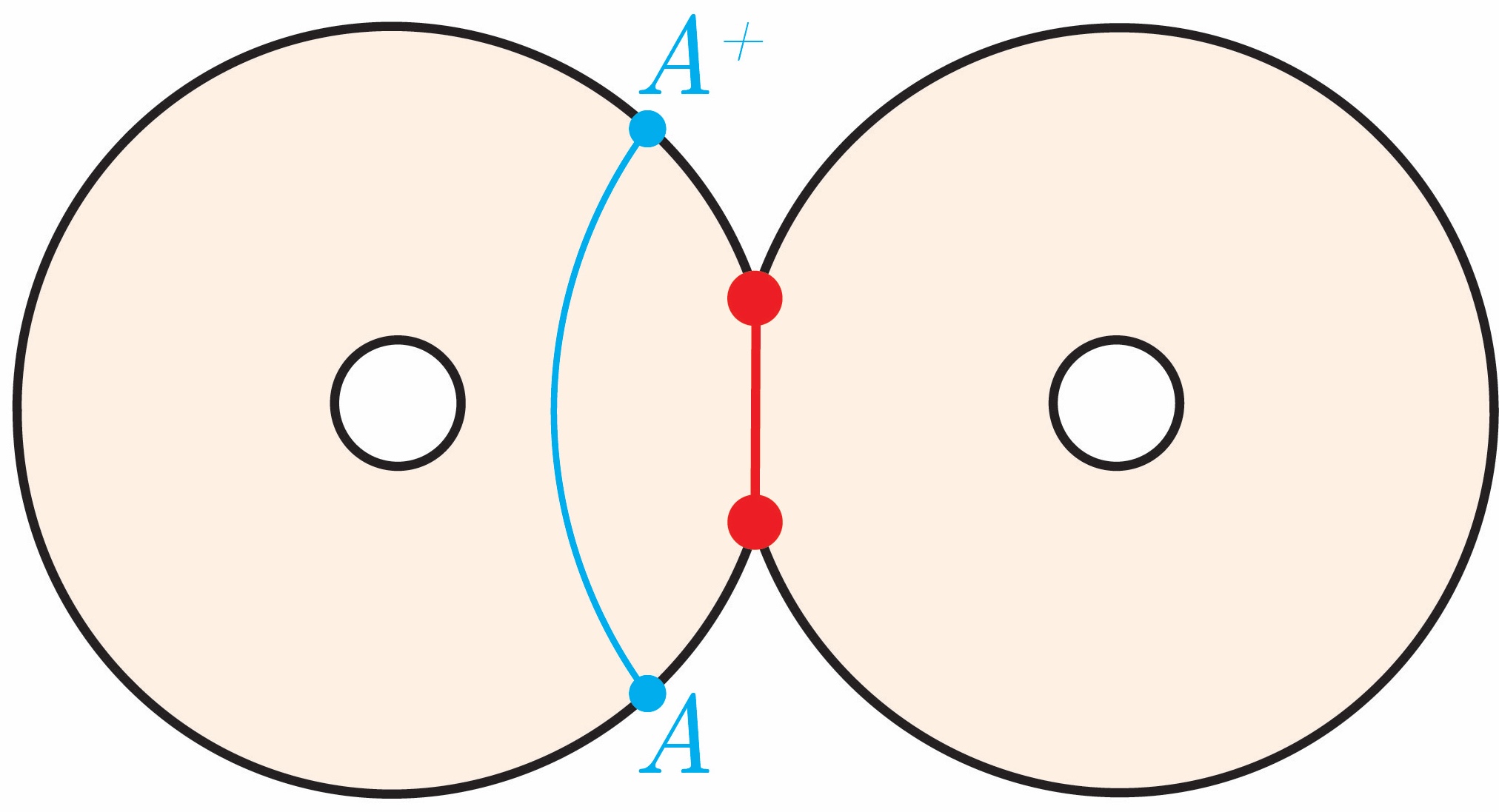}
    \caption{The saddle contributing to the two-point function. The on-shell gravitational action of the saddle yields a constant which cancels with the normalization of the state. The relevant quantity for the two-point function is the length of the blue geodesic $L(\alpha)$.}
    \label{fig:bulk2pf}
\end{figure}

We will first compute the two-point function using the gravitational path integral. In the (non-backreacting) particle { worldline} approximation, this is given by a geodesic connecting the insertions of $A$ and $A^\dag$, with an additional factor coming from the on-shell action of the saddle. This contribution is shown in Figure \ref{fig:bulk2pf}.

Because the on-shell gravitational action of the saddle is the same in the numerator and denominator, the two-point function reduces to
\begin{equation}
    \overline{G(\alpha)}=e^{-m_A L(\alpha)}
    \label{eq:gpi2pf}
\end{equation}
where the overline here means the quantity is computed using the gravitational path integral. Here $m_A$ is the mass of the bulk field dual to the primary operator, satisfying $m_A^2\ell^2=\Delta_A(\Delta_A-d)$ for a spin-$0$ primary, and $L(\alpha)$ is the renormalized geodesic length between the two boundary points where $A$ is inserted. The length is given by
\begin{equation}   L(\alpha)=2\ell\log\left[2\sinh\left(\frac{\alpha+\Delta\tau}{\ell}\right)\right],
\end{equation}
where $\Delta\tau$ is the amount of Euclidean time necessary for the dust shell to undergo half of its trajectory, depending on the shell's mass (see e.g. \cite{Sasieta:2022ksu,Antonini:2023hdh} for details). By varying the insertion point $\alpha$, this two-point function probes different regions of the closed universe. In fact, $\alpha$ is related to the radial location $r_0$ of the bulk particle in the closed universe at the reflection symmetric slice by
\begin{equation}
    r_0=\frac{\ell}{\sinh\left(\frac{\alpha+\Delta\tau}{\ell}\right)}.
\end{equation}
Notice that by increasing $\alpha$, we can probe regions further from the shell, all the way to the center of the closed universe in the limit $\alpha\to\infty$.

The two-point function computed with the gravitational path integral clearly contains information about the bulk geometry of the closed universe. Therefore, if we can measure this quantity for various $\alpha$'s from the CFT using a similar protocol to the one explained above, we can reconstruct the corresponding geodesic lengths, and then deduce the bulk metric from them (see e.g. \cite{Antonini:2022opp,Bilson:2010ff}). 

In order to reproduce this quantity from the CFT experiment, we write down the two-point function of interest
\begin{equation}
\begin{aligned}
    G(\alpha)&=\frac{1}{Z_i}\sum_{k,l,m,n}e^{-[\beta_\mathsf{R} E_k+\alpha E_n+(\beta_\mathsf{L}-2\alpha)E_m+\alpha E_l]}(\mathcal{O}^{(i)})^*_{{ lk}}A^*_{{ ml}}A_{mn}\mathcal{O}^{(i)}_{nk}\\
    &\to
  e^{2\alpha E_{gs}}\frac{\left|\langle gs|Ae^{-\alpha H}\mathcal{O}^{(i)}|gs\rangle\right|^2}{\left|\langle gs|\mathcal{O}^{(i)}|gs\rangle\right|^2}\,,
    \end{aligned}
\end{equation}
where in the second line we took the limit $\beta_\mathsf{L},\beta_\mathsf{R}\to\infty$ and we used the norm of the PETS in the same limit $Z_i=e^{-(\beta_\mathsf{L}+\beta_\mathsf{R})E_{gs}}|\langle gs|\mathcal{O}^{(i)}|gs\rangle|^2$. As we have already explained, the quantity in the denominator can be measured in the CFT. The quantity in the numerator can also be measured with a similar protocol combined with Euclidean evolution. Such Euclidean evolution can be simulated using a quantum computer; the computational resources to simulate it may be high, but we are not concerned with complexity limitations in this section. 

To make contact with a direct semiclassical description of the closed universe,  we must average these quantities over the ETH ensemble of shell operators $\{\mathcal{O}^{(i)}\}$.\footnote{Notice that in the gravitational path integral we computed the numerator and denominator separately and then took the ratio. Similarly, here we take the ratio of the averages of numerator and denominator.} We already encountered the average of the denominator, which is simply the variance $\sigma^2$ of the ensemble. The average of the numerator can again be obtained operationally by measuring $\left|\langle gs|Ae^{-\alpha H}\mathcal{O}^{(i)}|gs\rangle\right|^2$ for all operators $\mathcal{O}^{(i)}$ in our ensemble and then averaging the results. We thus obtain the averaged two-point function $\overline{G(\alpha)}$ purely from the statistics of CFT experiments. Identifying this with the two-point function \eqref{eq:gpi2pf} computed using the gravitational path integral, we can obtain the geodesic length $L(\alpha)$ for various $\alpha$'s and reconstruct the bulk geometry from our CFT data. 

To summarize, semiclassical physics (i.e. [QFT]) is encoded in the boundary theory in a rather unconventional way, namely in the statistics of boundary data instead of an approximately isometric bulk-to-boundary map between Hilbert spaces. Nonetheless, it can be probed from CFT experiments, and we gave an operational procedure for how to do so. As we have seen, as the entanglement between the closed universe and the AdS spacetimes grows, an approximately isometric encoding of the bulk Hilbert space into the dual CFT emerges, (approximately) recovering a more standard holographic dictionary.

One interesting open question is whether the ideas outlined in this section can be extended beyond the semiclassical approximation in the closed universe. Can the statistics of CFT data capture non-perturbative quantum gravity effects, such as the unitary evaporation of a small black hole in the closed universe or topology change effects? For an evaporating black hole, the answer is plausibly yes: an observer in the closed universe can collapse matter to form a black hole, collect radiation, and measure its entropy (provided the black hole is small enough to evaporate before the universe crunches) within the semiclassical regime, observing the Page curve. This is a complicated scattering experiment, whose outcome is captured by a number (the entropy of Hawking radiation) which could plausibly be computed using CFT data as explained above. In the future, it would be interesting to understand if and how topology change effects, and in general the microscopic description of quantum gravity in the closed universe, beyond [QFT] and [EFT], are related { to} the statistics of CFT data.

 In particular, one may ask if the closed universe description completely breaks down when one has precisely measured the microstate of the shell that forms the closed universe. From the CFT, we know that correlation functions would need to demonstrate an erratic behaviour which seems difficult to obtain from a smooth bulk geometry. However, the tensor network model suggests that the closed universe is well described by a smooth geometry everywhere away from the shell and the erratic behaviour comes precisely from the shell operator. In the bulk, this could potentially be realized by the interactions between light matter near the shell and the heavy shell itself. This is similar to what one would expect from the brane description in Kourkoulou-Maldacena states \cite{Kourkoulou:2017zaj}. 

Finally, we would like to briefly comment about the relationship of these ideas to models of observers in a closed universe recently proposed in \cite{Abdalla:2025gzn}. The modified rules for the gravitational path integral in the presence of an observer in \cite{Abdalla:2025gzn} can be interpreted as arising from a coarse-graining procedure over CFT data, which is similar in spirit despite some technical differences. The details of this idea will be discussed in upcoming work \cite{lucaupcoming}. The results of \cite{Abdalla:2025gzn} suggest that non-perturbative quantum gravity effects, including topology change, correcting the naive EFT (i.e. [QFT]) for an observer are captured by a Hilbert space larger than the simple CFT Hilbert space. Whether the same effects can be interpreted from the point of view of the coarse-grained CFT observables discussed in the present section is an interesting open question.

%%%%%%%%%%%%%%%%%%%%%%%
\section{Outlook}
\label{sec:outlook}
%%%%%%%%%%%%%%%%%%%%%%%%%%

We have spent quite some time exploring various perspectives and lessons stemming from the construction in \cite{Antonini:2023hdh}, so now we would like to pause and reflect { on} the broader picture. To recall the basic elements of the setup, there exist entangled states between two copies of a holographic CFT whose dual spacetime is formed by three disconnected pieces: two AdS regions and a crunching closed universe supported by a thin shell of dust. As discussed in Section \ref{sec:2}, one can arrive at this conclusion at least in two ways, either using tensor network models (and we developed a more detailed version of this in Appendix \ref{app:MERA}) or the perhaps more conventional Euclidean preparation via the gravitational path integral. Two key properties should be kept in mind. First, the size of the closed universe is controlled by the mass of the thin shell (in AdS units), and when we make this parameter large we get a very large closed universe. Curvatures away from the past and future singularities are small and there is no good reason to believe semiclassical physics should break down in such a situation. Second, the bulk states generically share entanglement between the AdS regions and the closed universe. In Section \ref{sec:ParameterSpace}, we discussed in detail the specific case of Type IIB string theory dual to $\mathcal{N} = 4$ SYM, showing that the entanglement entropy can scale with the local number of degrees of freedom $N^2$ of the CFT. In general, for a $(d+1)$-dimensional AdS, its entanglement entropy to the closed universe is upper bounded by $O(N^{2d/(2d-1)})$, corresponding to the entropy of the most entropic stable gas in AdS.

The two independent parameters of the setup correspond to the volume of the closed universe and its entanglement with the external AdS regions. Varying these parameters, in Section \ref{sec:encodinguniverse} we discussed the properties of the holographic encoding of the closed universe in the CFT in different regimes. When the entanglement entropy is sufficiently large, the situation is a rather conventional encoding, analogous to the situation for the black hole interior \cite{Akers:2022qdl}. In this case, there is an effectively isometric encoding (strictly non-isometric) of the closed universe in the CFT. A state-dependent reconstruction of the [QFT] in the closed universe background is possible using this encoding.

We have also described how the model relates to some recent related models in the literature. In particular, we have argued that the model presented in \cite{Harlow:2025pvj}, which relies on the definition of an ``observer'' and its ``clone'', is in essence a generalization of the setup in \cite{Antonini:2023hdh} to generic settings in which the Euclidean path integral does not prepare a state with [QFT] entanglement between the closed universe and an external reservoir (the clone). In both models, the closed universe is described within the Hilbert space of the clone, and its dimension is set by the entanglement between the observer and the clone. We have advocated that the physical definition of the observer, as a subsystem which decoheres the wavefunction in a basis of physical pointer states, was implicit in \cite{Antonini:2023hdh}. Moreover, although less general, the physical status of the clone is more transparent in the specific setup \cite{Antonini:2023hdh}, and it corresponds to a collection of entangled particles in the AdS regions.

We have also addressed the claims recently made in \cite{Engelhardt:2025vsp}, which question the validity of the AS$^2$ model and conclude that the closed universe cannot be semiclassical. In short, we argue that the SWAP operator discussed in \cite{Engelhardt:2025vsp}, which is supposed to give different results depending on whether or not a closed universe is present, is actually always just the boundary SWAP. Since, as a state of two CFTs, we are dealing with a pure state, the SWAP expectation value is always one, a fact that may be interpreted from the bulk perspective as saying that both the exterior AdS regions and the closed universe are swapped. The breakdown of semiclassicality advocated for in \cite{Engelhardt:2025vsp} follows only if the CFT inner product, which is not well-approximated by the bulk [QFT] in the regime of small entanglement, is the one relevant for the description of ordinary physics in the closed universe. We advocated this is not the case, and interpreted instead the large fluctuations (at low entanglement) in the [QFT] approximation of the CFT inner product as a breakdown of the ordinary holographic encoding of the closed universe.

In Section \ref{sec:intrinsic}, we considered the extremely non-isometric regime of zero entanglement and proposed a seemingly sensible and intrinsic description of the closed universe. Well-known arguments \cite{Harlow:2025pvj,Abdalla:2025gzn,McNamara:2019rup,Marolf:2020xie} indicate that the Hilbert space of the closed universe is one-dimensional in this regime. We have argued that this one-dimensionality simply represents the limitation of the external observer (CFT) to distinguish the quantum information in the closed universe, similar to the limitation of an external observer to describe an isolated quantum lab.

We argued that the difference, as signaled by the holographic encoding, is that there is a final state projection implemented in the universe.\footnote{This way it follows from \ieftp\, that the SWAP test of \cite{Engelhardt:2025vsp} always gives unit result in the closed universe description. After the final state projection, the bulk state of the external AdS regions is pure.} We proposed that the CFT determines the precise final state and provided a dictionary between its wavefunction and CFT data. We have motivated this dictionary from the tensor network model. An interesting feature of this proposal is that the quantum chaotic properties of the CFT make the final state quasi-random and thus locally thermalized. This is likely important if we expect the final state projection to leave the local physics in the closed universe unaltered \cite{Horowitz:2003he,Lloyd:2013bza}. Global quantum mechanics, on the other hand, is affected by the final state projection, irrespective of the amount of entanglement to the AdS spaces. This may sound problematic, but it is unclear if any final-state-sensitive observable could ever be physically measured by observers inside the universe, even given access to a widely distributed collection of communicating probes.

It is worth pointing out an important difference between our setup and the original discussion of the black hole final state proposal \cite{Horowitz:2003he}. In the black hole context, a consistent histories formulation with a final state projection was suggested as a way to describe the physics seen by an experimenter within the interior of the black hole. Applying the same logic to our setup in the regime of zero entanglement, i.e. allowing pure initial and final states of the closed universe, one does not recover conventional quantum mechanics: the decoherence functional for pure initial and final states is rank 1, and thus the standard quantum-mechanical measurement probabilities are not recovered, see e.g. \cite{hartle2020arrowstimeinitialfinal}.\footnote{We thank Hong Zhe (Vincent) Chen and Don Marolf for pointing this out to us and for discussing (the problems associated with) possible resolutions.} This conclusion is unchanged if the measurements are performed within sufficiently small subsystems and one starts from the global decoherence functional. However, it is unclear whether there is any observer in the closed universe that would rationally try to assign probabilities based on a pure initial and final state prescription. Said differently, the ``purpose'' of such a pure initial and final state decoherence functional is unclear.

One can consider different kinds of resolutions to these issues. One possible way to recover standard quantum mechanics for local subsystems is to instead use the consistent histories formulation directly for the subsystem, tracing out the complement both in the initial and final states. A related suggestion is to invoke the { observer's} ignorance, of the final state, of the precise UV completion, and so on, such that the pure state decoherence functional should be averaged over final states and Hamiltonians, thus effectively recovering a highly mixed final state. We showed that the CFT was able to yield such results in Section~\ref{sec:EFT}. Another is to declare that the experience of an observer is not directly affected by the final state, and just standard quantum mechanics with the usual decoherence via interactions with an environment is at play, at least until the singularity is approached. In this view, the final state projection has more to do with mapping some version of a closed universe state into CFT data. We explored such connections in Section~\ref{sec:intrinsic}. It is also possible that the final state enters in an unconventional way, such as in the arrow of time model in Section~\ref{sec:intrinsic}. In any case, the subject of how to make sense of the physics seen by an observer within the cosmology is a subtle one, and we leave for future work a more detailed exploration of the different possibilities.

It is worth emphasizing at this point that the final state projection is compatible with the reconstruction of operators in the closed universe via the non-isometric encoding when sufficient entanglement with exterior regions is present. In fact, the postselection in the non-isometric map is essentially equivalent to the final state projection discussed here. In the regime of large entanglement, the CFT can therefore probe the local semiclassical physics in the closed universe, which is preserved by the final state projection just like it is preserved by the projection arising in the non-isometric map, as it was discussed in \cite{Akers:2022qdl} for evaporating black holes after the Page time. We would like to point out that one advantage of the model introduced in \cite{Antonini:2023hdh} and considered here is that one can study questions analogous to the experience of interior observers in a black hole without having to deal with regions of large curvature where semiclassical physics would be expected to break down.\footnote{Here we mean the ``corner'' region of the Penrose diagram for an evaporating black hole.} 

We concluded in Section \ref{sec:EFT} explaining the relationship between the different bulk descriptions [QFT], [EFT], and \ieftp, and how [QFT] can be probed using the CFT independent of the amount of entanglement in the [QFT] description. The [QFT] description can be seen as arising as a maximally ignorant description of the details of the final state projection, or equivalently, after averaging over final states. From the CFT perspective, the statistics is taken over an ensemble of heavy microscopic operators compatible with the heavy shell, whose light matrix elements behave erratically. The [EFT] picture is recovered by performing a correlated average between different replicas for multiple-replica quantities. Most of this discussion can be rephrased in the language of $\alpha$-states in a baby universe Hilbert space \cite{COLEMAN1988867,GIDDINGS1988854,Marolf:2020rpm,Marolf:2020xie,Saad:2021uzi}. Finally, we explicitly showed an example of a CFT observable, together with sufficient statistics, which can probe [QFT] in the closed universe even in the absence of any entanglement in the [QFT] description.

We conclude this outlook with some future directions.

Taking the final state for granted, we have proposed a dictionary that relates this state in the AS$^2$ model to CFT data. It would be worth exploring whether a similar dictionary exists for more general cosmological singularities, and for the black hole interior.

{ In} a related direction, from the analysis of \cite{Horowitz:2003he} it seems natural to tie the final state projection to the presence of singularities. If constructions similar to those explored in \cite{antonini2022cosmologyvacuum,Antonini:2022ptt}---possibly entangled with exterior regions---can be found leading to eternally expanding universes with no singularity \cite{Betzios:2024oli,Betzios:2024zhf}, it would be interesting to understand the holographic encoding and whether or not a final state projection is implemented, possibly at the asymptotic future and past. From a Euclidean perspective, it seems possible to relate the final state to CFT data living at infinite Euclidean past and future in the AdS wormhole used to prepare the cosmological state, similar to the formal argument presented in Appendix \ref{app:gravargument}.\footnote{A similar relationship between CFT data at asymptotic Euclidean infinity and states of closed universes in the presence of an observer was discussed in \cite{Abdalla:2025gzn} and will be explored in more detail in upcoming work \cite{lucaupcoming}.}

Another interesting direction is to consider the holographic encoding at low entanglement, from the perspective of one-shot quantum information theory. To our knowledge, this subject was first discussed in a holographic context in \cite{Czech:2014tva} where it was observed that typically one has a thermodynamic limit which causes one-shot quantities to approach their many-copy versions. However, since then a comprehensive theory of one-shot quantities in holography has been developed \cite{Akers_2024}\footnote{See also \cite{Bousso:2023sya,Bousso:2024iry} for related work.} and it would be interesting to apply this to the closed universe model in \cite{Antonini:2023hdh} in the low entanglement regime where we do in fact expect a difference with many-copy results. Finally, we will report elsewhere on the possibility of using table-top quantum simulators to simulate the experience of observers in the closed universe, generalizing ideas discussed in~\cite{jafferis2021insidehologramreconstructingbulk,Gao_2022,deboer2022blackholeinteriorreconstruction}.

\section*{Acknowledgments} 

We thank Chris Akers, Jos\'e Barb\'on, Alex Belin, Horacio Casini, Hong Zhe (Vincent) Chen, Roberto Emparan, Netta Engelhardt, Elliott Gesteau, Daniel Harlow, Matt Headrick, Luca Iliesiu, Javier Mag\'an, Don Marolf, Geoff Penington, Moshe Rozali, Gonzalo Torroba, Misha Usatyuk, Mark Van Raamsdonk, and Ying Zhao for discussions. PR would like to dedicate this paper to his own new baby (universe), Gauri Jain Rath. Part of this work was performed at the Centro de Ciencias de Benasque, during the ``Gravity - New quantum and string perspectives'' conference.  SA is supported by the U.S.~
Department of Energy through DE-FOA-0002563 and by the UC Berkeley Department of Physics. PR is supported in part by the Leinweber Institute for Theoretical Physics; and
by the Department of Energy, Office of Science, Office of High Energy Physics under
Award DE-SC0025293. MS and BS gratefully acknowledge support from the Heising-Simons Foundation under grant number 2024-4849. { This work was supported in part by the U.S.~Department of Energy through the GeoFlow consortium (QuantISED Award DE-SC0019380).} AVL acknowledges support from the { Natural Sciences and Engineering Research Council of Canada} (NSERC) and the Simons Foundation via a Simons Investigator Award.
 
\appendix

%%%%%%%%%%%%%%%%%%%%%%%%%%%%%%%
\section{MERA tensor network of the closed universe}
\label{app:MERA}
%%%%%%%%%%%%%%%%%%%%%%%%%%%%%%%%%

\begin{figure}[h]
    \centering
    \includegraphics[width=0.85\linewidth]{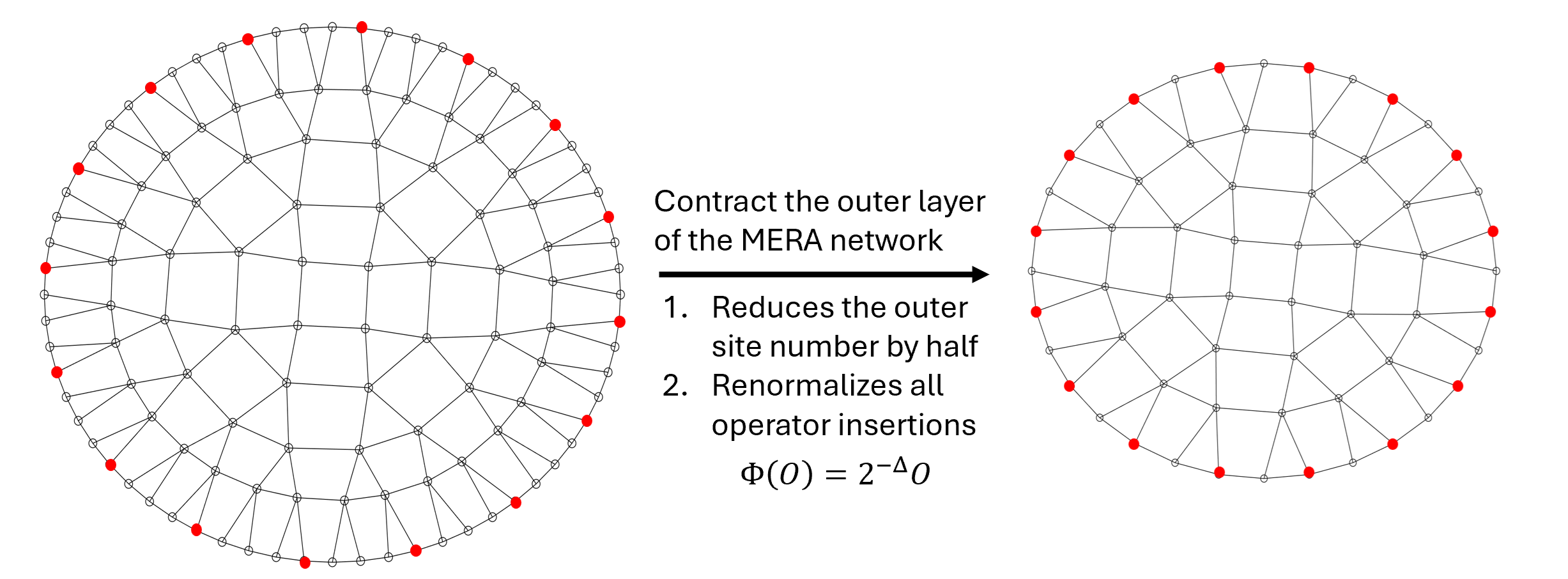}
    \caption{Left: the MERA computing the expectation value of the heavy shell operator, made out of primaries (red dots) with scaling dimension $\Delta$ inserted every $2^b$ sites in the outer layer of the MERA, which has $2^a$ sites. Right: Contracting the outermost layer of the MERA, the operator insertions are renormalized by a factor of $2^{-\Delta}$ and the number of sites in the outermost layer of the MERA is cut in half.}
    \label{fig:mera_shell}
\end{figure}

Here we present a more detailed model of the closed universe tensor network using MERA and the recent TMERA ansatz~\cite{sewell_tmera_2022}, which extends MERA from ground states to thermal states. To start, assume, as in~\cite{sewell_tmera_2022}, that the ground state has a MERA representation, as shown schematically in Figure~\ref{fig:mera_shell}. For concreteness, the RG scheme is a $2\to 1$ coarse-graining. 

The heavy operator shell is modeled as a placement of a primary operator of dimension $\Delta $ every $2^b$ sites. The UV layer of the MERA has $2^a$ sites. The continuum limit corresponds to $a\to \infty$ with $a-b$ fixed. Using the AdS/CFT mass-dimension relation, the total mass of the shell is 
\begin{equation}
    m\ell  \sim 2^{a-b} \Delta.
\end{equation}
The operator expectation value is obtained by sandwiching this product of local operators between a MERA bra and ket. The left side of Figure~\ref{fig:mera_shell} should be interpreted as the bra and ket folded onto each other, with the red operator insertions placed at the edge, corresponding to the UV layer.

This UV presentation is a perfectly valid representation of the expectation value, but we can further simplify the network by contracting the most UV layers. Since the operator insertions are primary, they have a very simple behavior under the contraction of one layer. In particular, they are mapped back to themselves at the next layer down, up to a multiplicative factor of $2^{-\Delta}$. This contraction can be safely continued until the operator insertions are adjacent to each other at the lattice scale. At this point, we can of course continue to contract the MERA, but the resulting network now has an increasingly complicated (and non-product) operator insertion at its outer layer, so there is a simple geometric description only if the process stops when the insertions are adjacent at the lattice scale.

At what scale does this occur? The spacing between insertions decreases by a factor of $2$ after each contraction of the current outer layer, so after $b$ layers are contracted, the insertions are adjacent to each other. Starting from layer $a$, the UV layer, the operator shell reaches layer $a-b$ before the process is halted. This corresponds to a scale in lattice units of $2^{a-b} \sim m\ell $.

Although we considered the case of one boundary spatial dimension for the above description, the result can easily be generalized to higher dimensions. Now the mass of the shell is
\begin{equation}
    m\ell \sim (2^{a-b})^{d-1} \Delta
\end{equation}
where $d-1$ is the spatial dimensionality of the CFT. The number of layers that can be contracted is still $b$, so the final length scale of the shell is
\begin{equation}
    2^{a-b} \sim (m\ell)^{\frac{1}{d-1}}.
\end{equation}

If we interpret the remaining uncontracted layers of the bra and ket MERA networks as the two AdS balls forming the closed universe, then we see they are glued at their boundary by a dense operator insertion. Moreover, the physical scale at which the gluing takes place precisely matches the mass-dependence of $R_*$ in \eqref{eq:Rstar}. Thus, we see that a very simple tensor network model of the CFT ground state and the operator shell insertion immediately produces a spatial geometry for the closed universe that qualitatively matches the semiclassical gravitational picture.

To form the bulk state $\ket{\Op_{\co}}$ alone, we view the isometries of the MERA network as unitaries with one bulk dangling leg. In more detail, suppose we want to prepare the two-sided CFT state \eqref{eq:microgluing},
\be \label{eq:AppendixCFTGluing}
 \ket{\Psi_{\mathsf{LR}}} \,\propto\, \bra{\Op_{BC}}\left(\ket{\Psi_{\mathsf{L}B}} \ket{\Psi'_{C\mathsf{R}}}\right)\,,
\ee
where $\ket{\Op_{BC}} = \Op_C\ket{\text{MAX}_{BC}}$ and there are corresponding bulk states mapped to the boundary via holographic isometries as $\ket{\Psi_{\mathsf{L}B}} = V_{\mathsf{L}B}\ket{\psi_{\mathsf{l}b}}$ and $\ket{\Psi'_{C\mathsf{R}}} = V_{C\mathsf{R}}\ket{\psi'_{c\mathsf{r}}}$. The bra $\bra{\Op_{BC}}$ glues the $B$ and $C$ CFT boundaries via a shell insertion. This insertion can be pushed down from the boundary to a radial position $R_{\star}$ using the structure of the tensor network, just like we described above. The bulk dangling legs that end up to the exterior of the shell are contracted between the bulk states in $b$ and $c$, with $b$ and $c$ denoting each of the AdS balls glued by the shell. To be more precise, divide the bulk EFT Hilbert space as $\mathcal{H}_b = \mathcal{H}_{b, \text{in}} \otimes \mathcal{H}_{b, \text{out}}$, where in legs have $r < R_{\star}$ and out legs have $r > R_{\star}$. Divide also the isometry as $V_B = V_B^{\text{(out)}} V_B^{\text{(in)}}$, with $V_B^{\text{(in)}}: \mathcal{H}_{b, \text{in}} \rightarrow \widetilde{\mathcal{H}}_{B}$ and $V_B^{\text{(out)}}:  \widetilde{\mathcal{H}}_{B} \otimes\mathcal{H}_{b, \text{out}} \rightarrow \mathcal{H}_{B}$ (analogous statements apply for $\mathcal{H}_c$). Pushing through the tensor network,
\begin{equation}
\bra{\mathcal{O}_{BC}} V^{\text{(out)}}_{B} V^{\text{(out)}}_{C} = \bra*{\tilde{\mathcal{O}}_{BC}} \bra*{\text{MAX}_{bc}^{\text{(out)}}} \, .
\end{equation}
Starting from \eqref{eq:AppendixCFTGluing}, we finally get a bulk to boundary map that still has the form \eqref{eq:bulktobdy}, 
\be 
 \ket{\Psi_{\mathsf{LR}}} \;\propto\; V_\mathsf{L} V_\mathsf{R} \bra{\Op_{\co}} \ket{\psi_{\mathsf{clr}}}\,,
\ee
where now
\begin{equation}
\ket{\mathcal{O}_{\co}} = V^{\text{(in)} \dagger}_{B} V^{\text{(in)} \dagger}_{C} \ket*{\tilde{\mathcal{O}}_{BC}} \, , \qquad \ket{\psi_{\mathsf{clr}}} = \bra*{\text{MAX}_{bc}^{\text{(out)}}} ( \ket{\psi_{\mathsf{lb}}} { \ket{\psi'_{c\mathsf{r}}}}) \, .
\end{equation}
Figure \ref{fig:TNMERA} shows this bulk state $\ket{\mathcal{O}_{\co}}$ (blue legs) by applying the isometry to $\ket*{\tilde{\mathcal{O}}_{BC}}$. 

Bulk entanglement can also be included following the TMERA prescription, which was shown to give a very efficient and accurate representation of the Gibbs state for the simple Ising CFT. One takes the closed universe tensor network with dangling bulk legs described above and adjoins two additional copies of the MERA network, also with dangling legs. These additional MERA copies represent the $\mathsf{l}$ and $\mathsf{r}$ AdS regions. Then we entangle the dangling bulk legs with thermal factors exactly as in TMERA. The resulting entanglement structure is precisely as shown in Figure~\ref{fig:setup}.

%%%%%%%%%%%%%%%%%%%%%%%%%%%%%
\section{Closed universe with a geometric bottleneck}
\label{app:UniverseBottleneck}
%%%%%%%%%%%%%%%%%%%%%%%%%%%%

Using multiple thin shells, we can construct closed universes that have geometric features not present in the ones discussed in the main text. In this section, we will study states with two shells that have a geometric bottleneck (a minimal area surface) in their time-reflection symmetric slice, showing how this affects the encoding map to the CFT. We work in the heavy-shell limit throughout ($G m / \ell^{d-2} \gg 1$). We start from a state in six copies of the CFT (labeled $\mathsf{L}, B, N_1, N_2, C, \mathsf{R}$) and glue them with shell operators to define a two-sided state as
\begin{equation}
\label{eq:BottleneckState}
\ket{\Psi} \equiv \ket{\Psi_{\mathsf{L R}}} = ( \bra{\mathcal{O}_{BN_1}} \bra{\mathcal{S}_{N_2 C}} ) (\ket{\Psi_{\mathsf{L}B}} \ket{\Phi_{N_1 N_2}} \ket{\Psi'_{C \mathsf{R}}}) \, ,
\end{equation}
where $\ket{\mathcal{O}_{BN_1}} = \mathcal{O}^{\dagger}_{N_1} \ket{\text{MAX}_{B N_{1}}}$ and $\ket{\mathcal{S}_{N_2 C}} = \mathcal{S}^{\dagger}_{C} \ket{\text{MAX}_{N_{2} C}}$, and we allow the shells to be different (e.g., by having different masses or, in CFT terms, different number of insertions). Much like in the main text, $\ket{\Psi_{\mathsf{L}B}}$ and $\ket{\Psi'_{C \mathsf{R}}}$ are states dual to two disconnected copies of AdS that share possibly entangled bulk states $\ket{\psi_{\mathsf{l} b}}$ and $\ket{\psi'_{c \mathsf{r}}}$, the canonical example being thermofield doubles below the Hawking-Page temperature. On the contrary, $\ket{\Phi_{N_1 N_2}}$ is a state dual to a two-sided black hole geometry. In the following, we will for concreteness work with a thermofield double state above the Hawking-Page transition, $\ket{\Phi_{N_1 N_2}} = \ket{\text{TFD}_{\beta_{\mathsf{N}}/2}}$, but other states with a similar dual work analogously (in particular, we could allow matter insertions in the preparation).

\begin{figure}[t]
    \centering
    \includegraphics[width=\linewidth]{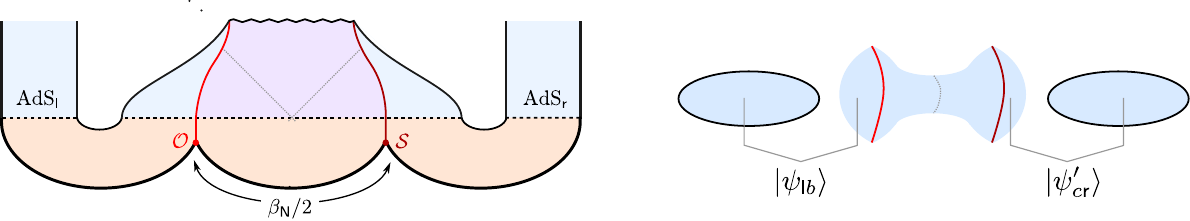} 
    \caption{\textbf{Left}: Euclidean preparation and Lorentzian evolution of the two-shell states with a bottleneck. \textbf{Right}: Geometry of the time-reflection symmetric slice.}
    \label{fig:BottleneckGeometry}
\end{figure}

The Euclidean geometry preparing these states and its Lorentzian evolution is shown in Figure \ref{fig:BottleneckGeometry}. The time-reflection symmetric slice has two hyperbolic balls that evolve into ${\rm AdS}_{\mathsf{l}}$ and ${\rm AdS}_{\mathsf{r}}$, together with a disconnected closed universe. This closed universe is supported by the two shells, it has a geometric bottleneck (i.e., a minimal $\mathbf{S}^{d-1}$ between the shells) and upon evolution the shells collapse to form a black hole.

\begin{figure}[b]
    \centering
    \includegraphics[width=0.6\linewidth]{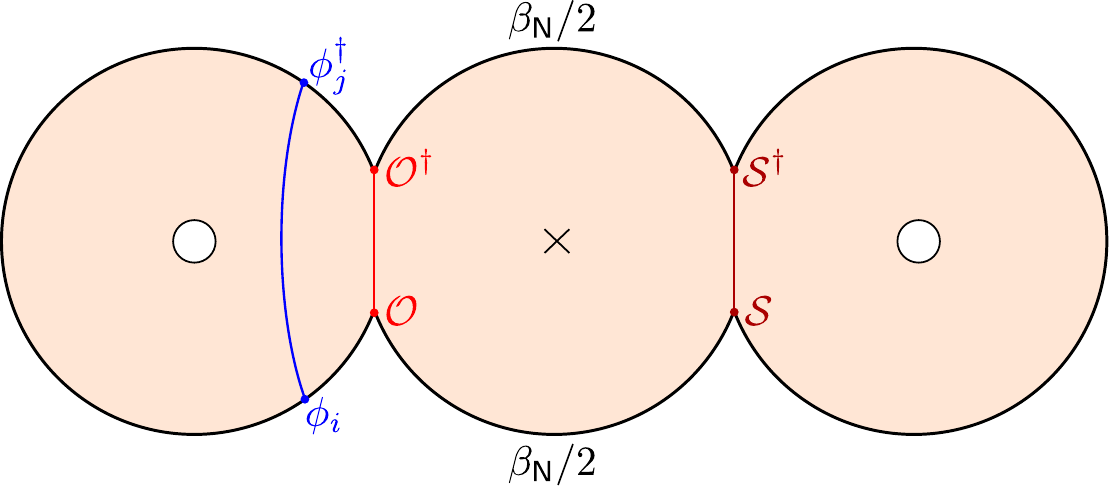} 
    \caption{Saddle computing the overlap $\overline{G_{ji}}$.}
    \label{fig:BottleneckOverlap}
\end{figure}

\begin{figure}[t]
    \centering
    \includegraphics[width=0.9\linewidth]{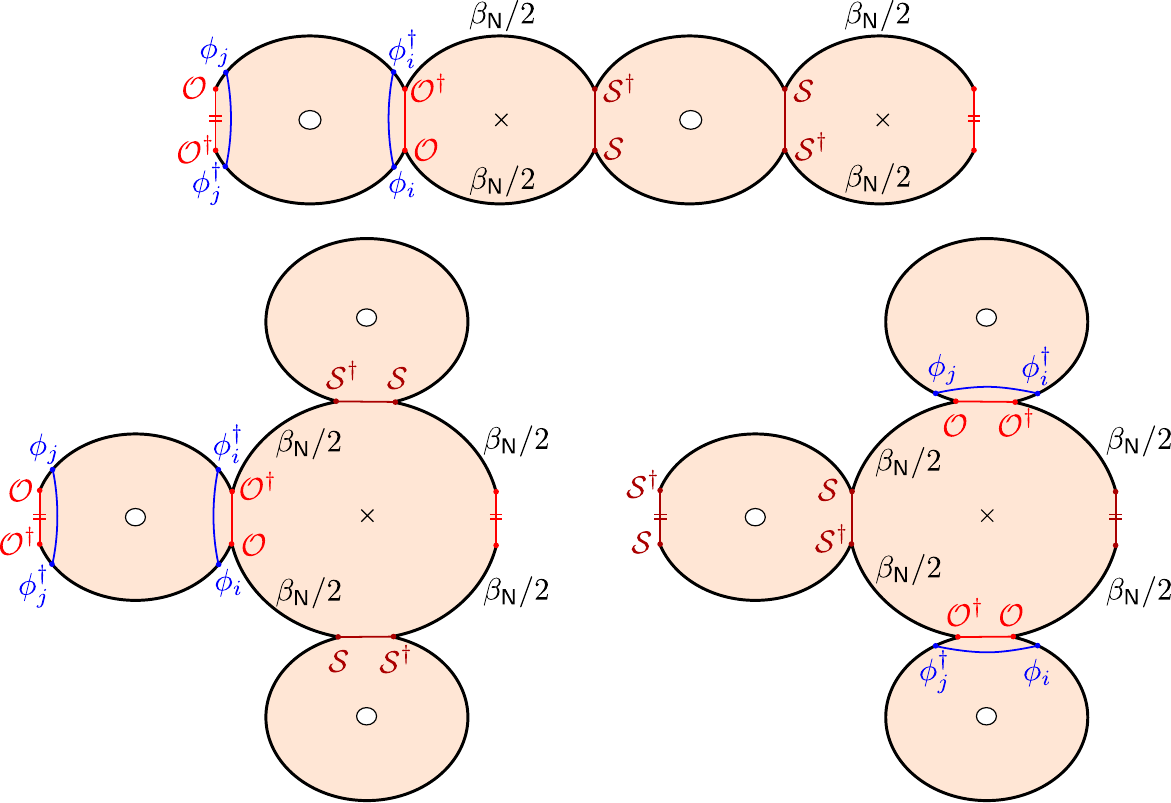} 
    \caption{The three two-boundary wormholes contributing to \eqref{eq:BottleneckConnectedSquare}. The blue geodesics have always the same length in the heavy shell limit in which the red shells pinch off.}
    \label{fig:BottleneckWormholes}
\end{figure}

Replicating the construction in Section \ref{sec:entanglobs}, we can consider a family of $K$ states by including heavy matter particles that cut the closed universe to the left of both shells (i.e., inserted appropriately in the preparation of $\ket{\Psi_{\mathsf{L} B}}$). We denote these states by $\ket{\Psi_i}$, and their microscopic overlaps are $G_{ji} = \bra{\Psi_j}\ket{\Psi_i}$. The gravitational path integral gives us coarse-grained information about these overlaps, e.g.,
\begin{equation}
\label{eq:BottleneckOverlap}
\overline{G_{ji}} = \delta_{ij} \bra{\psi_{\mathsf{l} b}} \ket{\psi_{\mathsf{l}b}} \bra{\psi'_{c \mathsf{r}}} \ket{\psi'_{c \mathsf{r}}} Z(\beta_{\mathsf{N}}) Z_{\mathcal{O}} Z_{\mathcal{S}} e^{- m_i L_i} \, ,
\end{equation}
where we are choosing the insertions to have orthogonal states in the bulk EFT. The saddle computing $\overline{G_{ji}}$ is shown in Figure \ref{fig:BottleneckOverlap}, the quantities involved are the same as in \eqref{eq:OverlapsSingleShell} with the extra ingredient $Z(\beta_{\mathsf{N}})$ coming from the middle disk with the Euclidean black hole geometry. The square of the overlaps has now, in addition to the disconnected contribution, three other saddle point geometries depending on how the thin shells are contracted. They are sketched in Figure \ref{fig:BottleneckWormholes}, producing as a result
\begin{align}
\label{eq:BottleneckConnectedSquare}
\nonumber \left. \overline{G_{ji} G_{ij}} \right|_{\rm conn.} = & {\rm Tr}(\rho_{\mathsf{l}}^2) {\rm Tr}(\rho_{\mathsf{r}}^2) Z(\beta_{\mathsf{N}})^2 Z^2_{\mathcal{O}} Z^2_{\mathcal{S}} e^{- m_i L_i - m_j L_j} + {\rm Tr}(\rho_{\mathsf{l}}^2) {\rm Tr}(\rho_{\mathsf{r}})^2 Z(2 \beta_{\mathsf{N}}) Z^2_{\mathcal{O}} Z^2_{\mathcal{S}} e^{- m_i L_i - m_j L_j} \\
& + \delta_{ij} {\rm Tr}(\rho_{\mathsf{l}})^2 {\rm Tr}(\rho_{\mathsf{r}}^2) Z(2 \beta_{\mathsf{N}}) Z^2_{\mathcal{O}} Z^2_{\mathcal{S}} e^{- m_i L_i - m_j L_j} \, .
\end{align}
The typical size of the variance of the inner products is thus controlled by the competition between the various terms,
\begin{align}
\nonumber \frac{\overline{\left|G_{ji} - \overline{G_{ji}} \right|^2}}{\overline{G_{ii}} \, \overline{G_{jj}}} & = \frac{{\rm Tr}(\rho_{\mathsf{l}}^2)}{{\rm Tr}(\rho_{\mathsf{l}})^2} \frac{{\rm Tr}(\rho_{\mathsf{r}}^2)}{{\rm Tr}(\rho_{\mathsf{r}})^2} + \frac{{\rm Tr}(\rho_{\mathsf{l}}^2)}{{\rm Tr}(\rho_{\mathsf{l}})^2} \frac{Z(2 \beta_{\mathsf{N}})}{Z(\beta_{\mathsf{N}})^2} + \delta_{ij} \frac{{\rm Tr}(\rho_{\mathsf{r}}^2)}{{\rm Tr}(\rho_{\mathsf{r}})^2} \frac{Z(2 \beta_{\mathsf{N}})}{Z(\beta_{\mathsf{N}})^2} \\
& = e^{-S_2(\rho_{\mathsf{l}}) - S_2(\rho_{\mathsf{r}})} + e^{-S_2(\mathsf{\rho_l}) - S_2(\beta_{\mathsf{N}})} + \delta_{ij} e^{-S_2(\mathsf{\rho_r}) - S_2(\beta_{\mathsf{N}})} \, ,
\end{align}
where, in addition to the second Rényi entropy of the gases, $S_2(\beta_{\mathsf{N}})$ is the second Rényi entropy of the thermal state at inverse temperature $\beta_{\mathsf{N}}$. Assuming $Z(2 \beta_{\mathsf{N}})$ is also in the black hole phase, we can estimate it using bulk black hole entropies to be $e^{-S_2(\beta_{\mathsf{N}})} \approx e^{- c_d S(\beta_{\mathsf{N}})}$, where $S(\beta_{\mathsf{N}})$ is the thermal black hole entropy and $c_d$ is an order $1$ dimension-dependent coefficient.

A quick way to determine the size of the Hilbert space in the closed universe that we can access from the CFT states above is to entangle them with an external auxiliary system and compute the second Rényi entropy of this external system. We will find a Page curve-like behavior that allows us to estimate the size of the Hilbert space we are entangling the external system with.\footnote{An alternative would be to replicate the resolvent computation presented in \cite{Antonini:2023hdh} for the two-shell cosmologies.} Consider then the entangled state
\begin{equation}
\ket{\chi} = \frac{1}{\sqrt{K}} \sum_{i=1}^K \frac{\ket{\Psi_i} \ket{i}}{(\overline{G_{ii}})^{1/2}} \, ,
\end{equation}
where the $\ket{i}$ are $K$ orthogonal states of the auxiliary system and we are using that the fluctuations of the norm of the CFT states are suppressed to normalize with the average (we assume for this that at least one of $S(\rho_{\mathsf{l}})$ or $S(\rho_{\mathsf{r}})$ is large). Computing the second Rényi of the auxiliary system we get, upon coarse-graining and using \eqref{eq:BottleneckOverlap} and \eqref{eq:BottleneckConnectedSquare}:
\begin{equation}
\overline{e^{-S_2(\chi)}} = \frac{1}{K^2} \sum_{i,j = 1}^K \frac{\overline{G_{ji} G_{ij}}}{\overline{G_{ii}} \, \overline{G_{jj}}} = \frac{1}{K} + e^{-S_2(\rho_{\mathsf{l}}) - S_2(\rho_{\mathsf{r}})} + e^{-S_2(\mathsf{\rho_l}) - S_2(\beta_{\mathsf{N}})} + \frac{1}{K} e^{-S_2(\mathsf{\rho_r}) - S_2(\beta_{\mathsf{N}})} \, .
\end{equation}
If we try to increase $K$ without bound, the second Rényi of the auxiliary system is limited by the minimum between $S_2(\rho_{\mathsf{l}}) + S_2(\rho_{\mathsf{r}})$ and $S_2(\rho_{\mathsf{l}}) + S_2(\beta_{\mathsf{N}})$. This is signaling that the size of the Hilbert space of CFT states $\mathcal{H}_K$ that we construct is
\begin{equation}
\log \, {\rm dim} \, \mathcal{H}_K \approx \min \left( \log K, S(\rho_\mathsf{l}) +S(\rho_{\mathsf{r}}), S(\rho_{\mathsf{l}}) +S(\beta_{\mathsf{N}})  \right) \, .
\end{equation}

This result shows how geometric features of the closed universe can affect the encoding of the states in the CFT by means of the entanglement with the external gases. To interpret it, let us first assume we decouple the left AdS, i.e., $S(\rho_{\mathsf{l}}) = 0$. The matter states we are creating are to the left of the bottleneck. Furthermore, to map to the right CFT we have to do it through the entanglement with the right AdS gas, which in the heavy shell limit is entangled mainly with degrees of freedom to the right of both shells. We thus have to overcome two bottlenecks: the geometric one (of size $S(\beta_{\mathsf{N}})$) and the one set by the entanglement $S(\rho_{\mathsf{r}})$. This explains the origin of both terms when $S(\rho_{\mathsf{l}}) = 0$. If $S(\rho_{\mathsf{l}}) \neq 0$, we are just adding to both cases above the possibility to map to the left CFT via the entanglement in the state $\rho_{\mathsf{l}}$. The two-shell cosmologies with a bottleneck provide in this way a useful model to reproduce many of the results obtained in 2d in \cite{Harlow:2025pvj}, giving a more transparent origin to the geometric suppression by $S(\beta_{\mathsf{N}})$ (equivalent to $S_0$ in the 2d models). That said, one should keep in mind that in higher dimensions it seems we are not allowed to have $S(\rho_{\mathsf{r}}) \gtrsim S(\beta_{\mathsf{N}})$, and thus keeping both terms above is in a sense futile. This is because in the black hole phase $S(\beta_{\mathsf{N}}) \sim \mathcal{O}(N^2)$, while as discussed in Section \ref{sec:2} the gases have at most $S(\rho_{\mathsf{l}}) \sim \mathcal{O}(N^{18/17})$.\footnote{Had we built the bottleneck with a small black hole by using some sort of microcanonical preparation, the gas should still have less entropy: the smallest stable black hole has larger entropy than the most entropic stable gas.}

A final interesting remark that can be done in the bottleneck cosmologies has to do with the encoding of the degrees of freedom in one side or the other, i.e., with the entanglement wedge of each CFT. Consider the state $\ket{\Psi}$ in \eqref{eq:BottleneckState}, now without matter insertions. In the language of observers and clones of \cite{Harlow:2025pvj}, we can consider the left gas (isometrically encoded in the standard way in the left CFT) as the clone $\mathsf{Ob}'$ of the corresponding observer entangled degrees of freedom in the cosmology, $\mathsf{Ob}$. Similarly, we can consider the right gas (encoded in the right CFT) as an auxiliary matter system $\mathsf{M}'$ entangled with some matter $\mathsf{M}$ in the closed universe (note the different interpretation we are now putting forward; previously both the right and left gases were interpreted as $\mathsf{Ob}'$). We can ask about the quantum extremal surface corresponding to the left CFT or, equivalently, to the clone of the observer. There are three candidates: one empty surface that cuts the entanglement between the left AdS and the cosmology, another empty surface that cuts between the closed universe and the right AdS, and the minimal surface that provides the bottleneck. We thus conclude that
\begin{equation}
S(\rho_{\mathsf{L}}) = {\rm min} \left( S(\rho_{\mathsf{l}}), S(\rho_{\mathsf{r}}), S(\beta_{\mathsf{N}}) \right) \, .
\end{equation}
where $\rho_{\mathsf{L}}$ is the state of the left CFT. When each of these terms dominates, the entanglement wedge of the left CFT respectively includes no portion of the cosmology, the whole closed universe, or just the left half of the closed universe up to the extremal surface (in addition to the left AdS in all cases). As emphasized before, $S(\beta_{\mathsf{N}})$ is much larger than the entropy of the gases, in which case the formula reduces to
\begin{equation}
S(\rho_{\mathsf{L}}) = {\rm min} \left( S(\rho_{\mathsf{l}}), S(\rho_{\mathsf{r}}) \right) \, .
\end{equation}
This is also the result one gets when no bottleneck is present.

%%%%%%%%%%%%%%%%%%%%%%%%%
\section{Gravitational argument { for the} final state wavefunction} 
\label{app:gravargument}
%%%%%%%%%%%%%%%%%%%%%%%

We now sketch a formal derivation of \eqref{eq:dictionary} in canonical quantum gravity. There are many issues with trying to make sense of this approach, which we will sweep under the rug in this appendix, and leave the clarification for future work.

In order to make the argument, we consider a closed $d$-dimensional manifold $\Sigma$ with Riemannian metric $h_{ij}(\bf{x})$ and matter fields $\phi_I(\bf{x})$, where we will drop the position label $\mathbf{x}$ from now on. We consider the closed universe wavefunctional $\Psi(h_{ij},\phi_I)$. The momentum constraint imposes the invariance of $\Psi(h_{ij},\phi_I)$ under diffeomorphisms on $\Sigma$. On the other hand, the Hamiltonian constraint imposes the invariance of the wavefunction under temporal diffeomorphisms
\be 
\Bigl[\dfrac{4\kappa^2}{\sqrt{h}}\,\left(\pi_{ij}\pi^{ij}- \dfrac{1}{d-1}\pi^{i}_i\pi^{j}_j\right) - \dfrac{\sqrt{h}}{4\kappa^2}(R(h) - 2\Lambda { -} 4\kappa^2 {T}_{00})
\Bigr]\;
\Psi[h_{ij},{ \phi_I}]
\;=\;0\,,
\ee
where $\kappa^2 = 4\pi G_N$ and $2\Lambda \ell^2 = -d(d-1)$. Here $\pi^{ij} = -\mathsf{i}\delta/\delta h_{ij}$ is conjugate momentum, and $T_{00}$ is the matter energy density.

\begin{figure}[h]
    \centering
    \includegraphics[width=.8\linewidth]{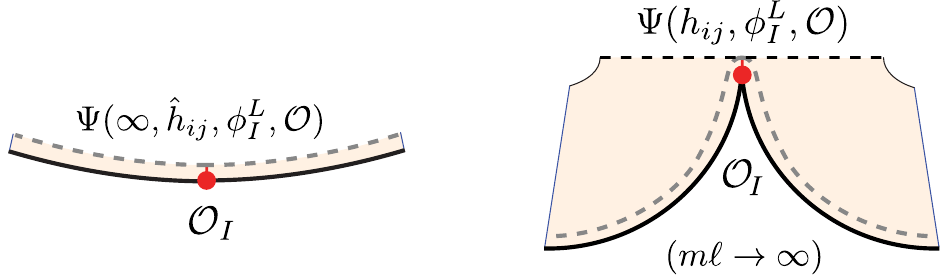} 
    \caption{On the left, the closed universe final state wavefunction for a large universe can be defined in terms of the matrix elements of the heavy matter operator computed by the CFT path integral. On the right, the wavefunction can be extended to the closed universes of this paper in the heavy shell limit, where the heavy matter operator stays in the asymptotic region.}
    \label{fig:la}
\end{figure}

We now divide $h_{ij} = { a^2} \,\hat{h}_{ij}$, where $a$ is some scale function and $\hat{h}_{ij}$ is a fixed volume metric. When the universe is very large, the constraints are solved by a wavefunctional of WKB form
\be 
\Psi \sim e^{-I_E}\,,
\ee 
where $I_E$ is the Euclidean action of the solution which connects the slice of the universe to the asymptotic Euclidean { boundary} where the Euclidean CFT lives. The solution has the form\footnote{Essentially \eqref{eq:solutionwhasym} is the metric that locally describes the asymptotic part of the Euclidean wormhole.}
\be\label{eq:solutionwhasym} 
\text{d}s^2 = \ell^2 \dfrac{\text{d}a^2}{a^2} + a^2 \left(\ell^2 \text{d}\rho^2 + r(\rho)^2 \text{d}\Omega_{d-1}^2\right)\,,
\ee 
where $r(\rho)$ is given by \eqref{eq:solution2} for some fixed $\rho_0$. In this limit, the Euclidean action $I_E$ has local divergences that can be renormalized with suitable counterterms using a conventional scheme of holographic renormalization. The renormalized final state wavefunction is then equal to the CFT  path integral
\be\label{eq:universelarge}
\lim_{a \rightarrow \infty}\Psi(a, \hat{h}_{ij},\phi_I)_{\text{ren}} = Z_{\text{CFT}}(\hat{h}_{ij},\phi^L_I, \mathcal{O}) \;\propto\; \mathcal{O}_I\,,
\ee 
where in the second step we have divided the bulk matter into light fields $\phi^L_I$ and the heavy shell $\mathcal{O}$. In the last definition, we have used that this CFT path integral computes the light matrix elements of the microscopic shell operator $\mathcal{O}$, up to proportionality factors arising from the conformal frame. We are thinking of the microscopic heavy operator $\mathcal{O}$ as being fixed, and looking at this wavefunctional as a function of the { light} field configuration $\phi^L_I$. This provides a proper non-perturbative definition of the closed universe wavefunction using AdS/CFT, only valid when the closed universe hugs the Euclidean asymptotic boundary, as shown in Figure \ref{fig:la}. However, from the gravity side, evaluating such a wavefunction $\mathcal{O}_I$ with the gravitational path integral by filling in the bulk will require some microscopic ingredient beyond the semiclassical description of the thin shell and it is not obvious that such a level of microscopic detail should be accessible semiclassically \cite{saad2021wormholesaveraging}.

Essentially, the argument is that in the limit of large mass $m\ell \rightarrow \infty$ for the heavy operator, the situation is equivalent to having a very large universe, where one can use \eqref{eq:universelarge} and define the final state wavefunction directly from CFT data. The matter that defines the background becomes so close to the asymptotic region that we can use this limit to also define the wavefunction at finite $a$, at the same level. Even if the rest of the universe also gravitates, it corresponds to a conventional vacuum AdS space, and all the microscopic data has been isolated to the thin shell close to the boundary. Thus, in this limit, we recover
\be 
\lim_{m\ell \rightarrow \infty }\Psi(h_{ij}, \phi_I,\mathcal{O})_{\text{ren}} \;\propto\; \mathcal{O}_I\,,
\ee 
which is essentially \eqref{eq:dictionary} from another perspective. Recall the importance of the heavy shell limit, which avoids specifying the wavefunction by evaluating the gravitational path integral for a fixed $\mathcal{O}$ operator; instead, it relates the wavefunction to a non-perturbatively defined object in the Euclidean CFT.

\bibliographystyle{ourbst}
\bibliography{bibliography,old_ass_bib}

\end{document}